\documentclass[useAMS,usenatbib]{mn2e}
\usepackage{graphicx}
\usepackage{color}
\usepackage{amsmath}
\usepackage{lscape}
\usepackage{pdflscape}
\usepackage{rotating}
\usepackage{amssymb}

\usepackage[T1]{fontenc} 
\usepackage{aecompl} 

\def\lsimeq{\hbox{\raise0.5ex\hbox{$<\lower1.06ex\hbox{$\kern-1.07em{\sim}$}$}}} 
\def\gsimeq{\hbox{\raise0.5ex\hbox{$>\lower1.06ex\hbox{$\kern-1.07em{\sim}$}$}}} 

\title[Ionized outflows in obscured QSOs]{X-shooter reveals powerful outflows in z$\sim 1.5$
  X-ray selected obscured quasi stellar objects}

\author[M. Brusa et al.]
{\parbox{\textwidth}{M. Brusa$^{{1},{2},{3}}$\thanks{E-mail:marcella.brusa3@unibo.it}, 
A. Bongiorno$^{4}$,
G. Cresci$^{5}$,
M. Perna$^{{1},{3}}$, 
 A. Marconi$^{6}$,
 V. Mainieri$^{7}$, 
 R. Maiolino$^{8}$,
 M. Salvato$^{2}$,
 E. Lusso$^{9,5}$,
 P. Santini$^{4}$, 
 A. Comastri$^{3},$ 
 F. Fiore$^{4}$, 
 R. Gilli$^{3}$, 
 F. La Franca$^{10}$,
 G. Lanzuisi$^{3,1}$,
 D. Lutz$^{2}$, 
 A. Merloni$^{2}$, 
 M. Mignoli$^{3}$,
 F. Onori$^{10}$,
 E. Piconcelli$^{4}$, 
 D. Rosario$^{2}$,
 C. Vignali$^{{1},{3}}$,  
 G. Zamorani$^{3}$
}
\vspace{0.4cm}\\
\parbox{\textwidth}{$^{1}$ Dipartimento di Fisica e Astronomia, Universit\`a di Bologna,
  viale Berti Pichat 6/2, 40127 Bologna, Italy \\
$^{2}$ Max Planck Institut f\"ur Extraterrestrische Physik, Giessenbachstrasse 1, 85748 Garching bei M\"unchen, Germany \\
$^{3}$ INAF - Osservatorio Astronomico di Bologna, via Ranzani 1, 40127 Bologna, Italy \\
$^{4}$ INAF - Osservatorio Astronomico di Roma, via Frascati 33,   00044 Monte Porzio Catone (RM) Italy \\
$^{5}$ INAF - Osservatorio Astronomico di Arcetri, Largo Enrico Fermi 5, 50125 Firenze, Italy \\
$^{6}$ Dipartimento di Astronomia e Scienza dello Spazio, Universit\`a degli Studi di Firenze, Largo E. Fermi 2, 50125 Firenze, Italy \\
$^{7}$ European Southern Observatory, Karl-Schwarzschild-str. 2,  85748 Garching bei M\"unchen, Germany \\
$^{8}$ Cavendish Laboratory, University of Cambridge, 19 J. J. Thomson Ave., Cambridge CB3 0HE, UK\\
$^{9}$ Max Planck Institut f\"ur Astronomie, K\"onigstuhl 17, D-69117, Heidelberg, Germany \\
$^{10}$ Dipartimento di Matematica e Fisica, Universit\`a degli Studi `Roma Tre', Via della Vasca Navale 84, I-00146 Roma, Italy\\
}}

\begin{document}

\date{Accepted 1988 December 15. Received 1988 December 14; in original form 1988 October 11}

\pagerange{\pageref{firstpage}--\pageref{lastpage}} \pubyear{2002}

\maketitle

\label{firstpage}

\begin{abstract}

We present X-shooter@VLT observations of a sample of 10 luminous, X-ray obscured QSOs at z$\sim1.5$ from the XMM-COSMOS survey, expected to be caught in the transitioning phase from starburst to AGN dominated systems. The main selection criterion 
is  X-ray detection at bright fluxes (L$_{\rm X}\gsimeq10^{44}$ erg s$^{-1}$) coupled to red optical-to-NIR-to-MIR colors. 
Thanks to its large wavelength coverage,  X-shooter allowed us to determine accurate  redshifts from the presence of multiple emission lines for five out of six targets for which we had only a photometric redshift estimate, with a 80\% success rate, significantly larger than what is observed in similar programs of spectroscopic follow-up of red QSOs. 
We report the detection of broad and shifted components in the [OIII]$\lambda\lambda$5007,4959 complexes
for 6 out of 8 sources with these lines observable in regions free from strong atmospheric absorptions. The FWHM associated with the broad components are in the range FWHM$\sim900-1600$ km s$^{-1}$, larger than the average value observed in SDSS Type 2 AGN samples at similar observed [OIII] luminosity, but comparable to those observed for QSO/ULIRGs systems for which the presence of  kpc scale outflows have been revealed through IFU spectroscopy. 
Although the total outflow energetics (inferred under reasonable assumptions) may be consistent with winds accelerated by stellar processes, we favour an AGN origin for the outflows given the high outflow velocities oberved (v$>1000$ km s$^{-1}$) and the presence of strong winds also in objects undetected in the far infrared.

\end{abstract}


\section{Introduction}

Since the seminal discovery, 15 years ago, of the
presence of Super Massive Black Holes (SMBH, M$>10^6$ M$_\odot$) in the
nuclei of virtually all galaxies \citep{Magorrian1998}, it has been
realised that Active Galactic Nuclei (AGN) are not exotic phenomena
occurring in a small fraction of galaxies, but rather a key ingredient
of their formation and evolution. 

The `Soltan argument' \citep{Soltan1982} posits that most galaxies went through phases of
nuclear activity in the past, the remnants of which are the quiescent SMBH in z=0 galactic nuclei.
During such active phases, a strong physical coupling (generally termed `feedback') could have established a
long-lasting link between hosts and SMBHs, leading to
the well-known local scaling relations (e.g., \citealt{Ferrarese2000,Gebhardt2000,Gultekin2009}).

While among less luminous sources (L$_{\rm bol}<10^{45}$ erg s$^{-1}$), nuclear activity and  Star Formation (SF) can be regulated 
by local processes such as accretion triggered gas
inflows or disk instabilities likely induced by minor mergers or galaxies encounters (see \citealt{Ciotti2007,Bournaud2011,Cen2012}),  among the most luminous sources  (L$_{\rm bol}>10^{46}$ erg s$^{-1}$),
major galaxy mergers are indicated as the culprit for such physical
coupling, since they can efficiently funnel a large amount of gas
into the nuclear region to feed (and obscure) the accreting
SMBH \citep{Menci2008}. 

Some AGN-galaxy co-evolutionary models indeed postulate, for the QSO population, a ``three stages''  phase, altogether lasting $<<500$ Myr, triggered by the funneling of a large amount of gas into the nuclear region  (e.g. \citealt{Menci2008,Hopkins2008}). 
The first phase (i) is associated with rapid SMBH growth and efficient SF, in a dust-enshrouded, dense environment.  Shortly later, (ii) the accreting BH should experience the so-called
`feedback' or `blow-out' phase (see e.g. \citealt{Hopkins2008}), during which it releases radiative and kinetic energy in the form of powerful, outflowing winds. At this point the accretion on the SMBH is expected to be at its maximum (a blue un-absorbed type 1 AGN or a red obscured Type 2 AGN depending on the line of sight through the torus). 
When the accretion stops, (iii) the galaxy then evolves passively to the massive early type
systems we observe today in local galaxies, with a central quiescent SMBH.

Some key features are common to most model realizations.
In particular, powerful winds from all the gas components (neutral,
ionized and molecular) are expected in the
`blow-out' phase, and AGN feedback should reveal itself in outflowing
material (see \citealt{King2010} and \citealt{Fabian2012} for recent reviews). 
In addition, the AGN luminosity peak is expected to lag the SF peak: the winds can expel most of the gas in the host galaxy,  which is therefore not  available anymore to substain SF and, later, the BH growth, explaining the termination of the two processes. 
However, differences in the details of the various physical conditions (such as the gas content and  consumption timescales) coupled with different model assumptions (such as timescale of the processes, fraction of the energy released, lag between  SF and AGN activity) translate into the fact that the 
SF properties of QSOs experiencing the outflow phase may indeed scatter substantially in the different models (see e.g. \citealt{Lamastra2013,Hickox2014}). In addition, the degree of obscuration of the QSO is also related to the gas content and depends on the viewing angle with respect to the molecular torus surrounding the accretion disk, the amount of gas available in the host galaxy, and the timescale of the feedback process (see e.g. \citealt{Sansigre2006}).

\begin{figure*}
\includegraphics[angle=0,scale=0.43]{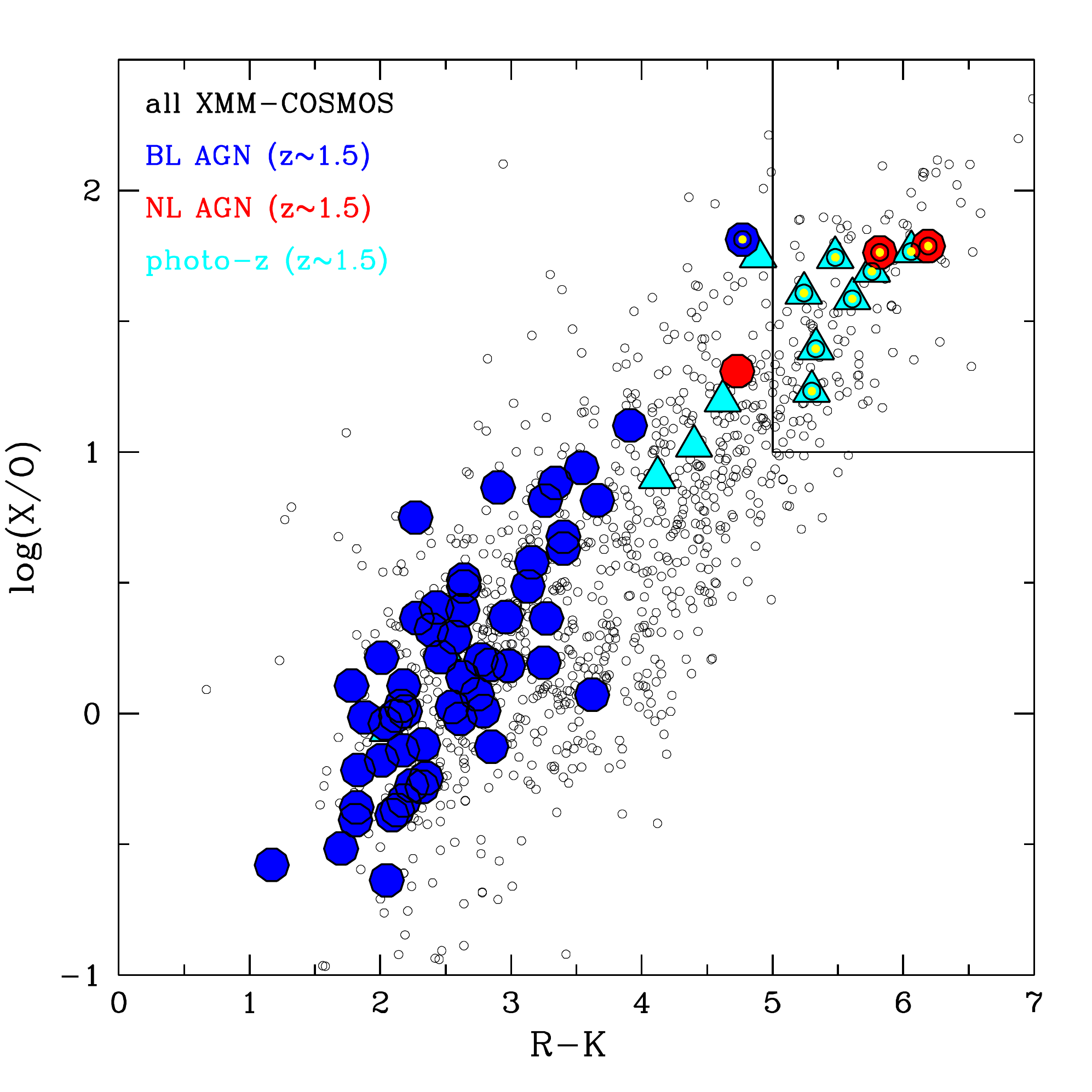}
\includegraphics[angle=0,scale=0.43]{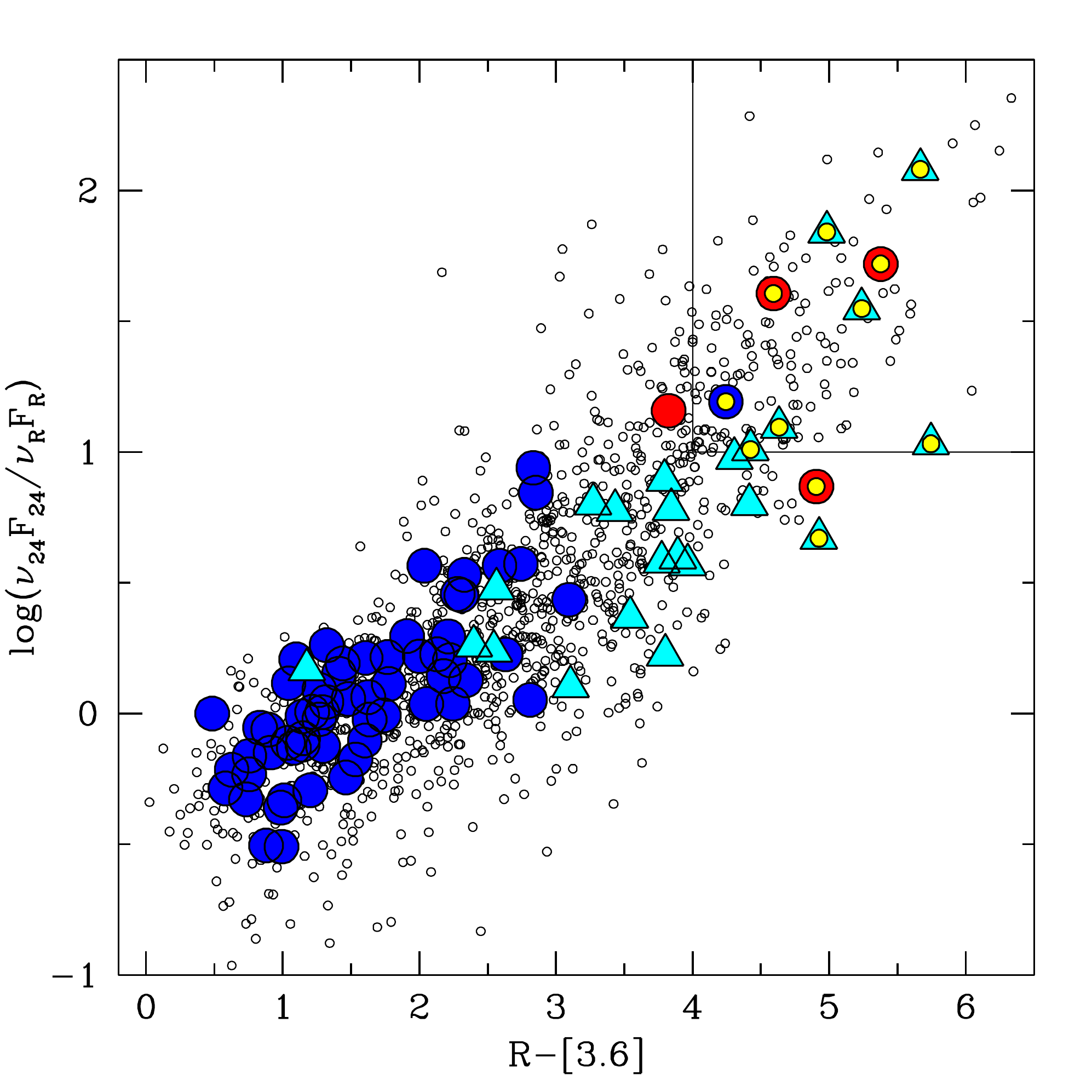}
\caption{Selection region for the X-shooter targets. (a): X--ray to
  optical flux ratio vs. R-K (Vega) color and (b) 24 $\mu$m to optical flux
  ratio vs. R-[3.6] color for all the XMM-COSMOS sources (small empty
  circles) and for those with redshift
  in the range z=1.25-1.72, K$<19$ and  F$_{\rm
    2-10,rest}>5\times10^{-15}$ erg cm$^{-2}$ s$^{-1}$ (large
  symbols).  Blue circles denote spectroscopically confirmed BL AGN,
  red circles spectroscopically confirmed NL AGN and cyan triangles objects
  with only photometric redshifts. The yellow circles are the obscured
  AGN candidates proposed for the X-shooter observations, e.g. sources with photometric redshifts which satisfy
  one (R-K$>4.5$ and X/O$>10$) or the other
  (log($\nu_{24}$F$_{24}$/$\nu_R$F$_R>$1 and R-[3.6]$>4$ ) selection criterion. 
\label{selection}}
\end{figure*}

From an observational point of view, the fact that vigorously star-forming galaxies (e.g. Ultra Luminous Infra Red galaxies, ULIRGs) are frequently associated with luminous, often obscured, quasars both in the local Universe and at high-z overall supports a coherent BH-SF growth (e.g. \citealt{Sanders1988, Alexander2005}),
although at moderate luminosity the coupling is more debated  \citep{Rosario2012,Page2012,Harrison2012a,Mullaney2012}.
Powerful outflows sustained by relativistic and collimated jets in the hosts of luminous radio-galaxies have been commonly observed out to z$\sim4$ (see e.g. \citealt{Nesvadba2008,Nesvadba2011,Fu2009}). 
Outflowing winds in radio-quiet objects, most likely to be radiatively driven, are instead less commonly observed. Only very recently spatially resolved optical,  far infrared and mm  spectroscopic studies convincingly showed evidences for the existence of such processes in the form of the predicted energetic outflows in  ULIRGs with Seyfert nuclei in the local Universe (e.g. \citealt{Feruglio2010,Fischer2010,Rupke2011,Villar2011a,Zaurin2013,Zhang2011,Westmoquette2012,Rupke2013, Harrison2014,McElroy2014};
see also \citealt{Elvis2000} for a discussion on the ubiquitous  presence of winds in the AGN structure). 

When moving to higher redshifts (z$>0.5-1$), several classes of objects (selected on the basis of well defined 
observed properties) have been proposed in the recent past as prototype 
of candidate sources in the outflowing phase.
For example, \citet{Lipari2006} first proposed that Broad Absorption Line Quasars (BAL QSOs, e.g. \citealt{Dunn2010}) may be objects in the transition phase between the ULIRGs and unobscured QSO phase, where the outflowing wind in the ionized component is seen directly in optical-UV spectra (e.g., \citealt{Dekool2001,Hamann2002}).  Extensive works in the past years uniquely contributed to our understanding of the winds physics, in terms of spatial location, ionization level and energetics involved (see, e.g., \citealt{Moe2009,Borguet2013} and \citealt{Arav2013}).
Recent works also suggest that  $\sim$40\% of  IR selected and SDSS QSOs are BAL QSOs,  pointing towards the fact that they are quite common \citep{Dai2008,Allen2011} and that the outflow and unobscured phases may be of comparable length, in agreement with the Hopkins scenario (see also \citealt{Glikman2012}). 
Moreover,  from the analysis of [OIII] luminosity matched obscured and unobscured samples of QSOs at z$\sim0.5$ presented in \citet{Liu2013} and \citet{Liu2014}, it has been proposed that high luminosities (L[OIII]$>10^{43}$ erg s$^{-1}$) are  characteristic of the peak of quasar feedback phase.

Given that the sources in the feedback phase are expected to be dusty and reddened (either by the host galaxy or the torus), another class of objects proposed to be in the transition phase are the ``red QSOs'', selected from large area, bright IR surveys such as 2MASS or UKIDSS, on the basis of a red color in the Near Infrared (NIR) band (e.g. J-K$\gsimeq2$) indicative of a steep SED due to obscured AGN \citep{Urrutia2009,Glikman2012,Banerji2012}. Detailed high resolution imaging, spectroscopy or morphological follow-up of carefully selected prototypes of this class of sources at various redshifts
convincingly favour such an interpretation (e.g. \citealt{Urrutia2012,Banerji2014}).  
Finally, another well studied class of objects are the submillimiter galaxies (SMG) associated with ULIRGs and luminous QSOs mentioned above. Only recently, with the advent of high resolution and sensitive NIR spectrographs, it has been possible to break the z$>1$ barrier and characterise the neutral and ionised kinematic components of  luminous high-z quasars by sampling the redshifted optical lines in the NIR band (e.g., \citealt{Alexander2010,CanoDiaz2012,Maiolino2012,Harrison2012}, hereinafter H12).

Luminous X-ray selected obscured AGN with red colors also represent optimal targets for objects where feedback from the AGN is expected to halt SF and to start `cleaning' out gas from the galaxy.  
Indeed, in the AGN evolutionary framework described above, the obscured phase of a quasar 
corresponds to a time when the BH is accreting mass very rapidly, implying that the SMBH should manifest itself as an X-ray quasar. 
In this respect, while most if not all of the studies at z$>1$ have been performed on QSO/ULIRGs and objects undergoing intense SF, a selection based on the X-ray emission offers the advantage of being independent of the SF properties, which as mentioned above cannot be predicted a priori for the QSOs in the feedback phase. 
In a previous work  based on XMM-Newton observations of the COSMOS field \citep{Scoville2007} we proposed a criterion to isolate such very rare objects, on the basis of their observed red colors and high X-ray to optical and/or mid-infrared to optical flux ratios (\citealt{Brusa2010}, hereinafter B10). The combined analysis of a high resolution Keck spectrum, morphology from HST/ACS (Advanced Camera for Survey) data and an accurate SED fitting of the prototype of this class of sources (XID 2028) convincingly showed that the proposed criteria appear robust in selecting luminous and obscured quasars in the ``blow-out'' phase discussed above (see Section 9 in B10).
In this paper we present the data reduction and analysis of X-shooter observations at the Very Large Telescope (VLT) of the 10 brightest obscured QSOs at z$\sim1.5$ in the XMM-COSMOS survey, and we will
focus on the detection of the broad and shifted components in [OIII] lines.  The measurement of the  BH masses from the same data are presented in a companion paper \citep{Bongiorno2014}.

The paper is organized as follows: 
Section 2 presents the sample selection and properties, Section 3 the X-shooter
observations and data reduction and Section 4 the data 
analysis and the results of the spectral fitting. Section 5 discusses the main result, i.e. the origin of the broad component and the outflows statistics. Section 6 discusses the energetic output associated with the outflow and finally we summarize our results and the implications in Section 7. In the appendix we also present the fit of the \citet{Urrutia2012} sample. 
All the rest frame wavelengths are given in the air, as quoted in 
http://www.sdss3.org/dr8/spectro/spectra.php.  Unless otherwise stated, uncertainties are quoted at the 68\% (1$\sigma$) confidence level.
Throughout the paper, we adopt the cosmological parameters $H_0=70$ km s$^{-1}$
Mpc$^{-1}$, $\Omega_m$=0.3 and $\Omega_{\Lambda}$=0.7 \citep{Spergel2003}.
In quoting magnitudes, the AB system will be used, unless otherwise stated.

\begin{table*}
\hspace{-0.9cm}
 \begin{minipage}{185mm}
\normalsize 
\caption{Main accretion and hosts galaxies properties for our 10 X-shooter targets.}
\normalsize 
\scriptsize
\footnotesize
\begin{tabular}{rccllrrrrrrrr}
\hline
(1) & (2) & (3) & (4) & (5) & (6) & (7) & (8) & (9) & (10)  & (11) & (12) & (13)\\
XID & RA & DEC & z & lg(L$_{\rm X}$) & lg(N$_{\rm H}$) & lg(M$_*$) & lg(M$_{\rm BH}$) & SFR & S$_{\rm 1.4 GHz}$  &  lg(L$_{\rm AGN}$) & L/L$_{\rm Edd}$ & specz \\
 & hms & dms & & erg/s & cm$^{-2}$ & M$_\odot$ & M$_\odot$ & M$_\odot$/yr & $\mu$Jy & erg/s & & \\
\hline
 18 & 10:00:31.93 & 02:18:11.8 &  1.598 & 44.9 & 22.5 & 11.39 & 8.68 & 4.9 & $<80$ & 46.2 & 0.5  & 1.6073  \\
60053 & 10:01:09.25 & 02:22:54.7 & 1.582 &  $>43.5$$\dag$ & [$^{**}$] & 11.17 & 8.65 & 740*  & 718 & 46.1 &  1.0 & 1.5812  \\ 
175 & 09:58:52.97 & 02:20:56.4 & 1.55(p) & 44.7 & 21.4 & 11.55 & 9.44 & 1.4 &  $<80$ & 45.6  &  0.10  & 1.5297 \\
2028 & 10:02:11.27 & 01:37:06.6 & 1.592 & 45.3 & 21.9 & 11.92 & 9.44 & 275* & $102$ & 46.3  & 0.05  &1.5927 \\ 
5321 & 10:03:08.83 & 02:09:03.5 & 1.27(p) & 45.7 & 21.6 & 12.22 & 9.81  & 230* & 180 & 46.3  &  0.01  & 1.4702 \\  
\hline
5053 & 10:01:29.03 & 01:57:11.6 & 1.374 & 44.3 & 23.2 & 11.03 &  $-$  & $<1$ & $<80$ & 45.3  & $-$ & 1.3735  \\ 
5325 & 10:02:18.83 & 02:46:04.3 & 1.43(p) & 43.1 & 20.0 & 11.19 &  7.50  & 28.8 & $<80$ & 44.6$^{a}$ &  1.0 & 1.3809 \\
5573 & 09:58:07.15 & 01:47:08.5 & 1.66(p) & 44.4& 22.2 & 10.91 & $-$ & 16.5 & $<80$ & 45.1  & $-$ & 1.1152  \\
54466 & 10:02:25.34 & 02:26:14.1& 1.47(p)  & 43.8$\dag$ & 23.2 & 10.93 & $-$ & 107* & 191 & 45.2  & $-$ & 1.0034 \\ 
31357 & 10:01:34.97 & 02:38:07.3 & 1.71(p)  & 44.5 & 23.2 & 11.24 & $-$   & $<1$ & $<80$ & 45.3  & $-$ &  $-$ \\ 
\hline
\end{tabular}
\hspace{1cm}
Notes and column description: The first 5 sources above the horizontal line are those with BH masses measurements presented in \citealt{Bongiorno2014}).
(1) XID from \citet{Brusa2010}; (2,3) optical coordinates (J2000); (4) redshift available before the X-shooter run; (p) marks photometric redshiift; 
(5,6) X-ray luminosities (L$_{\rm X}$, unobscured) and column densities are in the 0.5-10 keV rest frame range and they are obtained from proper spectral analysis if available \citep{Mainieri2011} or from rest frame flux corrected from absorption as inferred from the HR (sources marked with $\dag$, following \citealt{Merloni2014}). The source marked with a double asterisks (**) is a candidate Compton Thick AGN.
(7) Stellar masses are from Bongiorno et al. (2014; first 5 sources) or  Bongiorno et al. (2012; last 5 sources), computed from SED fitting and assuming a Chabrier IMF; 
(8) BH masses from \citet{Bongiorno2014}; (9) SFR are from \citet{Bongiorno2012} or  \citet{Bongiorno2014}. Sources marked with an asterisk (*) are those with detection in PEP. 
(10) The 1.4 GHz flux is taken from the VLA survey of COSMOS \citep{Schinnerer2010}. (11) Bolometric luminosity from SED fitting as derived from \citet{Lusso2012}; for the source marked with $^{a}$ L$_{\rm bol}$ is derived from L$_{\rm X}$ and a k$_{\rm bol}\sim20$; 
(12) Eddington ratio; (13) redshift as measured in this work.
\end{minipage}
\end{table*}

\begin{figure}
\includegraphics[angle=0, scale=0.43]{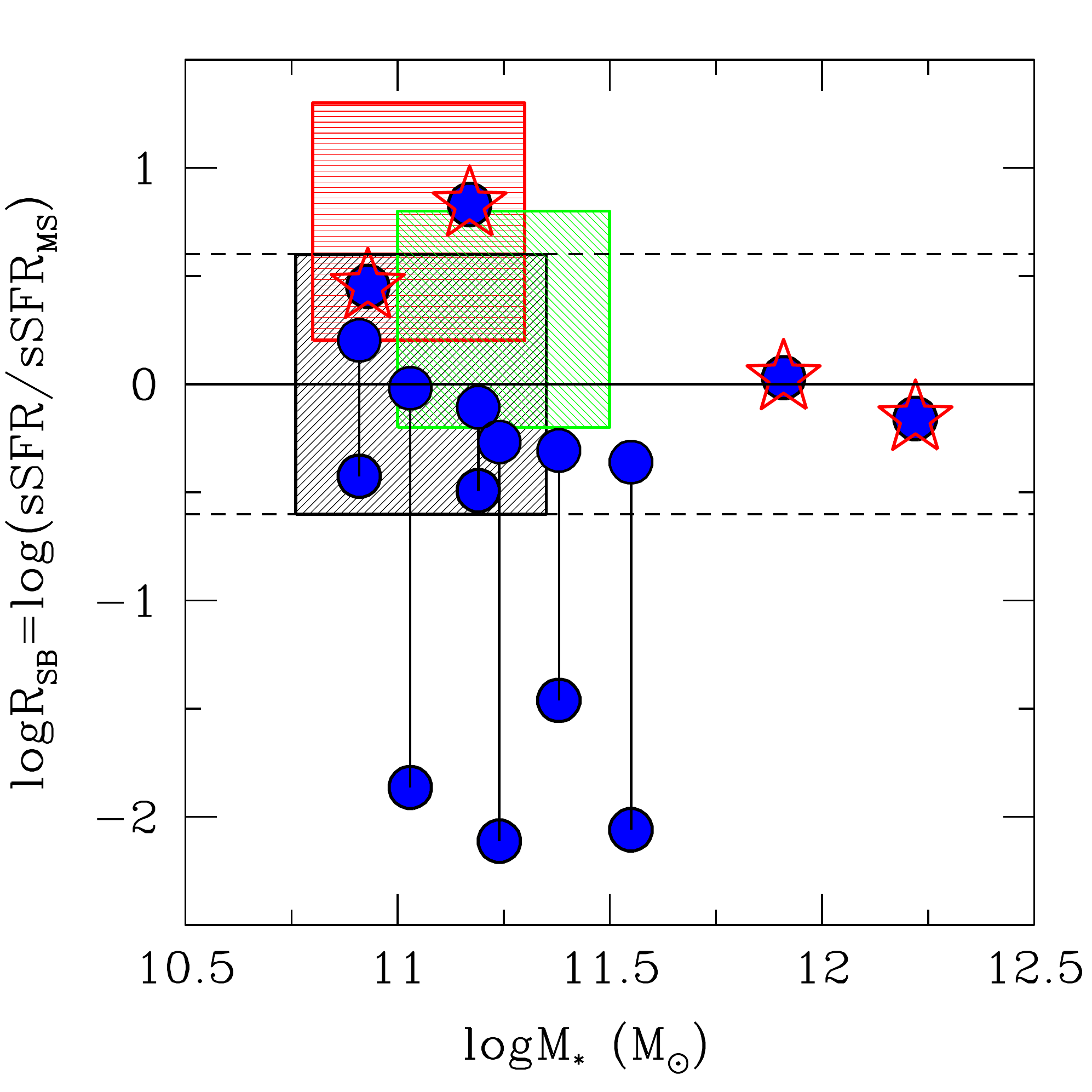}
\caption{ Ratio between the Specific Star Formation Rates (sSFR=SFR/M$_*$) of our targets with respect to those expected for Main Sequence galaxies (sSFR$_{\rm MS}$) at a given stellar mass, versus the host stellar mass M$_*$). The ratio R$_{\rm SB}$=sSFR/sSFR$_{\rm MS}$ is plotted in logaritmic scale. 
The sSFR values are normalized assuming the best fit of the galaxy main sequence as a function of redshift obtained by \citet{Whitaker2012}. The dashed horizontal lines mark the region of MS galaxies. Starred symbols mark objects for which the SFR is derived from PACS photometry. For the 6 sources undetected in PEP, we report as blue circles the value corresponding to the SFR derived from the SED fitting decomposition (lower paired points), and the R$_{\rm SB}$ value computed assuming a SFR$\sim70$ M$_\odot$ yr$^{-1}$ as derived from stacking of the PACS fluxes (upper paired points; see text for details). The shaded areas mark the loci occupied by the z$\sim2$ SMG/ULIRGs presented in H12 (red), the 8 massive star forming galaxies presented in \citet{FS2014} (green) and the MS galaxies presented in \citet{Kashino2013}  (black).  }
\label{hosts}
\end{figure}

\section{Sample selection and luminous obscured QSOs properties}

\subsection{Sample selection}
\label{sect_sel}
\par\noindent
The XMM-COSMOS survey  \citep{Hasinger2007,Cappelluti2009} consists of $\sim1800$ X--ray AGN selected over the entire 2 deg$^2$ COSMOS field, with complete multiwavelength data  from radio to UV, including spectroscopic and photometric redshifts, as well as morphological classification, accurate estimates of  stellar masses, SFR and infrared luminosities  (see: B10; \citealt{Salvato2011,Lusso2012,Bongiorno2012,Santini2012,Rosario2012}).  
In particular, the XMM-COSMOS survey has the combination of area and depth necessary to sample with the adequate statistics rare objects (such as those expected in the `feedback' phase discussed above) otherwise missed in pencil-beam surveys such as the Chandra Deep Fields (see e.g. \citealt{Alexander2012} for a review). In addition, the XMM-COSMOS survey sample obscured objects at lower intrinsic luminosity  (L$_{\rm  bol}\sim10^{45-46.5}$ erg s$^{-1}$)  
than the ``monsters'' accessible from larger area IR surveys  (e.g. WISE targets, L$_{\rm bol}>10^{47}$; \citealt{Weedman2012}; see also \citealt{Banerji2012}), more representative of the entire luminous QSO population.

In B10 we presented a sample of  $\sim170$ obscured AGN  isolated from the entire XMM-COSMOS sample, on the basis of their observed-frame 
mid--infrared (flux$_{24\mu m}$/O$>1000$), near--infrared (R-K$>5$ or R-[3.6]$>4$), optical and X-ray (X/O$>10$) properties (following \citealt{Fiore2003,Brusa2005,Mignoli2004,Fiore2009}, Melini et al. in preparation). 
We made use of the spectroscopic information and of the availability of spectral classifications, to assess the reliability of color cuts and flux ratios as diagnostics of the presence of obscured sources: of the 20 sources with optical spectra available, the vast majority (80\%) are classified as narrow line AGN, and most of them are X-ray obscured ($<HR>\sim-0.2$,  HR being defined as (H-S)/(H+S) where H and S are the counts in the 2-10 keV and 0.5-2 keV bands, respectively).  The adopted criteria  appear therefore robust in selecting additional $\sim150$  X-ray obscured objects lacking broad lines in the optical spectrum in the redshift range z$\sim1-3$. 
In Figure~\ref{selection}  we plot the selection regions in the 2 diagnostics described above that we used to select our X-shooter targets; the small empty circles in this plot represent the full XMM-COSMOS population. 

From the original sample of $\sim170$ objects, we selected all the sources with K$<19$ and deabsorbed rest-frame F$_{\rm 2-10}>5\times10^{-15}$ erg cm$^{-2}$ s$^{-1}$. We then extracted all the sources with photometric redshifts in the range z=1.25-1.72 (i.e. those for which  H$\alpha$ and [OIII] lines are expected to lie in regions free from strong atmospheric absorptions). This sums up to a total of 12 objects we proposed for our X-shooter observations, of which 10 were observed (see Section~\ref{sect_obs}). 
The 10 targets comprise: 3 spectroscopically confirmed obscured AGN  (XID 2028, XID 60053, XID 5053) selected on the basis of the lack of broad emission lines (CIV, MgII) in the  optical spectra, 1 BL AGN (XID 18), and 6 objects with only photometric  redshifts. In Figure~\ref{selection} objects marked with large symbols are those with redshift  in the range z=1.25-1.72 and satisfying the conditions on the K-band and the X-ray flux: blue circles denote spectroscopically confirmed BL AGN, red circles spectroscopically confirmed NL AGN and cyan triangles objects with only photometric redshifts. The yellow circles are the obscured  AGN candidates observed by X-shooter,  e.g. sources which satisfy  one (R-K$>5$ and X/O$>10$) or the other 
 (log($\nu_{24}$F$_{24}$/$\nu_R$F$_R>$1 and R-[3.6]$>4$) selection criterion.

Our sample is similar in size, but more homogeneous in the selection, with respect to the sample presented in H12 (10 objects), which represents one of the most recent analysis and study of [OIII] profiles in high-z (z$\sim2$) SMG/AGN systems.
The sample is also similar in size to the sample of heavily reddened quasars at z$\sim2$ selected from the UKIDSS Large Area Survey (ULAS; \citealt{Banerji2012}; 12 objects). We note that, although our sources on average show a  lower/bluer J-K color ((J-K)$_{\rm Vega}\sim1.5-2.5$ likely due to the on average lower redshift of the objects) than the color cut used in \citet{Banerji2012} (J-K$>2.5$), they actually populate the same region of the I-K vs. I band diagnostic, extending to faint magnitudes given the deeper limiting fluxes of COSMOS with respect to UKIDSS LAS survey. Indeed, our prototype source (XID2028) is  selected as a red quasar from ULAS (ULASJ1002+0137) and a SINFONI spectrum is presented in \citet{Banerji2012}.

The most important properties of the targets are reported in Table 1, and discussed below in detail.  
The multiwavelength properties, including the fluxes measured in the selection bands, can be retrieved via the XID from the B10 catalog publicly available at: http://www2011.mpe.mpg.de/XMMCosmos/xmm53\_release/.

\subsection{Host galaxies properties}
\label{sect_hosts}
For all the targets, host galaxies stellar masses (M$_*$) and SFR based on galaxy and AGN decomposition are already available\footnote{For those sources for which we could provide a new spectroscopic redshift (see Section~\ref{sect_spectroz}), we re-run the code and recomputed M$_*$ and SFR.  }
 (\citealt{Bongiorno2012,Bongiorno2014}).  
Four of the targets (XID5321, XID2028, XID60053 and XID54466) are detected  in the PEP (PACS Evolutionary Probe) survey by Herschel/PACS in at least one band \citep{Lutz2011,Santini2012}.
For these objects, L(FIR,8-1000 $\mu$m) were estimated 
by fitting monochromatic PACS fluxes with \citet{Dale2002} IR templates, using the same technique as in \citet{Santini2009},  
and  the luminosity was then converted into a SFR using the relation from \citet{Kennicutt1998}, assuming a Chabrier IMF. 

All our targets have M$_*$ in the range 10$^{11}$-10$^{12}$ M$_\odot$.
The SFR and stellar mass properties of our targets are shown in Figure~\ref{hosts}, where we plot the `starburstiness' R$_{\rm SB}$ = sSFR/sSFR$_{\rm MS}$ of the host galaxies versus M$_*$. The specific Star Formation Rate (sSFR=SFR/M$_*$) of our targets is normalized assuming the best fit of the galaxy Main Sequence (MS) as a function of redshift obtained by Whitaker et al. (2012).
Our targets are marked by blue symbols; those detected in the PEP survey have been highlighted by a red star.  From this figure it is possible to see that, with the exception of XID60053 which is above the MS,  
our targets lie in or below the MS of star forming galaxies at z$\sim1.5$ as defined by the black line \citep{Whitaker2012}.  
For the 6 PEP undetected sources, we report in Fig.~\ref{hosts} two values of R$_{\rm SB}$ (paired by solid lines): the lower one is obtained using the SFR obtained from SED fitting while the upper one is plotted assuming a SFR$\sim70$ M$_\odot$ yr$^{-1}$, corresponding to the PEP stacked signal detected at $\sim3\sigma$ level. 
Even if the  SFR from SED fitting were severely underestimated, the PEP stacked signal confirms that these systems are not actively forming stars with respect 
to their stellar mass.
In particular, the 4 objects with SFR from SED fitting $<10$ M$_\odot$ yr$^{-1}$ (logR$_{\rm SB}<-1$) may be in their way to be quenched (see also \citealt{Mignoli2004,Mainieri2011}).
In contrast, the SMG/ULIRG sample presented in H12 is on average above the MS of star forming galaxies at the same redshift (z$\sim2.2$; see red shaded area in Figure~\ref{hosts}). In the same plot we also show the loci occupied by two other samples we will use in the following as comparison: the green area marks the locus occupied by the 8 massive star forming galaxies presented in \citet{FS2014}, while the black area mark the mass range of MS star forming galaxies presented in \citet{Kashino2013}. 

Differently from H12, for which a radio detection was imposed to conduct the Integral Field Unit (IFU)  follow-up,  only 40\% of our objects (all those detected also in PACS) are detected in the radio band at 1.4 GHz in the Very Large Array (VLA) observations of the COSMOS field \citep{Schinnerer2010}. The radio power implied by the detections in the  survey (L$_{\rm 1.4 GHz}$=10$^{23}-10^{24}$ W Hz$^{-1}$) places these objects below the radio-loud class, and they are also one order of magnitude fainter than the high-z radio galaxies and the SMG/ULIRGs discussed in H12 (see their Figure 1b). In particular, the low level of radio emission (apart from 60053) assures 
little or marginal contribution from radio jets in the energetic of the systems. 

\begin{figure*}
\includegraphics[angle=0, scale=0.5]{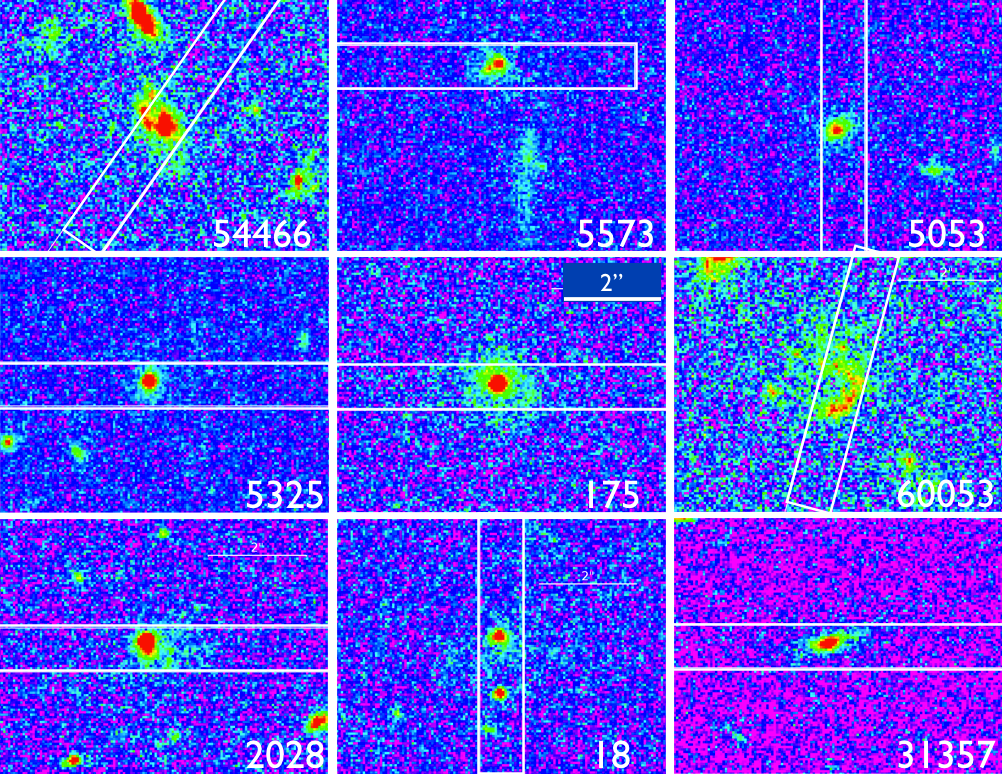}
\caption{I$_{814}$ band cutouts around the 9 targets with HST/ACS data (all but XID5321 which lies outside the HST COSMOS area), ordered by increasing redshifts (same as figure \ref{spectra}). The labels indicate the XID. In the central panel we report also the 2$\arcsec$ scale ($\sim17$ kpc at z$\sim1.5$)  common to all the figures, and the 0.9$\arcsec\times11\arcsec$ slit used in the X-shooter observations is marked in each panel. The intensity level of the images is logarithmic, with blue/purple marking the background level and red marking the peak emission, with an aveage contrast of about a factor of 10.} 
\label{slits}
\end{figure*}

\begin{figure}
\includegraphics[angle=0, scale=0.43]{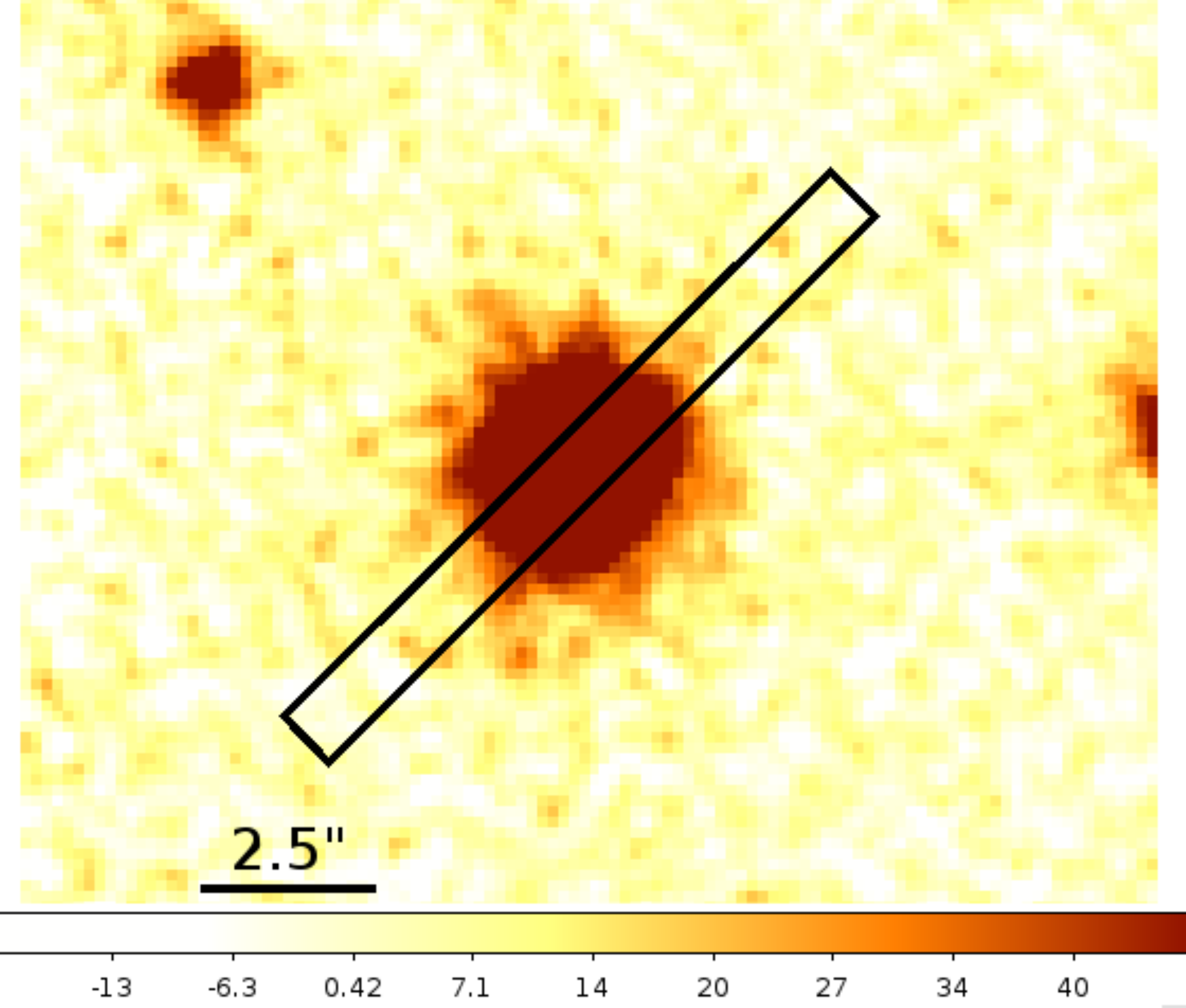}
\caption{UltraVista J-band cutout for source XID5321, the only one lacking HST/ACS data. The 2.5$\arcsec$  scale is shown on the bottom left, as well as the 0.9$\arcsec\times11\arcsec$ slit and position used for the NIR observations. }
\label{slit5321}
\end{figure}

Figure~\ref{slits} shows the HST/ACS  cutouts of the 9 sources for which HST data are available, with overplotted the slits positions and widths. 
The morphology of our prototype object (XID2028, bottom left in Fig.~\ref{slits}) was already presented in B10 (see their Figure 13). A clear pointlike nucleus is present associated with an extended, asymmetric emission.  
Similarly, XID175, XID54466, XID5573, XID5325 and XID18 present pointlike nucleus likely responsible for the X-ray emission, as well as residual diffuse or patchy components and/or close companions, which may trace the host galaxy, extending on scales comparable to or even larger than the slits apertures ($\sim8$ kpc, see next Section). Of the remaining three, XID60053 is a patchy irregular galaxy, XID31357 is an elongated, likely edge-on disk galaxy, while XID5053 may be an elliptical galaxy, consistent with the very low SFR derived from the SED fitting.  XID5321, unfortunately, lies outside the ACS/HST coverage of the COSMOS field. For this source we show in Figure~\ref{slit5321} the ground-based image with the best resolution available in the COSMOS, i.e. the J-band Ultra Vista image. The diameter of the source  corresponds to scales of $\sim30\times40$ kpc at z=1.47. This is our X-ray brightest (L$_{\rm X}$=5$\sim10^{45}$ erg s$^{-1}$) and most massive (M$_*\sim10^{12}$ M$_\odot$) target, with a total IR luminosity (AGN+SF) of the order of L$>10^{47}$ erg s$^{-1}$, similar to those discovered in the IR surveys (WISE, \citealt{Weedman2012,Banerji2012}).  

\subsection{Accretion properties}
\label{sect_agn}
Rest-frame, absorption corrected X-ray luminosities in the 0.5-10 keV band (L$_{\rm X}$) and absorbing column densities (N$_{\rm H}$) have been obtained from proper spectral analysis for objects with enough counting statistics ($>150$ counts; \citealt{Mainieri2011}). For the objects with low counting statistics, the N$_{\rm H}$ is inferred from the observed hardness ratio and it is used to K-correct the rest frame flux to derive the unobscured luminosity (see \citealt{Merloni2014}). Most of our sources have L$_{\rm X}>10^{44}$ erg s$^{-1}$ and moderate to large obscuring column densities (N$_{\rm H}>10^{21}$ cm$^{-2}$). We note that the N$_{\rm H}$ measured are based on relatively low counts data that cannot disentangle the complexity of the spectra and different components. 
In particular, the multiwavelength analysis of XID60053 suggests that for this source the column density is heavily underestimated (and therefore the unobscured L$_{\rm X}$) and this object is a candidate Compton Thick (CT) AGN (see  \citealt{Bongiorno2014}).

We estimated the total AGN bolometric luminosities (L$_{\rm bol}$)  for our sources following  the methods presented in \citet{Lusso2012} and \citet{Lusso2013} for the XMM-COSMOS sample. More in detail, the L$_{\rm bol}$  associated with the AGN component has been evaluated from SED fitting decomposition of the galaxy, torus and SF components, and has been derived by integrating the torus component only for Type 2 AGN or from the combined constraints from the disk and torus emission for Type 1 AGN. Overall, our targets have total AGN luminosities in the range L$_{\rm bol}$=10$^{45-46.5}$ erg s$^{-1}$, with median value L$_{\rm bol}$=10$^{46}$ erg s$^{-1}$.  When comparing the X-ray and the bolometric AGN luminosities, we note that these sources have bolometric corrections in the range  k$_{\rm bol}=L_{\rm bol}/L_{\rm X}\sim4-20$, lower than the value generally assumed for optically selected samples (k$_{\rm bol}\sim20-30$ e.g. H12; see discussion in \citealt{Lusso2012}). 

Finally, \citet{Bongiorno2014} published the BH masses measured from the X-shooter data for 5 out of 10 of our targets. XID5325 was not included in \citet{Bongiorno2014} because of its low N$_{\rm H}$ value ($<10^{21}$ cm$^{-2}$). However, a broad component is detected in the H$\alpha$ line (see Sect.~\ref{sect_model}) and it is possible to derive a BH mass following the same calibration presented in \citet{Bongiorno2014}. 
From the comparison between L$_{\rm bol}$  and the Eddington luminosity associated with the measured M$_{\rm BH}$, it is possible to infer also the Eddington ratios (L/L$_E$) for these 6 sources, in the range L/L$_{\rm Edd}$=0.01-1 (see Table 1). The source accreting at the Eddington level is XID60053, the candidate CT AGN.

\section{Observations and data reduction}
\label{sect_obs}
The XMM-COSMOS obscured QSOs targets have been observed  with the X-shooter  spectrograph
\citep{Dodorico2006,Vernet2011} on the ESO VLT-UT2 (Kueyen)
during the nights of February 8-10, 2013, as part of programme 090.A-0830(A). 
Due to scheduling constraints and time losses during the visitor mode
run (see below), only 10 targets have been
observed. 

X-shooter is an echelle spectrograph, with UV, visible
and near-infrared channels providing nearly continuous 
spectroscopy from 0.3$\mu$m to 2.48$\mu$m. Given the nature of our
sources (very red and optically obscured, with R-K$>5$), although
all the sources were observed with all 3 arms (UVB, VIS and NIR), they
returned signal only in the NIR arm (all targets) and in the VIS arm (with clear 
continuum and/or [O II]$\lambda 3727$ detection in all but one targets).
All the targets were acquired with acquisition images of 30s to 120s,
and with a blind offset from a USNO bright
star. Ad hoc position angles were set for all the sources in order to maximize the
efficiency of the observations (e.g. trying to remove contaminants  in the slit if at positions not suitable for the dithering; see slits position 
superimposed in Fig.~\ref{slits} and \ref{slit5321}). 

The exposure times range from 1hr to 2hrs.
We used the 0.9\arcsec width slit (corresponding to a spectral resolution R$\sim$5100 in the NIR and R$\sim$8800 in the VIS). In the NIR arm we adopted the JH filter (with the
K-band filter blocked): this solution 
reduced the background in
the J-band, essential for our faint targets. 
In the NIR we dithered 600s observations\footnote{In the VIS and UVB arm we dithered the observations in the same way, but we reduced the exposure times of each frame to 563 and 525 seconds, respectively, in order to gain in efficiency during the readout time for each object. }
 in an ABBAAB sequence (e.g. for a 1hr
observation) at positions +2.5\arcsec and -2.5\arcsec from the central
coordinates  along the slit long axis. 
Observing conditions were reported to be photometric, and  seeing
condition was 0.5-1.0\arcsec (FWHM).
We base our flux calibration on observations of the standard stars
LTT3218 and GD71 taken during the three nights with the same photometric conditions and
seeing. 
For most targets, a telluric standard of type B8V-B9.5V was observed
before and after our primary target, in order to create a
telluric absorption spectrum at the same airmass as the observations of our targets and to flux calibrate the data.

The data reduction of the three separate arms has been done with Reflex \citep{Freudling13}.
Previous versions of Reflex pipeline reported known problems with observations obtained 
with the JH blocking filter and flux calibration in the NIR arm. These problems have
been corrected in the newest version (v2.4).  We carefully checked the full data reduction and flux calibration obtained with Reflex v2.4 in the NIR arm by re-reducing
manually with {\it esorex} and the X-shooter recipes the standard stars observed in the three nights. 
Results on flux calibration were consistent within 10\%, and we therefore we adopted the response matrix obtained from the pipeline products.  
The X-shooter pipeline gives as an output the wavelength solution measured in air, and we will refer to this system when measuring redshifts.  
From the wavelength and flux calibrated 2D spectra, we manually extracted the 1D spectrum optimizing the extraction region (position along the slit and aperture)  with
the esorex task {\it xsh\_scired\_slit\_nod}. 
All the spectra have been extracted from an aperture of $\sim1\arcsec$, corresponding to physical sizes $\sim 8$ kpc.

\begin{figure*}
\includegraphics[scale=0.76]{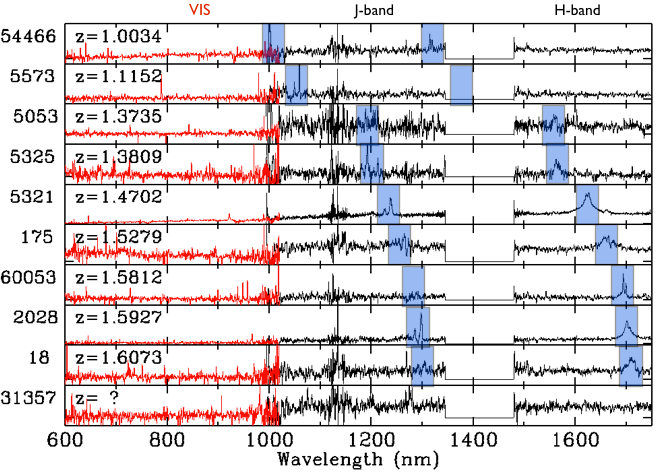}
\caption{X-shooter NIR spectra (black) of the 10 XMM-COSMOS targets sorted by increasing redshift. The XID and redshifts of the sources are given on the left of each panel. The range from $\sim 1\mu$m up to
  1.75$\mu$m is shown for the NIR spectrum. The bad region (due to low atmospheric transmission) defining the limits between the J and H filters is masked out. The shaded areas in each spectrum mark the wavelengths ranges of the redshifted [OIII] (left) and H$\alpha$ (right) lines. Zoom of these regions are given in Figure~\ref{spectroz}.
  The VIS spectra are also shown in red in the 6000\AA-1$\mu$m range. The fluxes are all flux calibrated in erg cm$^{-2}$ s$^{-1}$ \AA$^{-1}$, but the scale is not shown on the y-axis. 
\label{spectra}
}
\end{figure*}

Figure~\ref{spectra} shows the 10 X-shooter VIS+NIR spectra of our XMM-COSMOS targets, sorted by increasing redshift (as determined in Section~\ref{sect_spectroz}).  
The shaded areas in each spectrum mark the regions of the H$\beta$+[OIII] (left) and H$\alpha$+[NII] (right) lines. The flux calibration of the two arms has been done with the same standard star and the two spectra (VIS and NIR) for each target show an excellent match in absolute flux in the overlapping regions (few tens of {\rm \AA}).

\begin{figure*}
\medskip
\includegraphics[angle=0, scale=0.255]{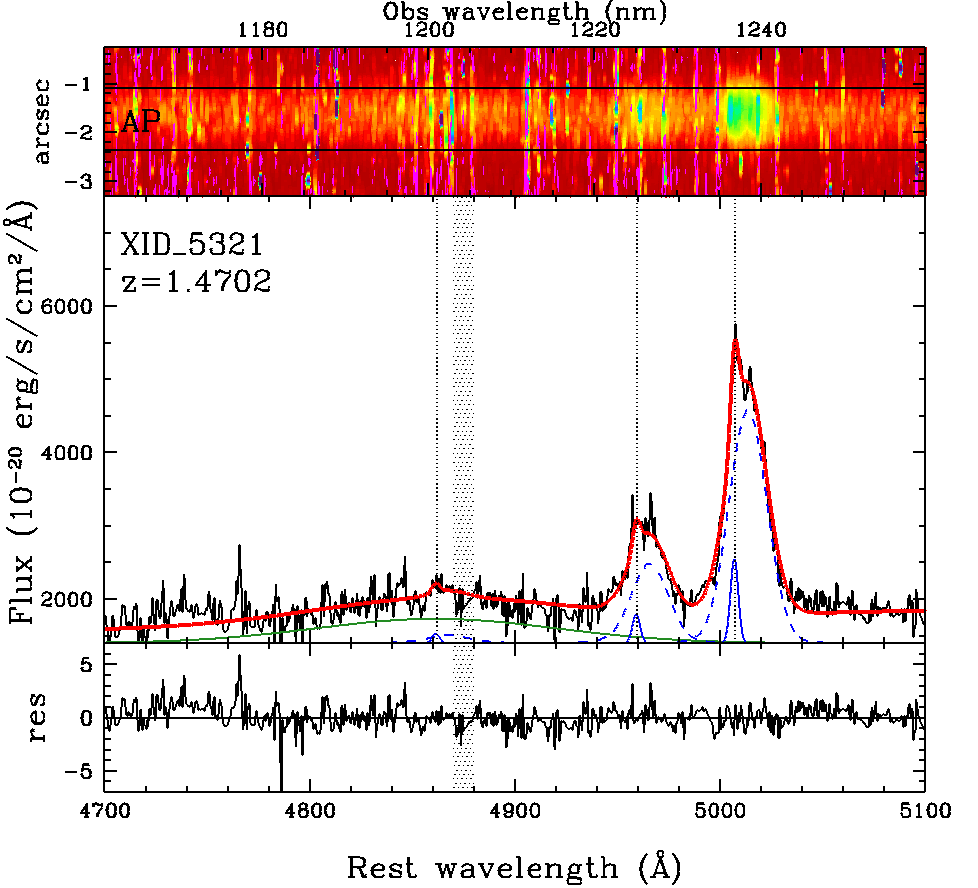}
\includegraphics[angle=0, scale=0.255]{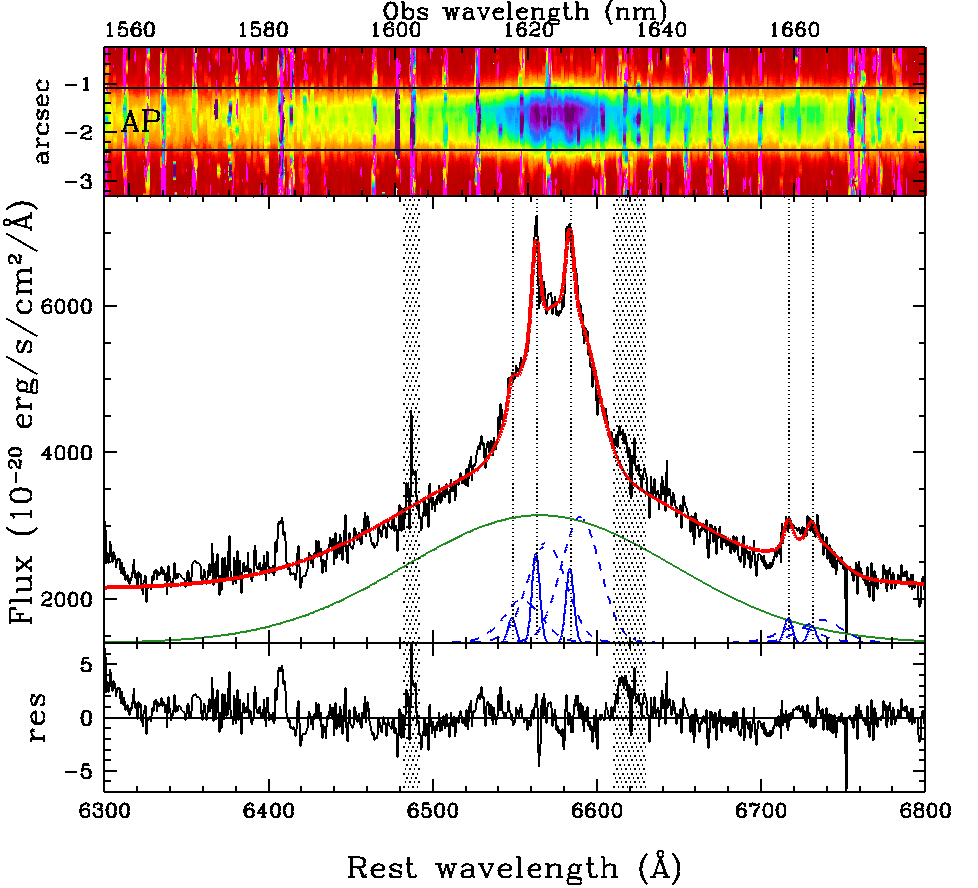}
\includegraphics[angle=0, scale=0.292]{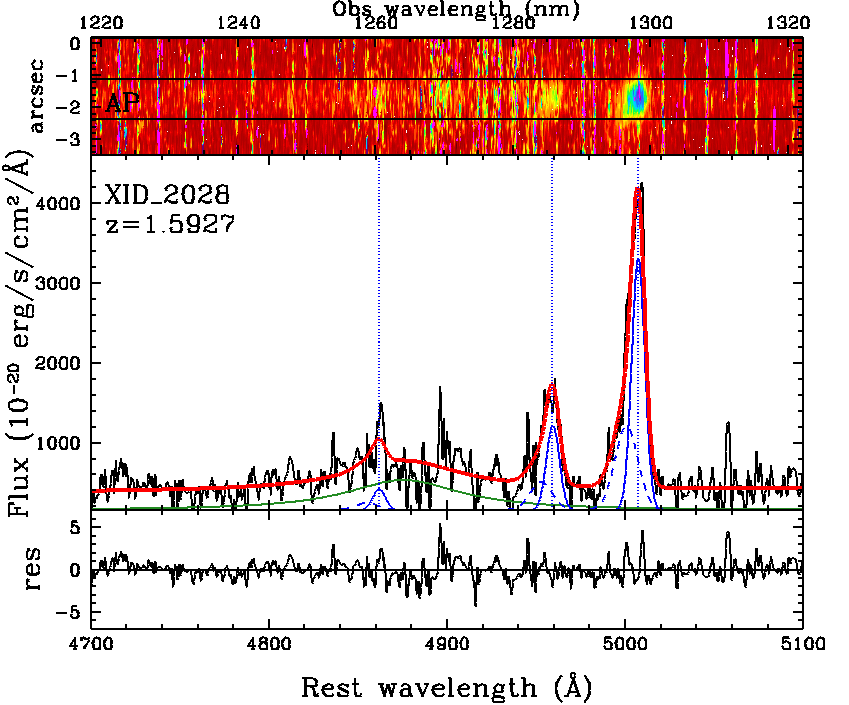}
\includegraphics[angle=0, scale=0.292]{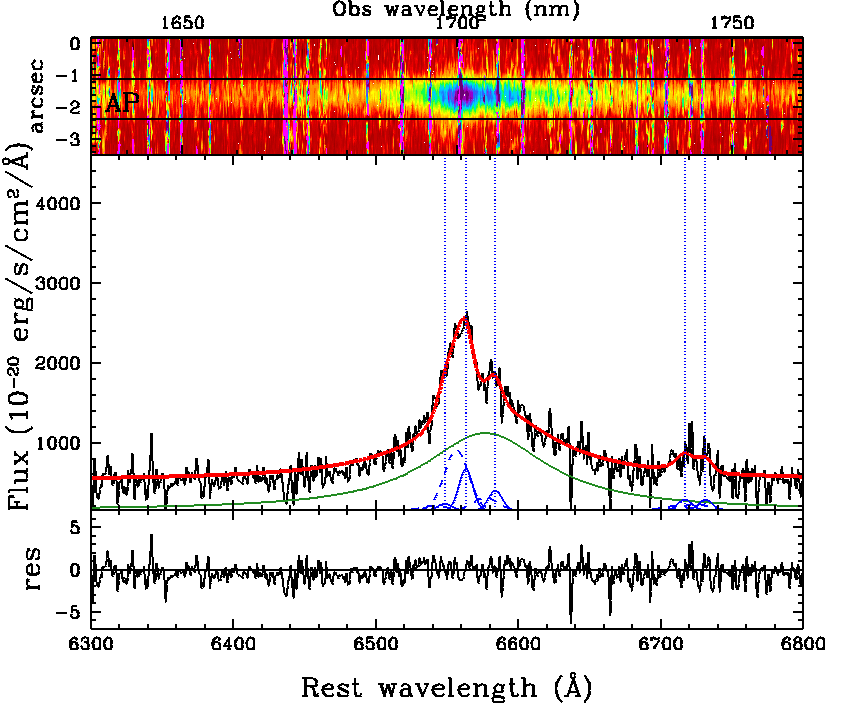}
\caption{Zoom in the regions of  [OIII] (left) and H$\alpha$ (right) lines for the 9 sources for which we could determine the redshifts. For each spectrum and each region, the upper panel shows  the observed X-shooter 2D spectrum with the 1\arcsec\ aperture used to extract the 1D spectrum labeled. The central panel shows in black the rest-frame 1D X-shooter spectrum (smoothed for plotting purposes with a binning factor of 3 to 11 depending on the spectrum). The flux scale has been multiplied by (1+z) in order to conserve the observed integrated flux in the rest-frame fit and is maintained the same for each source to ease the comparison of the relative strength of the emission lines. The dotted lines mark the wavelengths of H$\beta$, [OIII]4959, [OIII]5007 (left) and [NII]6548, H$\alpha$,[NII]6581, and the [SII] doublet (right), from left to right, respectively. The regions excluded from the fit below the gaussian components  and corresponding to the most intense sky lines)  are highlighted as shaded areas. Superimposed on the spectra are the best fit components presented in Section~\ref{sect_model},  with arbitrary normalization in order to ease the visualization: solid (blue) curves represent the systemic component (``S''); dashed (blue) curves the broad, shifted component (``B''). The solid (green) curves below H$\beta$ and H$\alpha$ with FWHM$\gsimeq2000$ km s$^{-1}$ represent the very broad component (``VB'').  The red solid curve shows the best-fit sum of all components (including the power-law). When only one component is needed, the fit is shown as red curve only. In the bottom panel of each fit the residuals with respect to the best fit are shown.  XID5321 and  XID2028 are shown here.
\label{spectroz}
\label{fit}}
\end{figure*}

\begin{figure*}
\includegraphics[angle=0, scale=0.31]{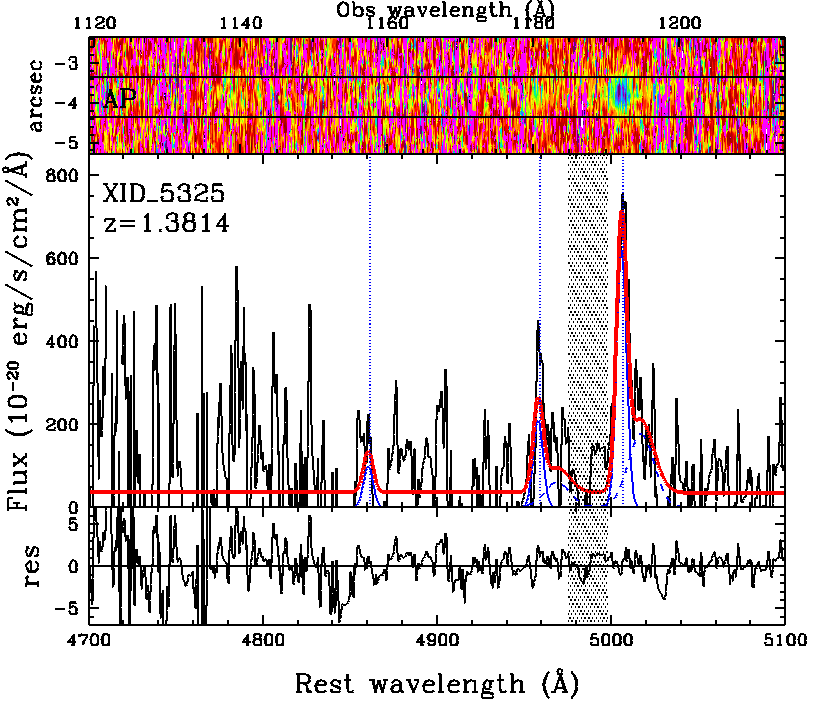}
\includegraphics[angle=0, scale=0.31]{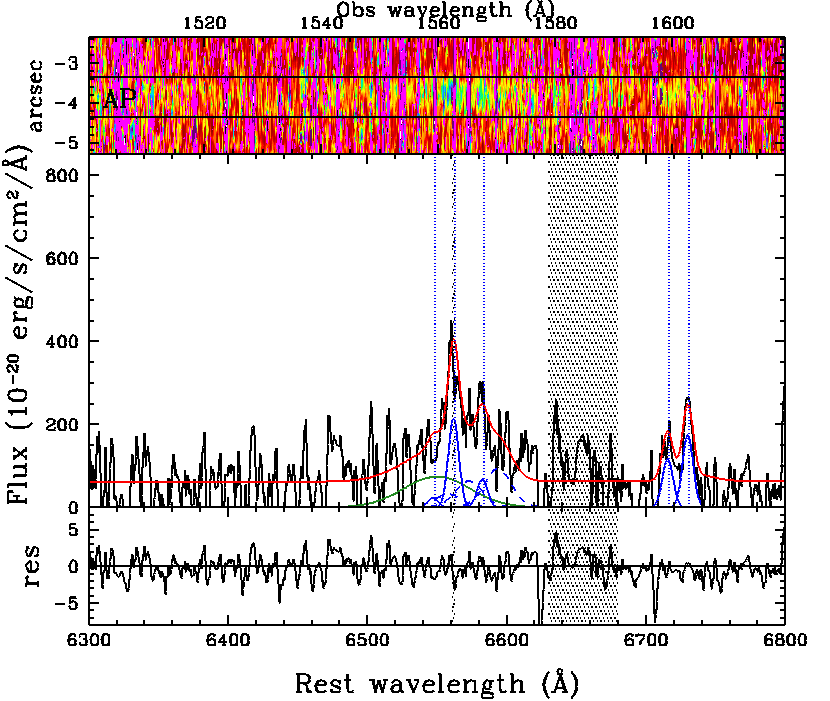}
\includegraphics[angle=0, scale=0.278]{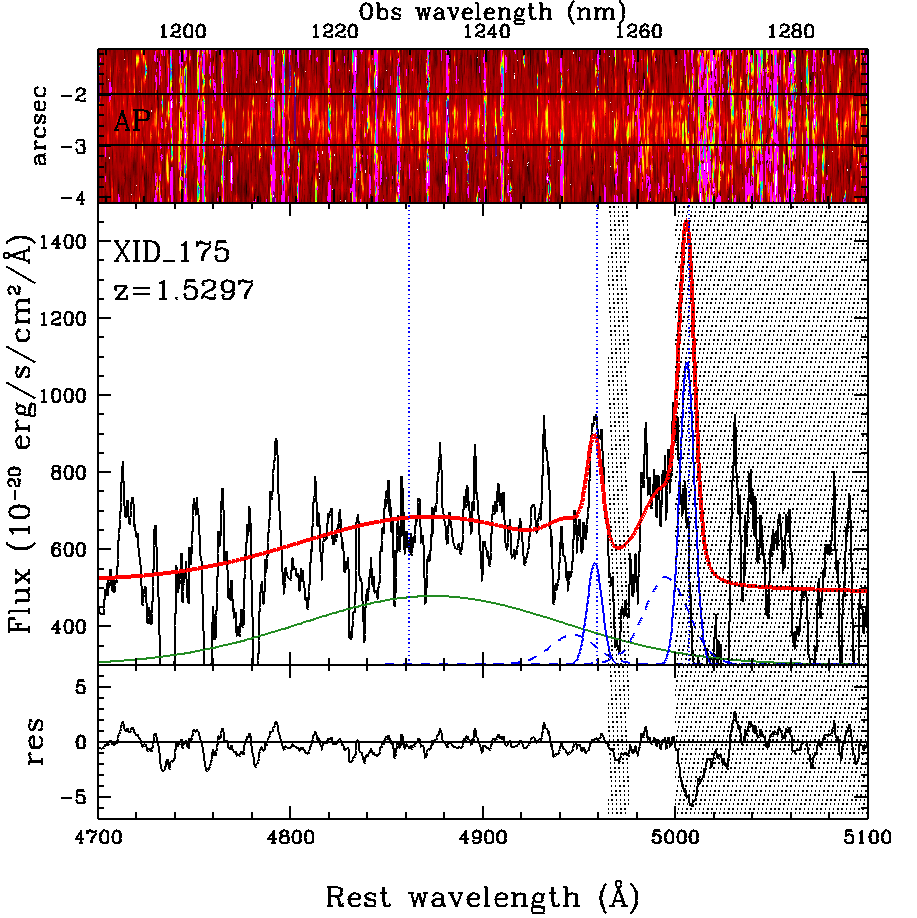}
\includegraphics[angle=0, scale=0.278]{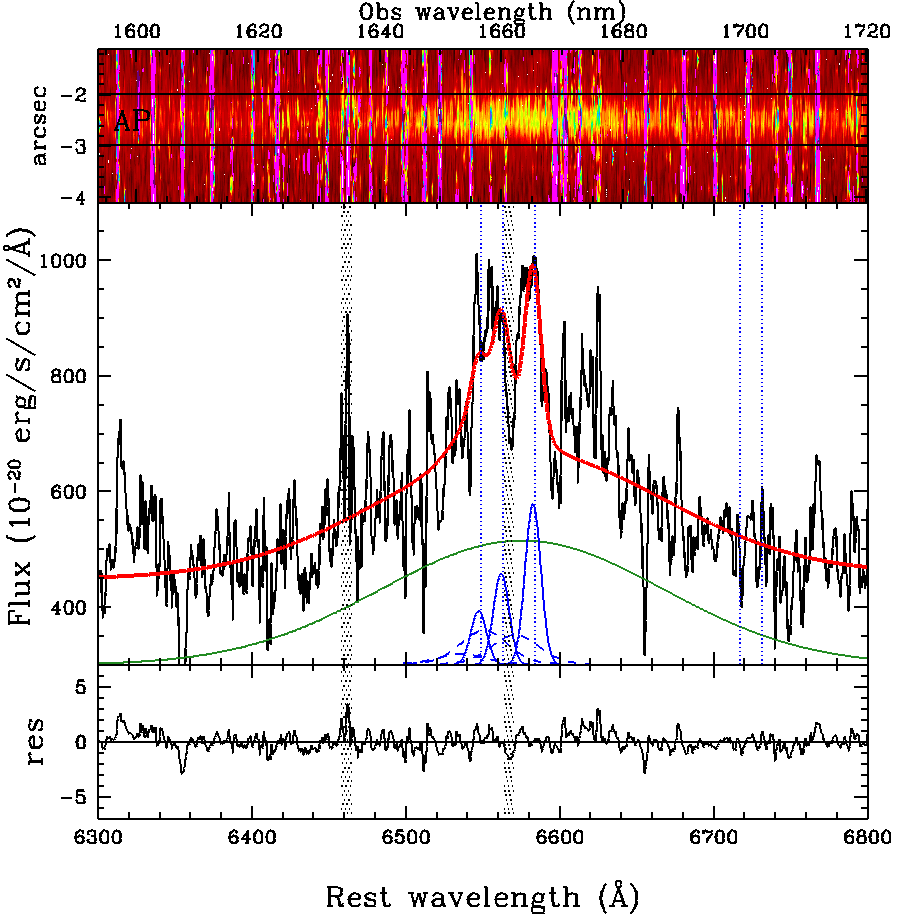}
\contcaption{$-$  Zoom in the regions of the [OIII] (left) and  H$\alpha$  (right) for XID5325 and XID 175. 
See previous page for description. }
\end{figure*}

\begin{figure*}
\includegraphics[angle=0, scale=0.3]{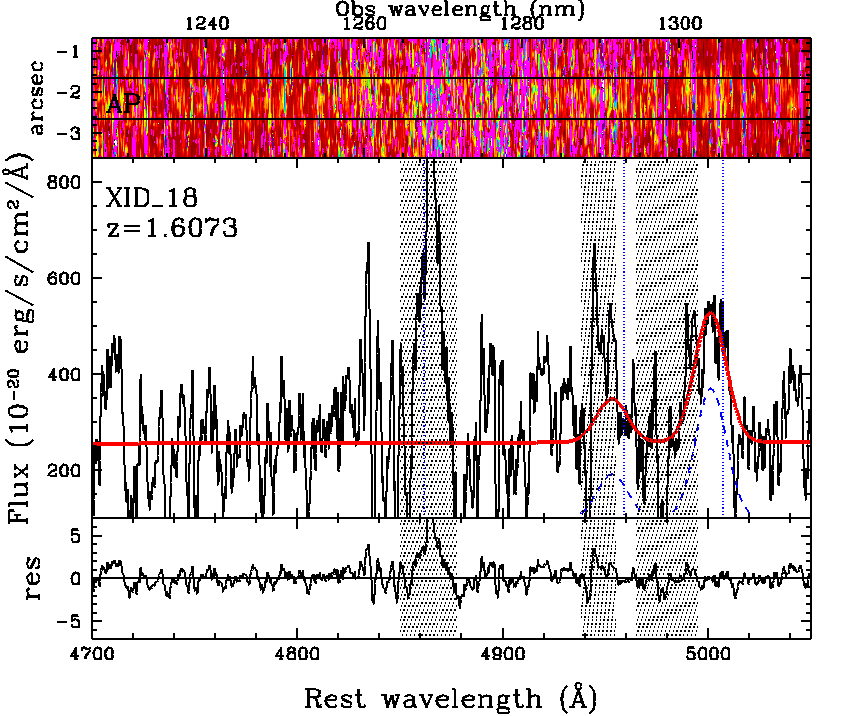}
\includegraphics[angle=0, scale=0.3]{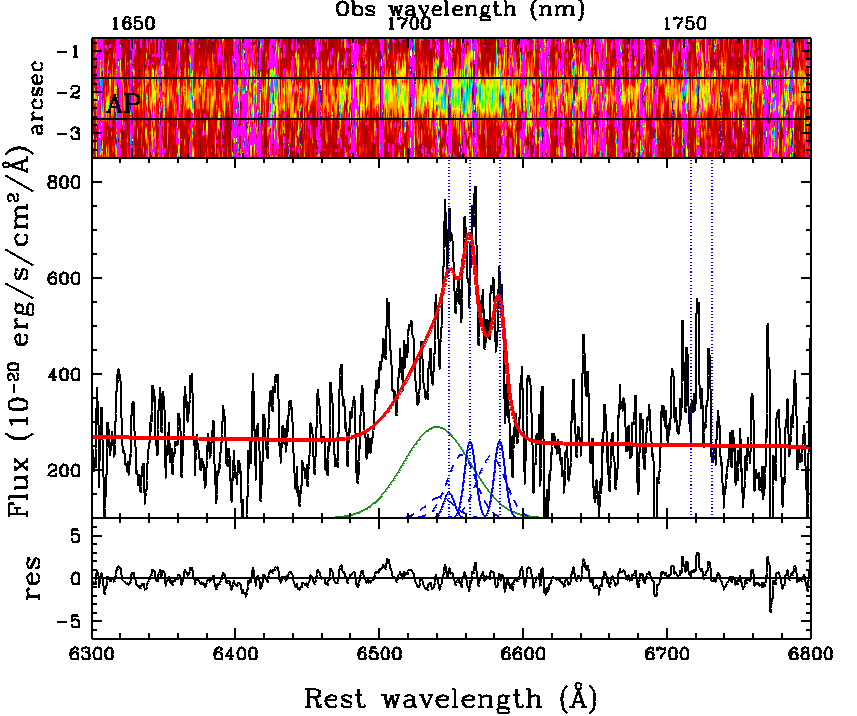}
\includegraphics[angle=0, scale=0.3]{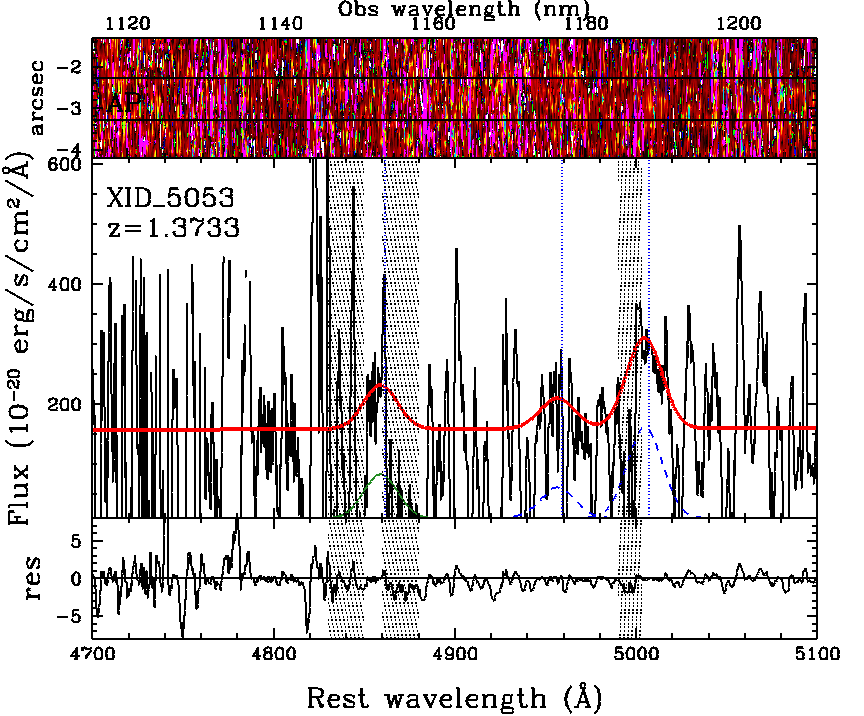}
\includegraphics[angle=0, scale=0.3]{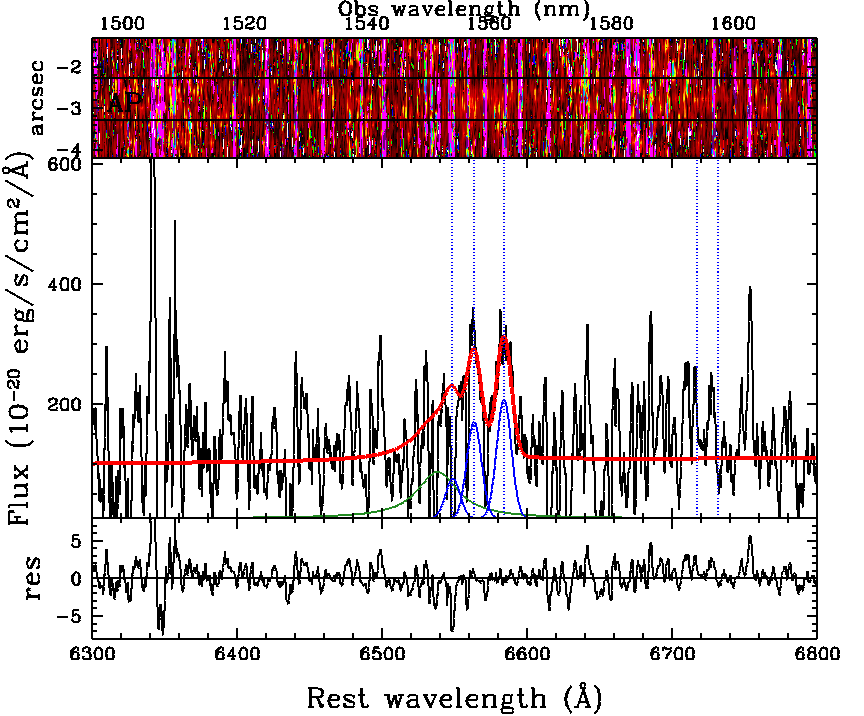}
\contcaption{ $-$ Zoom in the regions of the  [OIII] (left) and H$\alpha$ (right) for XID18 and XID5053.  See previous page for description. }
\end{figure*}

\begin{figure*}
\includegraphics[angle=0, scale=0.292]{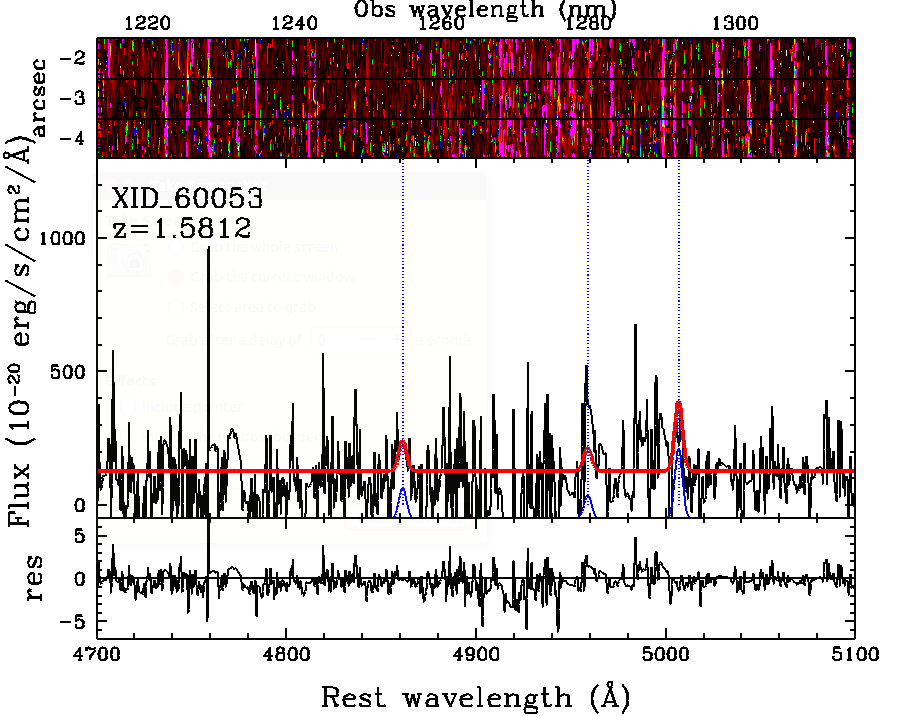}
\includegraphics[angle=0, scale=0.292]{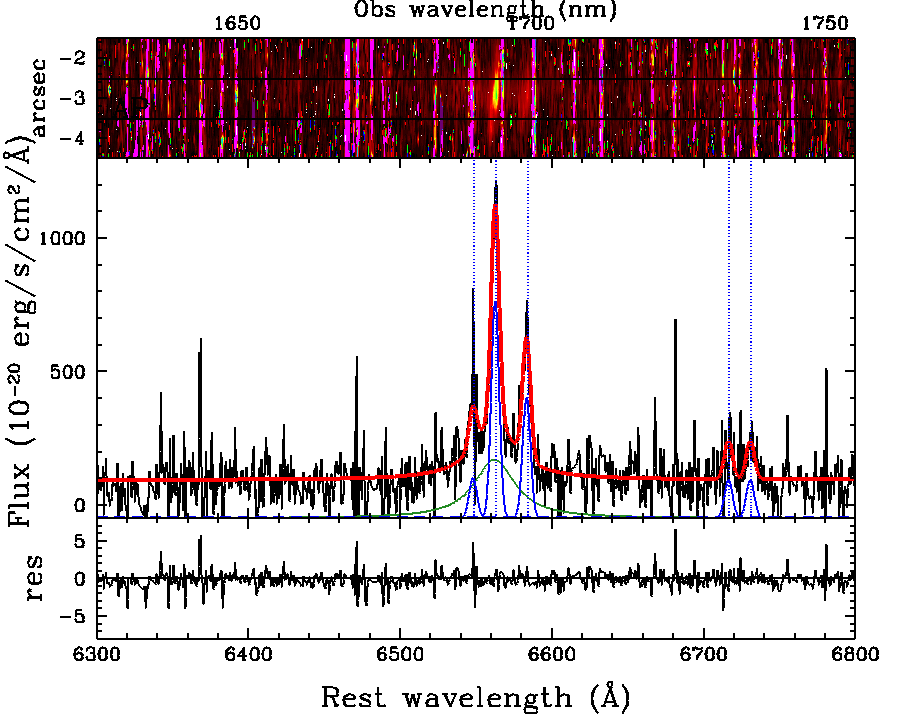}
\includegraphics[angle=0, scale=0.3]{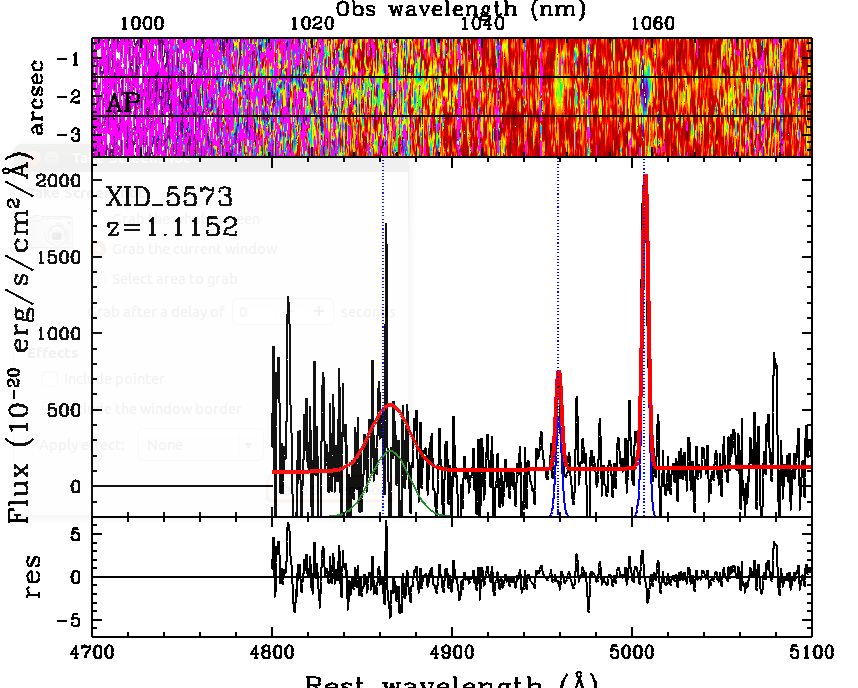}
\includegraphics[angle=0, scale=0.3]{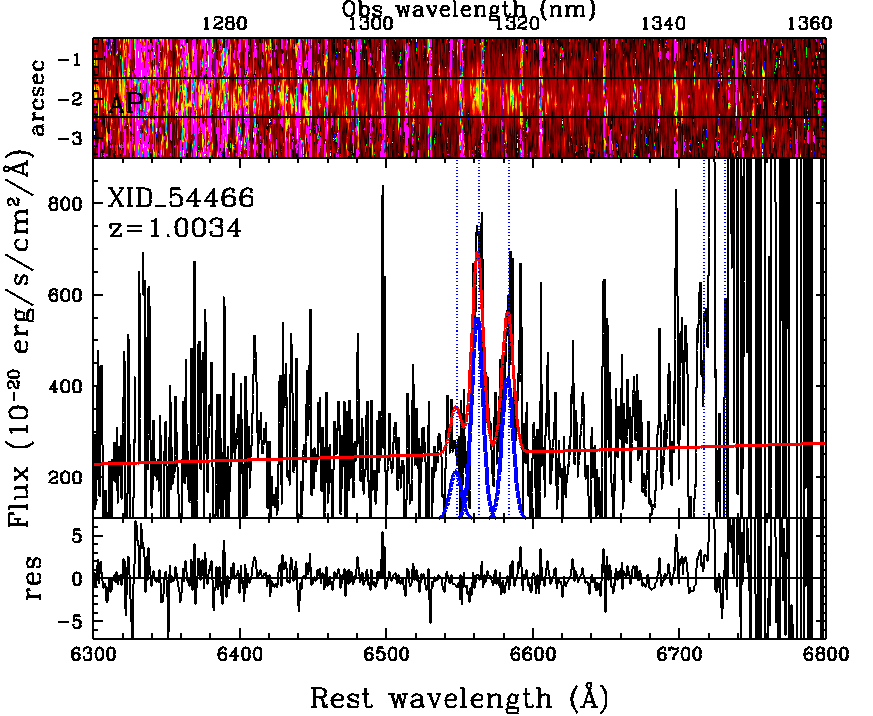}
\contcaption{Zoom in the regions of the  [OIII] (left) and H$\alpha$ (right) for XID60053 (first row). In the second row we show the [OIII] region for XID5573 (left) and the H$\alpha$ region for 54466 (right). See previous page for description }
\end{figure*}

\section{Results} 
\subsection{Spectroscopic redshifts}
\label{sect_spectroz}

We chose as best fit solution for the spectroscopic redshift the one which produces the best fit to the wavelengths of the narrow components of the [OIII]$\lambda\lambda$4959,5007, [NII]$\lambda\lambda$6548,6581 and H$\alpha$ lines (see next Section). We used these observed wavelengths to compute the systemic redshift. For the objects with significant signal in the VIS spectrum, we also imposed the solution to be consistent with the position of the resolved doublet of the [OII] line ($\lambda\lambda$3726.0, 3728.8 \AA)\footnote{
\citet{Vernet2011} quote a wavelength calibration accuracy of the order of 2 km s$^{-1}$s in the VIS arm and  4 km s$^{-1}$ in the NIR arm. We checked the wavelength calibration in our reduced spectra with the positions of known sky lines in the optical (e.g. $\lambda$5577.32,6300.30\AA, from \citealt{Gindilis1961}, http://www.star.ucl.ac.uk/\~msw/lines.html) and in the Infrared (from the table given in \citealt{Rousselot2000}) and they agree within 10 km s$^{-1}$. This is well below the resolution of the instrument in both arms ($\sim40$ and $\sim60$ km s$^{-1}$, respectively) and results in a internal velocity accuracy of the order of $\Delta$z = 0.0004.}. 

Figure~\ref{spectroz} shows  a zoom of the H$\beta$+[OIII] (left) and H$\alpha$+[NII] (right) regions for the 9 sources for which we could assign accurate redshifts. Each row corresponds to a different object, and in each panel we show the extracted spectrum (center), along with the corresponding 2D spectrum (upper inset). The observed frame wavelengths and the aperture used to extract the spectrum are also shown on the 2D image.
The proposed redshift solution is labeled in the left panel, and the corresponding wavelengths of H$\beta$, [OIII]$\lambda\lambda$4959,500 (left panel)  and  N[II]$\lambda$6548, H$\alpha$, [NII]$\lambda$6581 and [SII]$\lambda\lambda$6720,6735 (right panel) are superimposed on the 1D spectra. Two sources have only the H$\alpha$ line (XID54466) or the [OIII] lines (XID5573) in a good portion of the spectrum. For these 2 objects the redshift solution has been determined using only the accessible lines and they are shown on the same (last) row of Figure~\ref{spectroz}. Finally, for XID31357, due to the low S/N of the spectrum, the identification is less secure and only tentative. We note that the excess clearly visible in the NIR spectrum at just below 1.3$\mu$m (see Figure\ref{spectra}, lower panel) may be real, but cannot be identified as the [OIII] line, given that the H$\alpha$ then should appear at $\lambda\sim$1.65$\mu$m in a region free from strong atmospheric absorptions.

Overall, we were able to confirm the redshift of the 4 objects with optical spectra already available,  and to assign new spectroscopic redshifts to 5 out of 6 sources. 
Two of the 5 objects with photometric redshifts for which we were able to assign a spectroscopic redshifts (XID5573, XID54466) have $\Delta z/(1+z_{spec})>0.2$, considerably larger than the current precision  ($\sigma _{\Delta z/(1+z_{spec})}\sim 0.015$) 
to which photo-z for AGN have been computed for the XMM-COSMOS sample (see \citealt{Salvato2011}). 
Photometric redshifts  via SED fitting are always calibrated on the basis of a spectroscopic sample. This is true in particular for AGN where the relative galaxy/AGN contribution is unknown and the  libraries of templates change, depending on the type of sources we want to fit \citep{Salvato2011}. The 2 sources with $\Delta z/(1+z_{spec})>0.2$ are fainter than the objects used for the spectroscopic training sample for the XMM-COSMOS AGN and therefore  the discrepancy is not surprising. 
This discrepancy also caused the 2 targets to be the only two lacking information on both H$\alpha$ and [OIII] line complexes.

The success rate in assigning secure spectroscopic redshifts to color-selected objetcs (5/6, $\sim83$\%)  is higher than that reported in similar programs of spectroscopic follow-up of red quasars: for example, \citet{Banerji2012} were able to assign secure redshifts to 5 out of 13 objects  ($\sim40$\%) in their SINFONI H+K follow-up of red quasars (see also \citealt{Sarria2010,Bongiorno2014}). 
Our higher spectroscopic  success rate can be mainly ascribed to the larger X-shooter range covering both the visible and the entire NIR bands, down to the J filter. Indeed,  the 2 sources with $\Delta z/(1+z_{spec})>0.2$ mentioned above have spectroscopic redshifts $z=1-1.2$ and therefore H$\alpha$ is not sampled in the SINFONI H and/or K band used in the mentioned programs. Had these sources been observed with only the H or K band filters (or both) they would have turned out to be featurleess, dropping the success rate to 50\%.  
In addition, \citet{Banerji2012} speculate that the objects for which they could not derive redshifts are Luminous Red Galaxies at z$<1$, and therefore they do not show any feature in the NIR bands. The selection on the basis of the X-ray emission of our targets assures a negligible contamination by LRG despite the similar optical to IR colors.

\subsection{Modeling the H$\beta$+[OIII] and H$\alpha$+[NII] line complexes}
\label{sect_model}
To determine the dynamics and the outflow properties from the [OIII] line profile fitting, we proceeded as follows.
First, we brought all the spectra to the rest-frame by dividing the wavelength by (1+z). The rest frame wavelenghts are reported in the lower x-axis in each panel of Fig~\ref{fit}. In doing so we also multiplied the flux by (1+z) in order to conserve the observed integrated flux in the rest-frame fit. 
Prior to the modelling of the emission line profiles, the continuum was subtracted. We estimated the local continuum by fitting a power-law 
  to the spectra at both sides of the two regions (H$\beta$+[OIII], H$\alpha$+[NII]) using those wavelength ranges that are not affected by prominent emission or absorption features (e.g. 4200-4300 and 5050-5200\AA\ for the [OIII] lines).

Once the continuum has been subtracted, for all our 9 targets we fit the two regions with three sets\footnote{For sources with only the H$\beta$+[OIII] (XID5573) or H$\alpha$+[NII] (XID54466) lines in a good portion of the spectrum we modified the set of gaussians accordingly.} 
 of Gaussian profiles: 

\begin{itemize} 
\item[{\bf Set 1:}] (Systemic,``S''): 8 gaussian lines, one for each emission line (namely: H$\alpha$ and H$\beta$, and the [OIII], [NII] and [SII] doublets). We imposed the following constraints: (i) the flux ratios between [OIII]$\lambda4959$ and [OIII]$\lambda5007$ and between [NII]$\lambda6548$ and [NII]$\lambda6583$ were fixed at 1:2.99 \citep{Osterbrock1981}; (ii) the widths (FWHM) of the components of each line were set to be equal and $<550$ km s$^{-1}$; (iii) 
the relative wavelength of the lines was constrained to be equal to the laboratory differences. 

\item[{\bf Set 2:}] (Broad, ``B''): 8 gaussian lines to model the same emission lines as detailed in Set 1 (with the same constraints described above), but with no limit to the FWHM.

\item[{\bf Set 3:}] (Very Broad, ``VB''): 2 very broad (FWHM$\gsimeq2000$ km s$^{-1}$) gaussian (or lorentzian) functions, for the H$\alpha$ and H$\beta$ lines. For this component we only used the constraint that the widths of H$\alpha$ and H$\beta$ are forced to be equal.  This component is used only when needed/required.

\end{itemize}
\par\noindent
From a physical point of view, the first component should trace the systemic emission of the source associated with both the NLR and SF (when present); for this reason we limit the FWHM to $\lsimeq550$ km s$^{-1}$. The second component should trace the outflowing gas. Finally, we introduced the third component  to account for the possible presence of H$\alpha$ and H$\beta$ emission originated in the BLR.

We fit the gaussian profiles using a fortran code implementing the Minuit package \citep{James1975}, originally developed for high-energy physics. 
In the first run we fit only the prominent emission lines which have high S/N and are less affected by atmospheric features. Then, we initialize the parameters with the values obtained previously.  In the fitting procedure, the $\chi^2$ minimization is done using as error on single fluxes the variance evalutated in the continuum ranges previously indicated.
The need for the second set of lines was evaluated based both on the errors of the fits and a visual inspection of the residuals (see Section~\ref{sect_FWHM}). 

\begin{table*}
\begin{minipage}{175mm}
\footnotesize
\normalsize
\caption{Fit results}
\begin{tabular}{rccccccccl}
\hline
\multicolumn{9}{c}{Fit to the [OIII]5007 line in XMM-COSMOS obscured QSOs}\\
\hline
XID & specz & $\lambda$,S &FWHM(S) &  Flux(S) & $\lambda$,B & FWHM(B) & Flux(B)  & $\Delta$v  & $\chi^2_{\rm red}$ \\
&  &  nm &  km s$^{-1}$ & (10$^{-20}$) &nm &  km s$^{-1}$ & (10$^{-20}$) & km s$^{-1}$  & \\
\hline
      18 &   1.6073 &  (1305.52) &  (404 $\pm$85) &  (from H$\alpha$ fit)   &   1304.28 & 1065 $\pm$  409 &   4883  $\pm$   1875 &     -287  &  0.75\\ 
   $^{\ddag}$60053 &   1.5812 &  1292.35 &  272 $\pm$   11 &   782  $\pm$    65 &    $-$ &  $-$ &  $-$ & $-$ & 0.74 \\
     175 &   1.5297 &  1266.44 &  513 $\pm$   46 &  8167  $\pm$   301 &   1263.53 & 1652 $\pm$  150 &   7541  $\pm$   685 &     -688 & 0.50 \\ 
    2028 &   1.5927 &  1298.24 &  520 $\pm$   17 & 29188  $\pm$   658 &   1296.45 &  913 $\pm$   37 &  17127  $\pm$   694 &     -413 & 0.99\\ 
    5321 &   1.4702 &  1236.80 &  272 $\pm$   14 &  5556  $\pm$   149 &   1238.31 & 1306 $\pm$   14 &  74664  $\pm$   800 &      366 & 1.10\\ 
    5053 &   1.3735 &  (1188.48) &  (538 $\pm$   24) &   (from H$\alpha$ fit) &   1187.96 & 1372 $\pm$  616 &   3655  $\pm$   1641 &     -131 & 1.74 \\ 
    5325 &   1.3809 &  1191.85 &  403 $\pm$   12 &  4448  $\pm$    52 &   1194.45 & 1050 $\pm$  210 &   3313  $\pm$    662 &      653 & 1.76 \\ 
    5573 &   1.1152 &  1059.14 &  233 $\pm$   10 &  7982  $\pm$    93 &     $-$ &  $-$ &  $-$ & $-$ & 1.02\\

\hline
\end{tabular}
\hspace{3.2cm}
Notes: Fluxes are in units of 10$^{-20}$ erg cm$^{-2}$ s$^{-1}$. 
$\lambda$(S), FWHM(S) and Flux(S) denote the best fit parameters and errors for the ``systemic'' (S) component; $\lambda$(B), FWHM(B) and Flux(B) instead refer to the ``broad'' (B)  component. $\Delta$v is measured from the difference in centroids of the 2 measured components (B-S). Values in parenthesis refer to measurements constrained from the H$\alpha$ region fit. 
$^{\ddag}$: 2 components fit not significant. Narrow component consistent with only narrow lines seen in the H$\alpha$ region.
\end{minipage}
\end{table*}

The best fit solution of the modeling described above in the region of the H$\beta$+[OIII] is shown as a red curve in Figure~\ref{fit}. 
The best fit gaussian components needed to fit the full line profiles are also superimposed  with arbitrary normalization in order to ease the visualization: the ``S'' component, as solid/blue, the ``B'' component as dashed/blue, the ``VB'' component in solid/green.  The bottom panel shows the residuals ((data-model)/error, where the error is estimated in the local continuum) that, added in quadrature, determine the $\chi^2$. 
In the right panel for each spectrum we also show the corresponding fits for the  H$\alpha$+[NII]+[SII] region used as additional constraints on the FWHM of the [OIII] lines profiles: in particular, the ``S'', ``B'' and ``VB'' 
components, when present, are fixed at the same redshift and relative shifts as those present in the left panels. 
Table 2 summarizes the parameters obtained for the ``systemic'' (S) and ``broad'' (B) components for the [OIII]5007 line. The reduced minimum $\chi^2_{red}$ values obtained for the proposed best fit solutions are reported in the last column.

Before discussing the results on the broad component associated with the outflow, we note that the FWHM and fluxes of the ``VB'' component associated with the BLR are in agreement with those presented in \citet{Bongiorno2014}, despite the different modeling of the narrow components, and we refer to that paper for the estimate of the BH masses. 
We also note that in most of the sources where a ``VB'' component in the H$\alpha$ region is detected, originating from the BLR (XID5321, XID2028, XID18, XID175, XID60053 and XID5325; see also \citealt{Bongiorno2014}) the H$\beta$ is considerably extincted (with R(H$\alpha$/H$\beta)\sim$5-10). This is consistent with a Type 1.8-1.9 nature of the objects and with the moderate obscuration measured in the X--ray spectra (N$_{\rm H}\sim10^{21-22}$ cm$^{-2}$).

\subsection{Incidence of broad [OIII]5007 emission lines}
\label{sect_FWHM}

Four out of 8 sources (XID2028, XID5321, XID175, and XID5325) need all three sets of gaussians to reproduce simultaneously the H$\beta$+[OIII] and H$\alpha$+[NII] line profiles. The fit with a single set of gaussian lines to model the NLR emission, even without imposing an upper limit on the FWHM, produced a significantly larger $\chi^2$. This is particularly true for our highest S/N sources (XID5321 and XID2028), where the [OIII] emission is clearly asymmetric and a single component therefore cannot reproduce the observed emission \citep{Perna2014}.
In all cases, the normalizations of the ``S'' and ``B'' gaussians lines needed to fit the [OIII] profiles were not consistent with zero (while these may be the case for other lines of the same components at lower S/N, e.g. [SII]). In 4 cases, the ``B'' component shows a significant shift ($|\Delta$v$|>300$ km s$^{-1}$) from the systemic redshift of the galaxy: XID2028 and XID175 reveal a blueshift component, while XID5321 and XID5325 show a redshifted component. We note that the velocity shift measured for the [OIII] line gaussian decomposition for XID2028 ($\Delta$v$\sim -370$ km s$^{-1}$) is consistent with the value reported in B10 and measured from the shift of the MgII absorption lines  in the Keck spectrum ($\Delta$v$\sim -300$ km s$^{-1}$). A similar fraction of ``double'' to ``single'' line modeling has been found in H12:  4 out of the 8 targets presented in that work needed a multiple component fit (see their Figure 3). 

For the remaining 4 targets, in 2 cases  (XID18 and XID5053) the quality of the spectra was not such to allow a 2 gaussian decomposition below the [OIII] lines and the fit has been limited to a single, blueshifted broad component\footnote{In this case the definition of ``S'' and ``B'' is superseded, but we report these values in the column ``B'' in Table 2.}, whose centroid  has been constrained from the combined fit on the H$\alpha$ region. The detection of the broad component for these two sources is significant at the 2-3$\sigma$ confidence level. 
For XID5573, the best fit of the [OIII] line has been obtained with only a narrow component (FWHM$\sim230$ km s$^{-1}$).
Finally, for XID60053 only narrow components are detected in the H$\alpha$ region, in correspondence with two peaks observed also in the [OIII] lines (as labeled in Figure~\ref{fit}). The observed H$\alpha$ flux, if ascribed entirely to SF, translates into a SFR$\sim10$ M$_\odot$ yr$^{-1}$ \citep{Kennicutt1998}. From the comparison of the SFR from the FIR and the H$\alpha$, we infer a lower limit to the extinction of A$_{\rm V}\sim6$ mag. 
Such a high extinction would also suppress most of the [OIII] flux which indeed is only barely detected in the X-shooter spectrum.

Summarizing, for 6 out of 8 targets we report the presence of broad and shifted components with FWHM in the range 900-1600 km s$^{-1}$ (4/6 with high significance, while the remaining 2 at a $\sim2-3$ sigma confidence level), while in the remaining 2 sources  only narrow components (FWHM$\sim250$ km s$^{-1}$) are revealed in the [OIII] lines. The incidence of red-shifted lines in the sample with broad components is 33\% (2/6).

\begin{figure*}
\includegraphics[angle=0,scale=0.9]{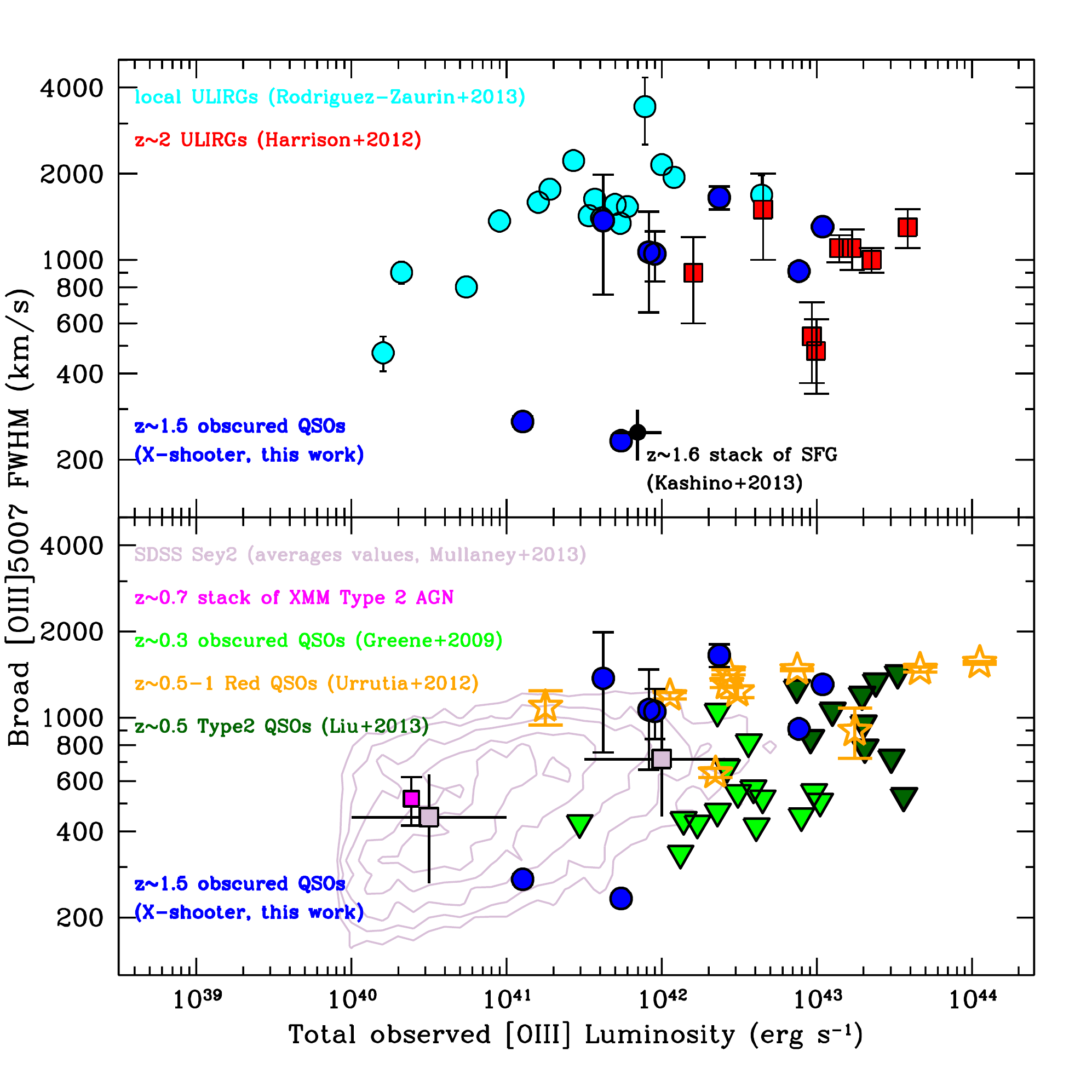}
\caption{FWHM (broad component) of the 5007\AA\ line against the total [OIII]5007 luminosity. The [OIII] luminosity is not corrected for extinction. In both panels our X-shooter targets are marked with blue circles. For completeness, we also plot the 2 sources  (XID5573 and XID60053), fitted by only a narrow component in the [OIII] lines, by using the FWHM of the ``S'' component. Our results are compared with known literature samples of Star forming systems (both ULIRGs and MS; {\it Upper Panel}) and Sey2 and Type 2 QSOs ({\it Lower Panel}), as labeled (see Section 5.1 and 5.2).  
The black filled circle in the upper panel represents the average for a population of 30 massive star forming galaxies at z$\sim1.6$ (\citealt{Kashino2013}, see text for details).
The magenta square in the lower panel represents the result from the stacked spectrum of $\sim110$ XMM-COSMOS Type 2 QSOs with spectra available from the zCOSMOS 20k survey in the range z=0.5-0.9,  without any preselection on their optical/IR  colors. 
Following H12, in this plot in case of fits with multiple gaussian components,  the FWHM from the ``B" component is plotted. Otherwise, the FWHM derived from a single component fit is adopted. For the objects in the sample of \citet{Liu2013} we use the velocity widths containing 80\% of the flux (W80). The [OIII] luminosity is instead derived from the total [OIII] flux for all samples. 
}
\label{FWHM}
\end{figure*}

\section{The origin of the broad component}
\label{sect_broad}

The detected FWHMs are far too high to be due to rotational motions in the host galaxy, for which the velocity dispersions rarely exceed 600 km s$^{-1}$ (see e.g. discussion in \citealt{Liu2013}). Similarly, we rule out that  the observed large and shifted velocities may be all ascribed to complex kinematics as a result of a merging system, as the deep ACS images of the sources do not show any obvious counterpart and sign of major mergers in the 1" aperture used for the extraction of the spectra (see Figure~3).

Emission from ionised gas in the forbidden lines, like [OIII], is suppressed by collisional de-excitation when produced in high-density environments. Therefore, the observed large FWHM of [OIII] cannot be ascribed to the BLR (usually confined to $<1$pc scale).  
Instead, forbidden ionised lines can be produced at scales of the NLR. 
Moreover, Type 1 and Type 2 quasars are also often associated with extended emission line regions (EELR) which can extend sometimes for tens of kpc 
(e.g. \citealt{Boroson1985,Fu2009,Villar2011a}), well beyond the size of the NLR.  
\citet{Fu2009} noted that the most likely explanation for the existence of EELR in radio loud samples is the presence of gas swept out of the host galaxy by a  blast wave accompanying the production of the radio jets. Similar conclusions have been reached by \citet{Matsuoka2012} and extended to the whole AGN population (see also \citealt{Villar2011a}).
The chosen apertures in our samples (corresponding to R$\sim4$ kpc) contain emission from both the NLR and,  when existing, the EELR. 
The observed line emission may be therefore broadened by bulk flow emission  likely ascribed to an outflowing wind in conditions similar to those observed in EELR (see \citealt{Matsuoka2012}). In this case the large FWHM observed may be directly related to the outflow velocity and the analysis of NIR integrated spectra of AGN at z$>$1 bring important information on the outflow properties (see for example \citealt{FS2014}).

Following H12, in Figure~\ref{FWHM} we plot as blue circles the FWHM of the broadest components (``B" or ``S" in case of single gaussian fits)
in our measurements versus the total  [OIII] luminosity, and we compare them with different literature samples, both local and at high redshift.  
We chose to plot the total {\it observed} [OIII] luminosity (i.e. not corrected for extinction) in order to ease the comparison with previous published works. 
The possible effects of reddening will be discussed when relevant and we will take this into accounts when referring to the energetic of the systems.

\subsection{QSOs in Starforming/ULIRGs hosts}
\label{sect_ULIRG}
As mentioned in the Introduction, most previously reported studies which showed convincing evidence of the existence of large scale outflows have been conducted on AGN-ULIRGs systems. Given the concomitant presence of on-going SF and BH activity, it is not obvious to determine which is the main driver of the observed outflowing wind. 
In the {\it bottom} panel of Figure~\ref{FWHM} we compare our results on the FWHM and integrated line flux with those obtained for  QSO-ULIRGs systems, at both low and high redshift. 

Rodriguez-Zaurin et al. (2013, hereinafter R-Z13) showed that in a nearly complete sample of local ULIRGs, 
those associated {\it with} Seyfert nuclei show evidence of broad and shifted lines on scale of $\sim$5kpc,  
while those {\it without} Seyfert nuclei do not show the complex and extended kinematics observed in their active siblings (see also \citealt{Zakamska2014}; see similar results on striking differences in the molecular gas outflow properties of AGN ULIRGs versus non-AGN ULIRGs presented in \citealt{Cicone2014}). 
This may be seen as an indication that the presence of the AGN rather than the on-going star-formation may be the major cause for the complex (extended) NLR kinematics in local ULIRGs (but see \citealt{Soto2012,SotoMartin2012} for 
different results).
The complex kinematic properties of the Sey-ULIRGs systems in the local Universe  (with up to 3 or more different components needed to fit the observed [OIII] profiles) emerges also in Figure~\ref{FWHM}  (upper panel):  most of the objects in the R-Z13 sample (cyan points)  have  FWHM$>1000$ km s$^{-1}$. 
The presence of outflowing ionised gas in these systems is confirmed by IFU data (see e.g. \citealt{Westmoquette2012}, where the fastest outflows are associated with systems that contain AGN), and by the fact that the broadest components measured in R-Z13 lie in the AGN-photoionized region in the BPT diagrams \citep{Baldwin1981}.

The red squares in the upper panel of Figure~\ref{FWHM}  are the 8 z$\sim2$ SMG/ULIRGs with AGN signatures presented in H12, with on average FWHM$\sim1000$ km s$^{-1}$. Thanks to the availability of IFU data, the presence of large scale outflows has been unambiguously traced  up to scales of 10-20 kpc  also in the majority of these systems.  
Similarly, \citet{FS2014} observed with SINFONI 8 massive (M$_*>10^{11}$ M$_\odot$) star forming galaxies at z$\sim2$, half of them with clear AGN signatures. On the basis of spatially resolved spectral analysis of the H$\alpha$ complex, they found evidence for powerful AGN-driven nuclear outflows with FWHM$\sim1500$ km s$^{-1}$ out to scales of 2-3 kpc in the stacked spectrum of these massive systems.\footnote{We do not report the results of \citet{FS2014} because we do not have information on the [OIII] flux.}.

Altogether, these  observational results  suggest that  FWHM larger than 1000 km s$^{-1}$ can be safely used to advocate the presence of kpc scale outflows, and that the AGN is likely the driving force. 
The 6 XMM-COSMOS obscured QSOs for which a broad component has been detected have FWHMs comparable to those measured in z$\sim2$ QSO/SF systems.  Different from these systems, though, as detailed in Section~\ref{sect_hosts}, our 6 targets have not been preselected on the basis of their SF properties 
and span a quite large range of SFR, from $\sim500$ M$_\odot$ yr$^{-1}$ to basically passive systems. 
Overall, this may be another indication that the AGN rather than the on-going star-formation sample may be the major driver for the presence of the observed broad and shifted components.

In order to compare our results with pure star forming systems, we plot in Figure~\ref{FWHM}  the average FWHM ($\sim4.2$\AA\, corresponding to $\sim 250$ km s$^{-1}$ in the velocity space) as a function of the uncorrected [OIII] luminosity measured on the stacked spectrum of a sample of 30 massive (logM$_*$=10.76-11.35) star forming galaxies selected to be on the MS at z$\sim 1.6$ and observed with FMOS in the COSMOS survey (\citealt{Kashino2013}, black filled circle).  In this case, the stacked spectrum has been constructed by carefully excluding AGN from the sample. All our targets have measured FWHM of the broad (and shifted) component well above the average value of star forming galaxies at the same redshift (see also \citealt{Newman2012}).  If outflows driven from SF winds were common in MS galaxies at z$\sim1.6$, these would translate in a broadened FWHM in the stacked spectrum, which instead is not observed. 

The only source above the MS in our sample with properties comparable to the SMG/ULIRGs presented in H12 and for which we have the [OIII] spectrum is XID60053  (SFR$\sim900$ M$_\odot$ yr$^{-1}$). This object shows only narrow (``S'') components in the combined fit of the H$\alpha$ and [OIII] lines. 
The possible CT nature for this source, coupled with the  other observed properties (high SFR, high extinction, irregular morphology, and accretion rate at the Eddington level; see Sections~\ref{sect_agn} and \ref{sect_FWHM}) point towards the interpretation that XID60053 may be caught in the ``dust-enshrouded'' phase of rapid black hole growth which should occur before the feedback phase. This would naturally explain the non detection of strong broad components (as observed in XID5321 and XID2028) despite the similar intrinsic AGN luminosity (see next Section).

\begin{figure}
\includegraphics[angle=0,scale=0.43]{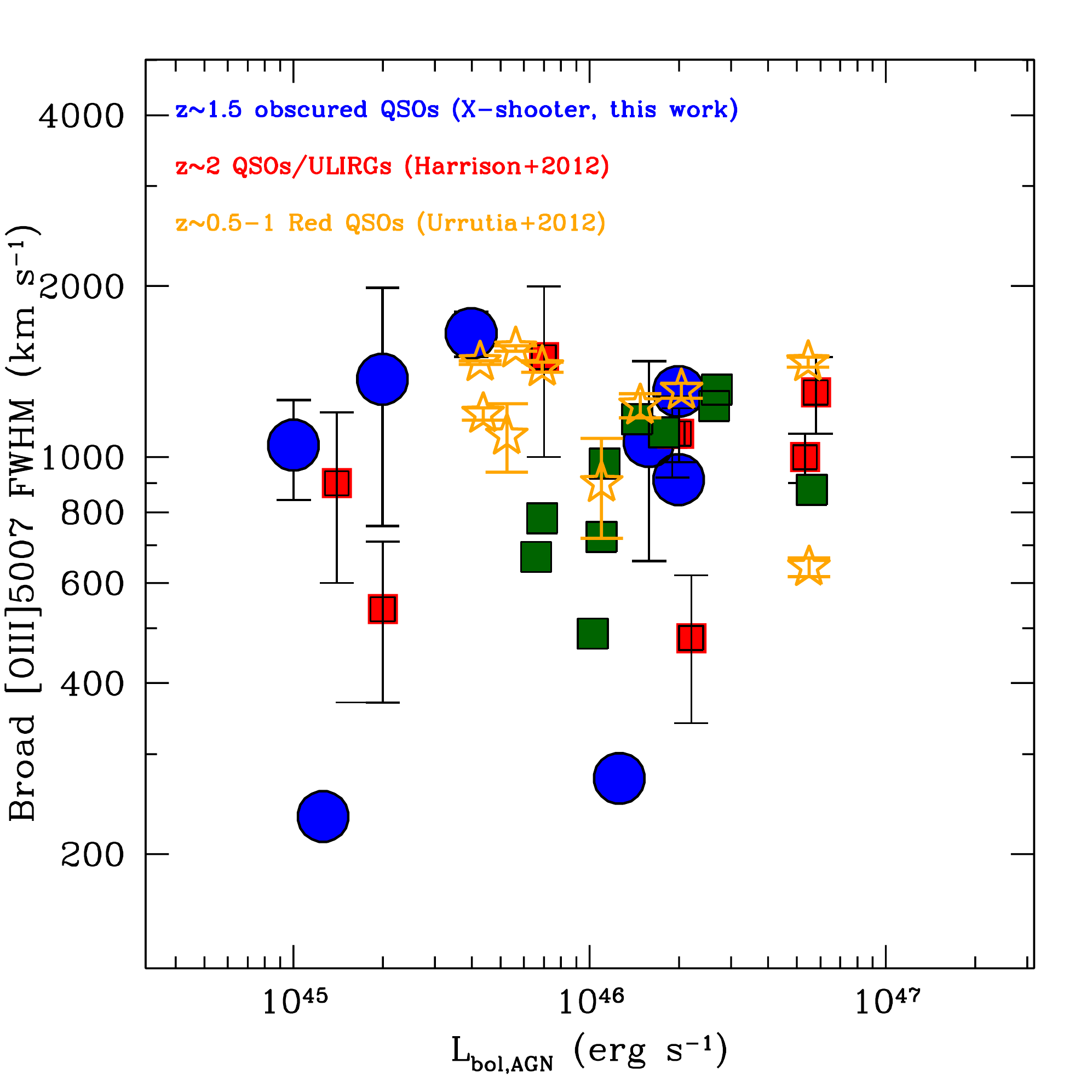}
\caption{FWHM (broad component) of the 5007\AA\ line against the AGN bolometric luminosity for our targets and other 2 comparison samples for which AGN bolometric luminosities from SED fitting are available, as labeled (see text for details).}
\label{notrend}
\end{figure}

\subsection{Type 2 AGN samples}
\label{sect_AGN2}
We now compare our results with those reported in the literature for objects selected on the basis of a purely AGN classification.  


\citet{Mullaney2013} presented the analysis from a multicomponent (allowing for the presence of a broad component) line fit of the [OIII]5007 line in the SDSS population.  In the {\it lower} panel of Fig.~\ref{FWHM} we report the contour levels extracted for the Type 2 AGN only, at observed total [OIII] luminosities larger than 10$^{40}$ erg s$^{-1}$. The pinkish-grey squares represent the average values of the ``broad'' FWHM  in two luminosities ranges (L[OIII]$\sim10^{40.5}$ erg s$^{-1}$ and L[OIII]$\sim10^{42}$ erg s$^{-1}$).
All but two of our X-shooter targets have FWHM $>716$ km s$^{-1}$, which represents the average FWHM of the broadest component in the SDSS Type 2 AGN sample at  L[OIII]$\sim10^{41.5-42.5}$ erg s$^{-1}$. We note that objects with FWHM $>900$ km s$^{-1}$ at L[OIII]$\gsimeq10^{42}$ erg s$^{-1}$ are rare in the SDSS sample ($\sim$2\%, see \citealt{Harrison2014IAU}) while all of our 5 targets with observed L[OIII] larger than this luminosity threshold revealed a broad component with FWHM larger than the SDSS average. 
We also do not find a clear trend of the broad FWHM with the L[OIII] in our sample, as already pointed out in \citet{Harrison2014IAU}. 

In order to verify the efficiency of  selection criteria applied to X-ray sources in detecting objects with large FWHM, we constructed the stacked spectrum of all XMM-COSMOS Type 2 AGN at z=0.5-0.9  for which [OIII] is visible in the zCOSMOS spectra and without imposing any preselection on their optical/IR  colors ($\sim110$ objects).
We measured the FWHM in the average spectrum and the fit is consistent with a single and symmetric line component with FWHM$\sim 540$ km s$^{-1}$ (magenta square in Fig.~\ref{FWHM}). This value is consistent  with the average value of the broadest component observed in the SDSS sample at comparable observed [OIII] luminosities (L[OIII]=$10^{40-41}$ erg s$^{-1}$, FWHM$\sim450$ km s$^{-1}$; see also \citealt{Heckman1981}).
We note that both the SDSS Type 2 and the XMM-COSMOS Type 2 samples may contain also objects in the feedback phase (which occur at different L and redshift due to the downsizing) and therefore with individual large FWHM associated with blueshifted (or redshifted) [OIII] lines, but they are washed out in the average stacking.

The higher average FWHM measured in our sample with respect to the z$\sim0.7$ XMM-COSMOS Type 2 AGN may be due to the larger luminosity of our sample, and may be in principle simply ascribed to the fact that more luminous systems are on average larger and therefore the NLR extends at larger radii (e.g. \citealt{Netzer2004}; see also R-Z13, \citealt{Greene2011,Hainline2013}; see also the higher average FWHM in SDSS Type 2 AGN at high L[OIII]).  
In the lower panel of Fig.~\ref{FWHM} we also plot the results for 15 Type 2 QSOs from the SDSS studied in Greene et al. (2009,2011), with total observed L[OIII]$\gsimeq10^{42}$ erg s$^{-1}$ (light green triangles), therefore more directly comparable to our targets. 
In this case no further selection in addition to line ratio diagnostic has been applied. 
Although the authors indicate outflows on scales extending from few up to 10 kpc as a possible origin for the observed broad widths,
we notice that, on average, their values (average FWHM$\sim525$ km s$^{-1}$) are consistent with those observed in the SDSS Seyfert 2 sample, and a factor of $\sim2$ lower than the average observed in our sample. As a comparison we also plot the radio-quiet Type 2 QSOs at L[OIII]$\gsimeq10^{43}$ erg s$^{-1}$ (dark green triangles, from \citealt{Liu2013}), for which the existence of large scale ($\sim10$ kpc) outflows over most of the extent of the gas emitting region has been convincingly demonstrated via IFU spectroscopy. In this case we plot the W80 non-parametric measure of the line width (see discussion in \citealt{Liu2013}, section 2.3), which at a very first order can be used  as representative of the FWHM of the lines. Also in this case, we found an average value of the line width comparable to that measured in our targets, despite their one order of magnitude larger observed [OIII] luminosities.

Overall, the comparison of the results obtained in our sample and the other  samples discussed above may be seen as an indication that the color selection applied to our X-ray sample is effective in picking up objects with FWHM larger than the average values. 
\citet{Urrutia2012} presented a sample of 13 dust-reddened QSOs at z=0.5-1, selected solely on the basis of a red J-K color. The intrinsic E(B-V) $\sim0.5$ of these targets is similar to what is observed in our targets. 
We have analysed 11 out of 13 objects from the \citet{Urrutia2012} sample using the spectra provided by \citet{Glikman2012} and applying a model similar to that described in the previous Section~\ref{sect_model}, limited only to the H$\beta$+[OIII] region but including a proper modeling of the FeII lines. The details of the fits and the results are reported in the Appendix, and the FWHMs associated with the broad component are shown as orange stars in the lower panel Figure~\ref{FWHM}.
In the hypothesis that the broad components can be ascribed to outflowing winds (see previous subsection), the high incidence of very broad lines in ours and the \citet{Urrutia2012} sample (13/19 have FWHM$>1000$ km s$^{-1}$) in such dust reddened QSOs may be an additional evidence that the blow-out phase is indeed heavily obscured, on the entire galactic scale.  

We finally consider all the samples  for which the bolometric  luminosity has been derived in a similar way from a multiwavelength SED fitting (the H12 sample, the \citealt{Urrutia2012} sample and our sample), and therefore an estimate of the {\it intrinsic} AGN bolometric luminosity can be obtained. In this way we remove the effect of the reddening on the [OIII] luminosities.
The FWHMs as a function of the intrinsic (bolometric) AGN luminosities for this combined sample of 27 objects are plotted in Fig.~\ref{notrend}.  We span a 2 orders of magnitude range in L(AGN) (10$^{45}$-10$^{47}$ erg s$^{-1}$) and no clear trend between these two quantities is seen in these data. This result is at odds with the results recently presented in \citet{Zakamska2014}, where a trend of the [OIII] width with the IR luminosity is seen in SDSS luminous Type 2 quasars, as expected for outflows driven by the radiation pressure of the quasar (e.g., \citealt{Menci2008}).

\section{Mass outflows rates and energetic}
\label{sect_kinetic}
In the previous Section we have reported compelling evidences that the color selection applied to our X-ray sample is effective in picking up objects with disturbed kinematics, which can be likely ascribed to the presence of outflowing wind. 
In order to unveil the mass and energy involved in the wind component,  the bulk outflow velocity, the knowledge of  the distribution (geometry), the density of the gas ($n_{e}$), and the spatial scale (e.g. the radius R of the emitting volume) are needed. We can derive an order of magnitude estimate of the expected outflow power under reasonable assumptions on these quantities, as detailed below. 

Following the arguments presented in Cano-Diaz et al. (2012, see their Appendix B),   
a lower limit on the kinetic power ($ \dot{E}^{ion}_{K}$) associated with the outflows  can be given by: 
\begin{equation}
\rm \dot{E}^{ion}_{K} = 5.17~10^{43}
  \frac{\mathcal{C}~L_{44}([OIII])~v_3^3}{\langle n_{e3}\rangle ~10^{[O/H]}~R_{kpc}}~erg~s^{-1}.
         \label{eq_edot}
\end{equation}
where $\rm L_{44}([OIII])$ is the [OIII] luminosity of  the broad component in units of 10$^{44}$ erg s$^{-1}$, $\rm n_{e3}$ is the electron density in units of 1000 cm$^{-3}$, $\mathcal{C}$ is the condensation factor ($\approx 1$), $\rm v_3$ is the outflow velocity in units of $\rm 1000~km~s^{-1}$, 10$^{[O/H]}$ is the metallicity in units of solar metallicity, and $\rm R_{kpc}$ is the radius of the outflowing region, in units of kpc. 

The spatial scale sampled by the extraction window ($\sim1\arcsec$; see upper panels in Fig.~\ref{spectroz}) corresponds to about 8 kpc at z$\sim1.5$. Therefore, 
for  the radius we assumed $\rm R_{kpc}=5$, consistent with most of the outflow measured in our data being confined in the near-nuclear region (see also R-Z13).

We can determine an estimate of the density of the outflowing gas from the ratio of the flux in the broad components of the [SII] doublet (r=I(6717)/I(6731)) obtained in the fit
only for our highest S/N target (XID5321). We measure a ratio of r$\sim0.65\pm0.15$, which translates into 1000-3000 cm$^{-3}$  for reasonable temperatures assumptions \citep{Osterbrock1989, Stanghellini1989}. However, in the absence of a measurement for all the sources, we decided to adopt a value of n$_{e}=100$ cm$^{-3}$, as routinely adopted in similar studies of the ionized components based on H$\beta$ fluxes (e.g. \citealt{Liu2013,Harrison2014}). This choice is also justified by the slit-resolved spectral analysis of XID5321 presented in \citet{Perna2014}, where we infer n$_{e}=120$ cm$^{-3}$ from  the [SII] ratio at the outflow position and outside the central 1$\arcsec$ extraction. 
In the equation described above, we therefore adopted for $\rm n_{e3}=0.1$. 

Finally, for the estimate of the outflow velocity we used the maximum velocity  inferred from the [OIII] profile (v$_{\rm max}$), which is probably representative of the average outflow velocity (see Cano-Diaz et al. 2012), as confirmed by our independent analysis on the two brightest targets based on X-shooter slit-resolved spectroscopy \citep{Perna2014} and SINFONI data (Cresci et al. in preparation). For our targets this assumption leads to bulk velocities of the order of $\sim1200-1800$ km s$^{-1}$. 
Further assuming a spherical geometry and solar metallicity, using for $\rm v_3$=v$_{\rm max}$/1000  and computing L[OIII] from the integrated flux in the broad component only, we can estimate the outflow kinetic powers for our targets associated with the {\it ionized gas component}.  The values obtained from Equations 1 for our 6 sources with a broad components are listed in Table 3, along with the velocity measurements used in the computation of the outflow powers (v$_{max}$).

\begin{figure*}
\includegraphics[angle=0,scale=0.43]{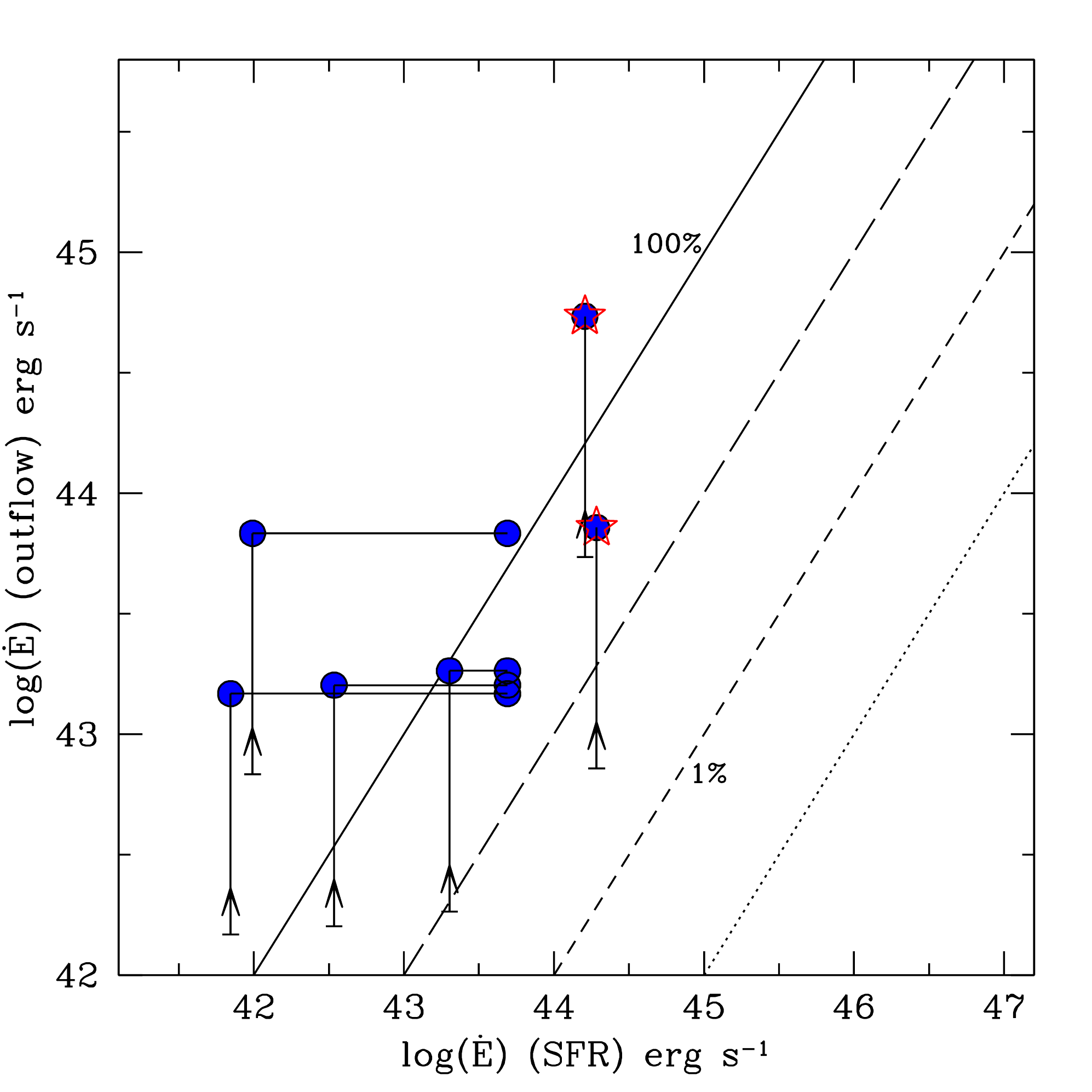}
\includegraphics[angle=0,scale=0.43]{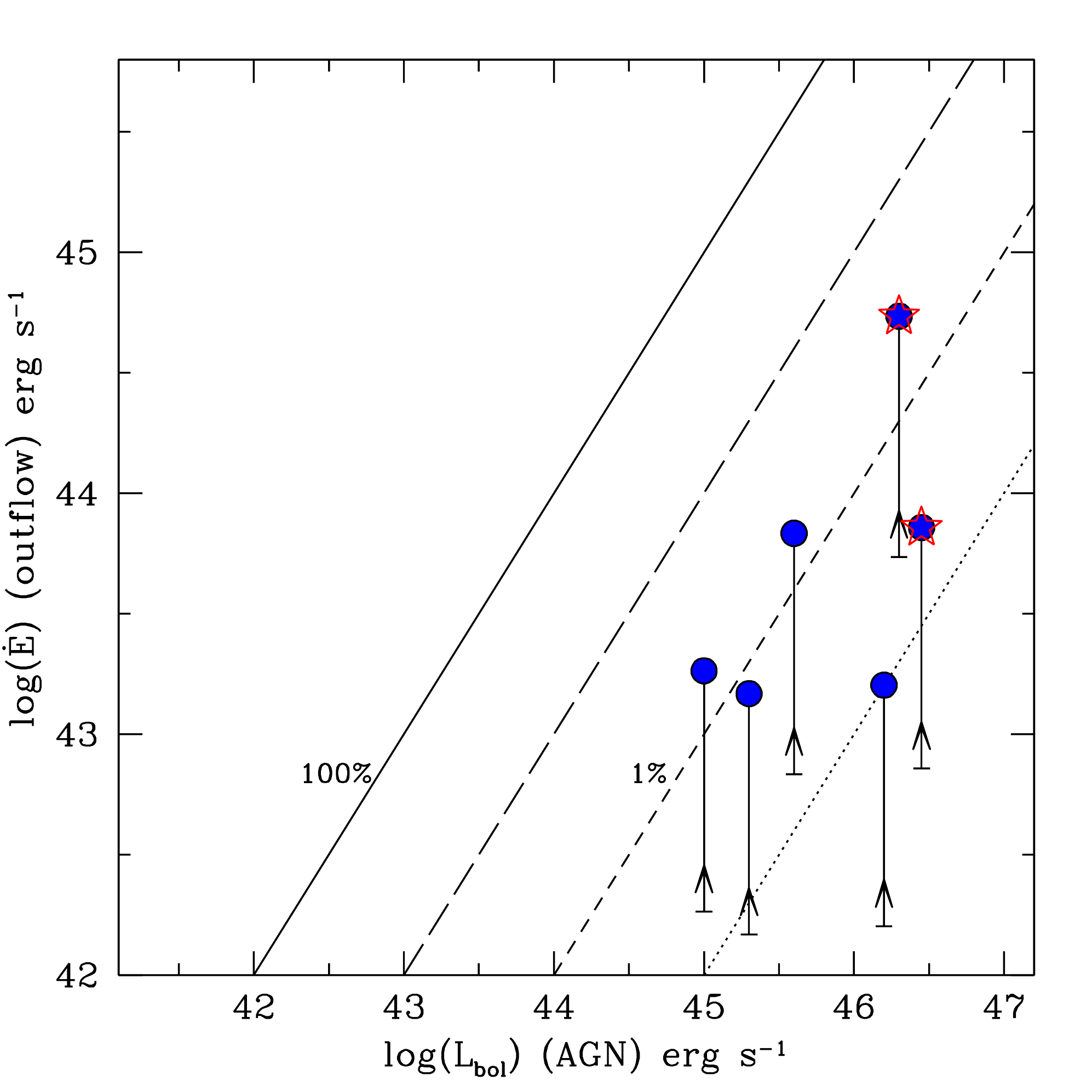}
\caption{ ({\it Left Panel}): Our inferred values of the  outflow kinetic energy injection rate (i.e. last column of Table 3) against the predicted energy input rate from SF. Both quantities are derived as described in Section~\ref{sect_kinetic}. The solid, long-dashed, dotted and short-dashed lines represent the 100\% (one to one correlation), 10\%, 1\% and 0.1\% ratios, respectively.  Starred symbols are the sources detected in PACS (XID2028 and XID5321). 
For sources not detected by Herschel, the predicted energy input from SF estimated from the SFR measured from the SED fitting (left symbol) and from the PACS stacked signal (right symbol) are connected by a solid line. The outflow kinetic powers associated with the the ionised component  and derived using Equation 1 are shown as lower limits (upward arrows). The value for XID2028 (leftmost starred symbol) has been shifted in the horizontal axis by 0.15 dex for clarity in the plot.
({\it Right Panel}): our inferred values of the outflow kinetic energy injection rate against the bolometric AGN luminosity as derived from SED multicomponent fitting (see Section~\ref{sect_agn}). The lines and symbols have the same meaning as in the left panel. 
\label{edot}}
\end{figure*}

\begin{table}
\footnotesize
\normalsize
\caption{Outflow kinetic powers}
\begin{tabular}{rccc}
\hline
XID & v$_{max}$ & log($\dot{E}^{ion}_{K}$) & log($\dot{E}^{tot}_{K}$)    \\ 
(1)  & (2) & (3)  & (4) \\
\hline
   18 & 1230 &    42.20  & 43.20 \\
 2028 & 1350 &    42.86 & 43.86 \\
  175 & 1800 &    42.83  & 43.83 \\
 5321 & 1730 &    43.73  & 44.73\\
 5053 & 1500 &    42.17  & 43.17 \\
 5325 & 1660 &    42.26  & 43.26 \\
\hline
\end{tabular}
\hspace{20.2cm}
Notes: (1) XID; (2)  maximum velocity inferred from the line profile, in km/s; (3)  outflow kinetic power of the {\it ionized} gas component as derived from Equation 1 (lower limits);
(4) Inferred total outflow rates (see Section 6 for discussion). 
All values are in log scale, units are erg s$^{-1}$. 
\end{table}

Admittedly, the values listed in Table 3 suffer from large uncertainties, given that the exact measurements of the densities, emissivities (depending on the temperature of the medium), metallicity and [OIII] ionization status are not possible with our data.
However, we are confident that the estimates obtained through Equation~1 should be safely regarded as absolute lower limits to the total outflow rates, for several reasons.

First, we note that 
we do not correct the [OIII] luminosity for the extinction and therefore we may underestimate the ionizing flux. Using the intrinsic [OIII] luminosity (corrected for reddening) in Equation~1 the estimate of $\dot{E}^{\rm ion}_{\rm K}$ may increase by a factor 2-8, depending on the source.

Second, in deriving Equation~1 Cano-Diaz et al. (2012)  assumed that all the oxygen ions are in the O$^{2+}$ state and that oxygen is a good tracer of the ionised gas. H$\beta$ may provide a better tracer of the gas, given that it is independent from the assumptions on the metallicity and of the ionisation state. Unfortunately, in all but our two highest S/N targets, the broad and shifted H$\beta$ component remains undetected. In the 2 cases where it is detected, the H$\beta$ flux is a factor $\sim10$ lower than the [OIII] flux (see also \citealt{Perna2014}). This would translate into a factor of  $\gsimeq5-10$ larger kinetic powers, when adopting routinely used calibrations from H$\beta$ \citep{Liu2013,Harrison2014}, and depending on temperature assumptions. A difference of a factor of 10 is further confirmed by our spatially resolved analysis of XID2028 based on deep SINFONI J-band data (Cresci et al. 2014). 

Third,  and most important, the ionized wind is most likely coupled to outflowing neutral and/or molecular gas components, of the same order of magnitude of the ionised one. These additional gas components easily sum up to one order of magnitude more massive material present in the outflow, as observed in the prototype outflowing QSO Mrk231 in the local Universe, or even more, given the typical SFR or our targets (see \citealt{Harrison2014}). 

Following the discussion above, 
the total outflow rates of the sources in our sample can be reasonably inferred by applying a correction factor of 10 to the measured quantities (e.g. the lower bounds of the corrections described above).  We stress again that these inferred values are only meant to be representative of the order of magnitude of the expected {\it total outflow power} associated with the broad components measured in our sample. These values are reported in the last column of Table 3, and are plotted in the y-axes of Figure~\ref{edot}.

\subsection{Origin of the wind}

In order to investigate the possible origin of the wind, we need to compare our inferred values of the {\it total outflow kinetic power} with the AGN bolometric luminosities and the kinetic  power expected to be ascribed to  stellar processes.  

The kinetic output expected from stellar processes ($\dot{E}$(SF)) has been assumed to be proportional to the SFR, and is at most $\sim7\times10^{41}\times$(SFR/M$_\odot$ yr$^{-1}$) following \citet{Veilleux2005}. 
For the AGN luminosity we use the values derived in  Section~\ref{sect_agn}.

Figure~\ref{edot} (left panel)  shows  our inferred total outflow kinetic powers versus the predicted energy input rate from SF. Starred symbols are objects with PEP detections. For all the others, the expected $\dot{E}$(SF) assuming a SFR=70 M$_\odot$ yr$^{-1}$ (e.g. the value corresponding to the stacked PACS signal, see Section~\ref{sect_hosts}) is also shown (rightmost point in the x-axis) paired with the value obtained from the SFR derived from SED fitting (leftmost point). The right panel shows instead the inferred total outflow kinetic power against the AGN bolometric luminosity.  The different lines in both panels refer, from left to right to 100\%, 10\%, 1\% and 0.1\% energy output ratios. 
As a comparison, in the both panels we also plot the lower limits to the outflow rate as derived from Equation 1 (third column in Table 3; upward arrows) paired with the inferred total outflow powers. 
We notice that the uncertainties associated with the measurements of the outflow kinetic output and related to our spectral modeling (e.g. the uncertainties in the measured velocities and broad line fluxes) can be neglected when compared to the much larger uncertainties coming from the systematics and assumptions discussed above.

Taking the inferred estimates of the total outflow kinetic power,  from Figure~\ref{edot} (left) we see that a $>50$\% coupling between the stellar processes and the wind energy is required in order to explain the inferred energetics, corresponding to mass loading factors of 0.5 (XID2028) to 5 (XID5321), and up to $>10$ for the objects undetected by PACS. Mass loading factors close to unity or even above can be easily produced by SF induced outflows (e.g. \citep{Martin1999,Newman2012}).
 
On the other hand, the AGN luminosities in our systems are largely enough to sustain the inferred outflows powers: the value of $\dot{E}$ is between 0.1\%-5\% of the AGN bolometric luminosity,  in agreement with models which predict   a reasonable coupling between the energy released by the AGN and the one needed to drive the outflow (e.g. \citealt{King2005}).
An AGN origin for the outflow is indeed favoured by the high velocities observed in the winds ($>1000$ km/s). Such high velocities are not commonly reproduced in feedback models of ``pure'' starburst galaxies (e.g. without an AGN at the center),  for which generally velocities larger than 500-600 km/s are not expected  (e.g. \citealt{Murray2005,Ceverino2009,Lagos2013}; but see \citealt{Diamond2012} for a different conclusion, which, however, applies only to galaxies with very high SFR surface densities). The arguments based on the wind velocity would discard a SF origin for the observed outflow even if we do not correct the power observed in the ionised gas for its associated molecular component (see similar arguments also used in \citealt{Genzel2014}).
In addition, we note that the broad FWHMs measured in our targets undetected at 100 and 160 micron  in the PEP survey provide observational evidence of the presence of such strong winds in the ionised gas component in QSOs without high SFR. Overall, this may be another indication that the AGN rather than the on-going star-formation  may be the major driver for the presence of the observed broad and shifted components (see also \citealt{Genzel2014}, for similar conclusions on lower-luminosities AGN). \\

We finally note that an estimate of the {\it total} outflow kinetic power by assuming an energy conserving bubble in a uniform medium and a spherical outflow (covering factor = 1) was  first proposed by \citet{Heckman1990}, and adopted in the recent past by several authors, including Harrison et al. (2012b) for their sample of z$\sim2$ SMG galaxies to which we compare our sample. 
According to models of AGN feedback (e.g. \citealt{King2005}), however,  the assumption of an energy conservation seems to overpredict the local scaling relation and therefore would be basically discarded by local constraints.

\section{Conclusions}
The large body of XMM-COSMOS state-of-the-art  multiwavelength information allowed us to devise a robust
method to isolate candidate objects transitioning from being starburst dominated to AGN dominated, i.e. exactly in the phase when powerful outflows driven by the SMBH are expected. In order to study the physical properties of these systems and shed light on their origin, we obtained follow-up observations with X-shooter at VLT of 10 objects, representative of the entire population of luminous, obscured QSOs at z$\sim1-3$ and with bolometric luminosities of LAGN$\sim10^{45-46.5}$ erg s$^{-1}$. 
This sample is similar in size, but more homogeneous in the selection, with respect to the sample presented in H12, which represents one of the most recent analysis and study of [OIII] profiles in z$\sim2$ AGN-ULIRGs systems at similar spectral resolution ($\Delta$v$\sim$50 km s$^{-1}$). Our sample also shares the same AGN bolometric luminosities of the H12 sample, and of other few additional QSO2 samples of similar size (10-20 objects) at z$<1$ we used as comparison samples.

The main results of the X-shooter observations, presented in this paper, are summarized below:

\begin{itemize}
\item
Thanks to its large and unique wavelength coverage,  X-shooter allowed us to sample simultaneously the observed frame where emission lines are redshifted (H$\alpha$ and [OIII] and H$\beta$ in the NIR spectrograph,  [O II]$\lambda 3727$ in the VIS spectrograph), and determine accurate  redshifts from the presence of multiple emission lines for all but one of the 6 targets for which we had only a photometric redshift estimate, with a success rate of 80\%, significantly larger than what is observed in similar programs of spectroscopic follow-up of red QSOs (see Section~\ref{sect_spectroz}). 
Although photometric redshifts are accurate enough to define AGN samples for cosmological studies, spectroscopic confirmation is mandatory for follow-up observations to probe, in a complementary way, AGN feedback effects.

\item 
In addition to the broad components with FWHM$>2000$ km s$^{-1}$ needed to model the H$\alpha$ and H$\beta$ emission from the BLR in the majority of our targets (``VB'' components in our fits; see \citealt{Bongiorno2014}), we found compelling evidence for the presence of broad components in the fits of the ``narrow'' line profiles of the [OIII]+H$\beta$ and H$\alpha$+[NII]+[SII] regions.  In particular,  four out of eight objects required two sets of gaussians with different widths (FWHM$\lsimeq 510$ km s$^{-1}$ and FWHM$\gsimeq1000$ km s$^{-1}$) to model the 6 forbidden transitions ([OIII], [NII] and [SII] doublets) and the associated narrow components of the Balmer lines on the scale of the NLR and likely beyond.  
In two of the remaining 4 targets for which we fit a single set of gaussian lines we also found a best fit solution with a FWHM$>1000$ km s$^{-1}$, although at a lower S/N ratio. 

\item
Thanks to the information available from the H$\alpha$ line which helped us in better constraining the systemic (``S'') component even in case of low S/N data in the [OIII] region, we were able to measure or infer a shift of the measured broad components in the [OIII] lines from the systemic wavelengths. of the order of and/or shifted $|\Delta v|\sim300-700$ km s$^{-1}$; two out of 6 (33\%)  of the sources for which broad components have been revealed turned out to be red-shifted (see Section~\ref{sect_FWHM}). H12 also reported a similar fraction of redshifted lines (25\%, one out of 4 objects in their sample with double components, see RGJ0302+0010 in their Figure~3). Deep NIR spectroscopy of obscured sources can therefore start to unveil in a much unbiased way (with respect to, e.g. BAL QSOs which favour only one line of sight) the ubiquitous presence of outflows in the full AGN population.

\item
All the observed properties of the only source above the MS in our sample (XID60053: high SFR, high extinction, no detection of broad components) coupled with the irregular morphology, an accretion rate at the Eddington level and the possible CT nature for this source; see Section~\ref{sect_agn}) point towards the interpretation that XID60053 may be caught in the first, still heavily obscured phase of rapid black hole growth. 

\item
We compared the observed large FWHMs in our sample with literature results on pure Type 2 QSOs and Seyferts (lower panel of Figure~\ref{FWHM}) and ULIRGs/AGN systems  (upper panel of Figure~\ref{FWHM}). 
We found that the objects for which we detect a broad component have FWHM larger than the average value observed in SDSS Type 2 AGN samples at similar observed [OIII] luminosity (\citealt{Mullaney2013,Greene2011}, see pinkish-grey and green points in Fig.~\ref{FWHM}).  This may be seen as an indication that the color selection applied to our X-ray sample is effective in picking up objects with FWHM larger than the average values. 

\item
The similarity we observe in the integrated flux profiles and average FWHM of our targets with those derived in other samples  for which the presence of kpc scales outflows likely driven by the AGN have been confirmed by IFU data (e.g. \citealt{Villar2011a,Harrison2012,Liu2013,FS2014} can be considered a clear evidence  that the proposed selection does efficiently work  in order to pick up objects experiencing outflowing winds. 
\item
When performing a similar analysis on a sample of z$\sim$0.5-1 AGN at comparable bolometric luminosity as our targets and selected purely on the basis of a IR color cut \citep{Urrutia2012}, we also found a high incidence of broad lines (with FWHM$>1000$ km s$^{-1}$). This may be an additional evidence that the blow-out phase is indeed heavily obscured, on the entire galactic scale, as predicted in  evolutionary models of AGN. This is also confirmed in our targets by their moderate to high X-ray obscuration and accretion rates (Section 2). 

\item
The main differences with respect to the previous samples at z$\sim2$ is the {\it lower} starburstiness and radio luminosity of our targets: on average, our sources lie on or below the MS of star forming galaxies at z$\sim1.5$ while both \citet{Harrison2012} and \citet{FS2014} targets lie on average on the upper part or above the MS even when considering the redshift evolution of the sSFR  \citep{Whitaker2012,Karim2011}.  

\item 
In systems with substantial SF ongoing among our targets (SFR$\sim100-500$ M$_\odot$ yr$^{-1}$) the kinetic power predicted from stellar winds may in principle be  enough to sustain the kinetic energy associated with the outflows, estimated under reasonable assumptions on the gas conditions (our inferred total outflows kinetic powers; see Section~\ref{sect_kinetic} and left panel of Figure~\ref{edot}). However,  arguments related to the high observed winds velocities ($>1000$ km s$^{-1}$) and to the much lower coupling required to the QSO to drive the outflow  (right panel of Figure~\ref{edot}) seem to suggest that the central luminous QSO is the most likely mechanism responsible for the launch of the wind.

\end{itemize}

\section{Perspectives}

Although based on observations of a small sample of sources, our X-shooter follow-up demonstrated that the adopted selection based on X-ray and optical to MIR red colours may be effective in isolating luminous obscured QSOs in the crucial ``feedback'' phase predicted in  galaxy-AGN coevolution model. 
Large area X-ray surveys at bright X-ray fluxes with associated moderate depth multiwavelength follow-up in the IR bands, such as XXL \citep{Pierre2012} and Stripe-82 \citep{Lamassa2013} can be exploited  to collect larger samples of such rare objects at comparable AGN luminosities. The forthcoming eROSITA survey \citep{Merloni2012} instead will sample the brightest end of the AGN bolometric luminosity and will yeld samples of few thousands X-ray obscured QSOs at z$\sim1-3$ and at L$_{\rm X}>10^{45}$ erg s$^{-1}$. These very powerful systems, still basically unexplored, will be those objects in which the outflows will be routinely discovered and could be studied with unprecedented details. 

Although integrated NIR spectroscopy can be very powerful in detecting the presence of outflows and getting order of magnitudes estimates of the involved kinetic power, slit resolved spectroscopy and the study of the spatial distribution and intensity of the velocity field over the largest possible field of view (obtained through IFU observations such as SINFONI or KMOS)  will be critical in assessing the true energetics associated with the outflows and the corresponding spatial scales.
IFU spectroscopy may also be critical to map the spatial distribution of the SFR (as traced by the narrow component of H$\alpha$) and verify if SF is heavily suppressed in the region with the strongest velocity component (as done in, e.g., \citealt{CanoDiaz2012}).
This can provide a direct observational proof of quasar feedback quenching SF at high-z, measured for the first time on radio-quiet, X-ray selected obscured QSOs, more representative of the full AGN population than radio-loud and/or very luminous unobscured QSOs. 

Finally, finding lower gas mass reservoirs in objects in the ``blow-out'' phase, such as those presented in this work, than that measured in normal star forming galaxies at the same redshift (e.g. \citealt{Tacconi2013}) would constitute another way to assess the effect of  AGN feedback in diminishing the cold gas mass in the hosts galaxies of these ``transition'' objects and will give unique insights on the time scale of the gas consumption rate and the effect of AGN feedback in stopping SF.  These can be achieved with follow-up observations with ALMA, PdBI/NOEMA and JVLA observations of CO transitions (see e.g. \citealt{Feruglio2014}).

\section*{Acknowledgments}
This work is based on observations made at the European Southern Observatory, Paranal, Chile (ESO program 090.A-0830(A)) and on observations obtained with XMM-{\it Newton} and {\it Herschel},  two ESA Science Missions with instruments and contributions directly funded by ESA Member States and the USA (NASA).
MB and AB are grateful to A. Mehner and R. Wesson for the support at the
telescope. MB, AB and GC acknowledge useful discussion with N. Neumeyer, A. Modigliani, M. Romaniello and the ESO USD team, for Reflex support, and are grateful to  A. Mucciarelli and S. Piranomonte for help on  X-shooter data reduction and flux calibration. 
MB gratefully thank J. Mullaney, D. Kashino and T. Urrutia for providing us their datasets, and Chiara Feruglio, Francesca Civano, Mark Sargent and Kirpal Nandra for useful discussions. 
MB and MP acknowledge support from the FP7 Career Integration Grant ``eEASy'' (``SMBH evolution through cosmic time: from current surveys to eROSITA-Euclid AGN Synergies", CIG 321913). AB work is supported by the INAF-Fellowship Program.
Support for this publication was provided by the Italian National Institute for Astrophysics (INAF) through PRIN-INAF 2011 (``Black hole growth and AGN feedback through the cosmic time'') and PRIN-INAF-2012 (``The life cycle of early black holes''), and by the Italian ministry for school, university and reasearch (MIUR) through PRIN-MIUR 2010-2011 (``The dark Universe and the cosmic evolution of baryons: from current surveys to Euclid'').
We gratefully acknowledge the unique contribution of the entire COSMOS collaboration for making their excellent data products publicly available; more information on the COSMOS survey is available at \verb+http://www.astro.caltech.edu/~cosmos+. 

\bibliography{angi}{}

\begin{thebibliography}{125}
\expandafter\ifx\csname natexlab\endcsname\relax\def\natexlab#1{#1}\fi

\bibitem[{{Alexander} \& {Hickox}(2012)}]{Alexander2012}
{Alexander} D.~M., {Hickox} R.~C., 2012, \nar, 56, 93

\bibitem[{{Alexander} {et~al}\mbox{.}(2005){Alexander}, {Smail}, {Bauer},
  {Chapman}, {Blain}, {Brandt}, \& {Ivison}}]{Alexander2005}
{Alexander} D.~M., {Smail} I., {Bauer} F.~E., {Chapman} S.~C., {Blain} A.~W.,
  {Brandt} W.~N., {Ivison} R.~J., 2005, \nat, 434, 738

\bibitem[{{Alexander} {et~al}\mbox{.}(2010){Alexander}, {Swinbank}, {Smail},
  {McDermid}, \& {Nesvadba}}]{Alexander2010}
{Alexander} D.~M., {Swinbank} A.~M., {Smail} I., {McDermid} R., {Nesvadba}
  N.~P.~H., 2010, \mnras, 402, 2211

\bibitem[{{Allen} {et~al}\mbox{.}(2011){Allen}, {Hewett}, {Maddox}, {Richards},
  \& {Belokurov}}]{Allen2011}
{Allen} J.~T., {Hewett} P.~C., {Maddox} N., {Richards} G.~T., {Belokurov} V.,
  2011, \mnras, 410, 860

\bibitem[{{Arav} {et~al}\mbox{.}(2013){Arav}, {Borguet}, {Chamberlain},
  {Edmonds}, \& {Danforth}}]{Arav2013}
{Arav} N., {Borguet} B., {Chamberlain} C., {Edmonds} D., {Danforth} C., 2013,
  \mnras, 436, 3286

\bibitem[{{Baldwin} {et~al}\mbox{.}(1981){Baldwin}, {Phillips}, \&
  {Terlevich}}]{Baldwin1981}
{Baldwin} J.~A., {Phillips} M.~M., {Terlevich} R., 1981, \pasp, 93, 5

\bibitem[{{Banerji} {et~al}\mbox{.}(2014){Banerji}, {Fabian}, \&
  {McMahon}}]{Banerji2014}
{Banerji} M., {Fabian} A.~C., {McMahon} R.~G., 2014, \mnras, 439, L51

\bibitem[{{Banerji} {et~al}\mbox{.}(2012){Banerji}, {McMahon}, {Hewett},
  {Alaghband-Zadeh}, {Gonzalez-Solares}, {Venemans}, \&
  {Hawthorn}}]{Banerji2012}
{Banerji} M., {McMahon} R.~G., {Hewett} P.~C., {Alaghband-Zadeh} S.,
  {Gonzalez-Solares} E., {Venemans} B.~P., {Hawthorn} M.~J., 2012, \mnras, 427,
  2275

\bibitem[{{Bongiorno} {et~al}\mbox{.}(2014){Bongiorno}, {Maiolino}, {Brusa},
  {Marconi}, {Piconcelli}, {Lamastra}, {Cano-D{\'{\i}}az}, {Schulze},
  {Magnelli}, {Vignali}, {Fiore}, {Menci}, {Cresci}, {La Franca}, \&
  {Merloni}}]{Bongiorno2014}
{Bongiorno} A. {et~al.}, 2014, \mnras, 443, 2077

\bibitem[{{Bongiorno} {et~al}\mbox{.}(2012){Bongiorno}, {Merloni}, {Brusa},
  {Magnelli}, {Salvato}, {Mignoli}, {Zamorani}, {Fiore}, {Rosario}, {Mainieri},
  {Hao}, {Comastri}, {Vignali}, {Balestra}, {Bardelli}, {Berta}, {Civano},
  {Kampczyk}, {Le Floc'h}, {Lusso}, {Lutz}, {Pozzetti}, {Pozzi}, {Riguccini},
  {Shankar}, \& {Silverman}}]{Bongiorno2012}
{Bongiorno} A. {et~al.}, 2012, \mnras, 427, 3103

\bibitem[{{Borguet} {et~al}\mbox{.}(2013){Borguet}, {Arav}, {Edmonds},
  {Chamberlain}, \& {Benn}}]{Borguet2013}
{Borguet} B.~C.~J., {Arav} N., {Edmonds} D., {Chamberlain} C., {Benn} C., 2013,
  \apj, 762, 49

\bibitem[{{Boroson} {et~al}\mbox{.}(1985){Boroson}, {Persson}, \&
  {Oke}}]{Boroson1985}
{Boroson} T.~A., {Persson} S.~E., {Oke} J.~B., 1985, \apj, 293, 120

\bibitem[{{Bournaud} {et~al}\mbox{.}(2011){Bournaud}, {Dekel}, {Teyssier},
  {Cacciato}, {Daddi}, {Juneau}, \& {Shankar}}]{Bournaud2011}
{Bournaud} F., {Dekel} A., {Teyssier} R., {Cacciato} M., {Daddi} E., {Juneau}
  S., {Shankar} F., 2011, \apjl, 741, L33

\bibitem[{{Brusa} {et~al}\mbox{.}(2010){Brusa}, {Civano}, {Comastri}, {Miyaji},
  {Salvato}, {Zamorani}, {Cappelluti}, {Fiore}, {Hasinger}, {Mainieri},
  {Merloni}, {Bongiorno}, {Capak}, {Elvis}, {Gilli}, {Hao}, {Jahnke},
  {Koekemoer}, {Ilbert}, {Le Floc'h}, {Lusso}, {Mignoli}, {Schinnerer},
  {Silverman}, {Treister}, {Trump}, {Vignali}, {Zamojski}, {Aldcroft},
  {Aussel}, {Bardelli}, {Bolzonella}, {Cappi}, {Caputi}, {Contini},
  {Finoguenov}, {Fruscione}, {Garilli}, {Impey}, {Iovino}, {Iwasawa},
  {Kampczyk}, {Kartaltepe}, {Kneib}, {Knobel}, {Kovac}, {Lamareille},
  {Leborgne}, {Le Brun}, {Le Fevre}, {Lilly}, {Maier}, {McCracken}, {Pello},
  {Peng}, {Perez-Montero}, {de Ravel}, {Sanders}, {Scodeggio}, {Scoville},
  {Tanaka}, {Taniguchi}, {Tasca}, {de la Torre}, {Tresse}, {Vergani}, \&
  {Zucca}}]{Brusa2010}
{Brusa} M. {et~al.}, 2010, \apj, 716, 348

\bibitem[{{Brusa} {et~al}\mbox{.}(2005){Brusa}, {Comastri}, {Daddi},
  {Pozzetti}, {Zamorani}, {Vignali}, {Cimatti}, {Fiore}, {Mignoli}, {Ciliegi},
  \& {R{\"o}ttgering}}]{Brusa2005}
{Brusa} M. {et~al.}, 2005, \aap, 432, 69

\bibitem[{{Cano-D{\'{\i}}az} {et~al}\mbox{.}(2012){Cano-D{\'{\i}}az},
  {Maiolino}, {Marconi}, {Netzer}, {Shemmer}, \& {Cresci}}]{CanoDiaz2012}
{Cano-D{\'{\i}}az} M., {Maiolino} R., {Marconi} A., {Netzer} H., {Shemmer} O.,
  {Cresci} G., 2012, \aap, 537, L8

\bibitem[{{Cappelluti} {et~al}\mbox{.}(2009){Cappelluti}, {Brusa}, {Hasinger},
  {Comastri}, {Zamorani}, {Finoguenov}, {Gilli}, {Puccetti}, {Miyaji},
  {Salvato}, {Vignali}, {Aldcroft}, {B{\"o}hringer}, {Brunner}, {Civano},
  {Elvis}, {Fiore}, {Fruscione}, {Griffiths}, {Guzzo}, {Iovino}, {Koekemoer},
  {Mainieri}, {Scoville}, {Shopbell}, {Silverman}, \& {Urry}}]{Cappelluti2009}
{Cappelluti} N. {et~al.}, 2009, \aap, 497, 635

\bibitem[{{Cen}(2012)}]{Cen2012}
{Cen} R., 2012, \apj, 755, 28

\bibitem[{{Ceverino} \& {Klypin}(2009)}]{Ceverino2009}
{Ceverino} D., {Klypin} A., 2009, \apj, 695, 292

\bibitem[{{Cicone} {et~al}\mbox{.}(2014){Cicone}, {Maiolino}, {Sturm},
  {Graci{\'a}-Carpio}, {Feruglio}, {Neri}, {Aalto}, {Davies}, {Fiore},
  {Fischer}, {Garc{\'{\i}}a-Burillo}, {Gonz{\'a}lez-Alfonso},
  {Hailey-Dunsheath}, {Piconcelli}, \& {Veilleux}}]{Cicone2014}
{Cicone} C. {et~al.}, 2014, \aap, 562, A21

\bibitem[{{Ciotti} \& {Ostriker}(2007)}]{Ciotti2007}
{Ciotti} L., {Ostriker} J.~P., 2007, \apj, 665, 1038

\bibitem[{{Dai} {et~al}\mbox{.}(2008){Dai}, {Shankar}, \& {Sivakoff}}]{Dai2008}
{Dai} X., {Shankar} F., {Sivakoff} G.~R., 2008, \apj, 672, 108

\bibitem[{{Dale} \& {Helou}(2002)}]{Dale2002}
{Dale} D.~A., {Helou} G., 2002, \apj, 576, 159

\bibitem[{{de Kool} {et~al}\mbox{.}(2001){de Kool}, {Arav}, {Becker}, {Gregg},
  {White}, {Laurent-Muehleisen}, {Price}, \& {Korista}}]{Dekool2001}
{de Kool} M., {Arav} N., {Becker} R.~H., {Gregg} M.~D., {White} R.~L.,
  {Laurent-Muehleisen} S.~A., {Price} T., {Korista} K.~T., 2001, \apj, 548, 609

\bibitem[{{Diamond-Stanic} {et~al}\mbox{.}(2012){Diamond-Stanic}, {Moustakas},
  {Tremonti}, {Coil}, {Hickox}, {Robaina}, {Rudnick}, \& {Sell}}]{Diamond2012}
{Diamond-Stanic} A.~M., {Moustakas} J., {Tremonti} C.~A., {Coil} A.~L.,
  {Hickox} R.~C., {Robaina} A.~R., {Rudnick} G.~H., {Sell} P.~H., 2012, \apjl,
  755, L26

\bibitem[{{D'Odorico} {et~al}\mbox{.}(2006){D'Odorico}, {Dekker}, {Mazzoleni},
  {Vernet}, {Guinouard}, {Groot}, {Hammer}, {Rasmussen}, {Kaper}, {Navarro},
  {Pallavicini}, {Peroux}, \& {Zerbi}}]{Dodorico2006}
{D'Odorico} S. {et~al.}, 2006, in Society of Photo-Optical Instrumentation
  Engineers (SPIE) Conference Series, Vol. 6269, Society of Photo-Optical
  Instrumentation Engineers (SPIE) Conference Series

\bibitem[{{Dunn} {et~al}\mbox{.}(2010){Dunn}, {Bautista}, {Arav}, {Moe},
  {Korista}, {Costantini}, {Benn}, {Ellison}, \& {Edmonds}}]{Dunn2010}
{Dunn} J.~P. {et~al.}, 2010, \apj, 709, 611

\bibitem[{{Elvis}(2000)}]{Elvis2000}
{Elvis} M., 2000, \apj, 545, 63

\bibitem[{{Eracleous} \& {Halpern}(1994)}]{Eracleous1994}
{Eracleous} M., {Halpern} J.~P., 1994, \apjs, 90, 1

\bibitem[{{Fabian}(2012)}]{Fabian2012}
{Fabian} A.~C., 2012, \araa, 50, 455

\bibitem[{{Ferrarese} \& {Merritt}(2000)}]{Ferrarese2000}
{Ferrarese} L., {Merritt} D., 2000, \apjl, 539, L9

\bibitem[{{Feruglio} {et~al}\mbox{.}(2014){Feruglio}, {Bongiorno}, {Fiore},
  {Krips}, {Brusa}, {Daddi}, {Gavignaud}, {Maiolino}, {Piconcelli}, {Sargent},
  {Vignali}, \& {Zappacosta}}]{Feruglio2014}
{Feruglio} C. {et~al.}, 2014, \aap, 565, A91

\bibitem[{{Feruglio} {et~al}\mbox{.}(2010){Feruglio}, {Maiolino}, {Piconcelli},
  {Menci}, {Aussel}, {Lamastra}, \& {Fiore}}]{Feruglio2010}
{Feruglio} C., {Maiolino} R., {Piconcelli} E., {Menci} N., {Aussel} H.,
  {Lamastra} A., {Fiore} F., 2010, \aap, 518, L155

\bibitem[{{Fiore} {et~al}\mbox{.}(2003){Fiore}, {Brusa}, {Cocchia}, {Baldi},
  {Carangelo}, {Ciliegi}, {Comastri}, {La Franca}, {Maiolino}, {Matt},
  {Molendi}, {Mignoli}, {Perola}, {Severgnini}, \& {Vignali}}]{Fiore2003}
{Fiore} F. {et~al.}, 2003, \aap, 409, 79

\bibitem[{{Fiore} {et~al}\mbox{.}(2009){Fiore}, {Puccetti}, {Brusa}, {Salvato},
  {Zamorani}, {Aldcroft}, {Aussel}, {Brunner}, {Capak}, {Cappelluti}, {Civano},
  {Comastri}, {Elvis}, {Feruglio}, {Finoguenov}, {Fruscione}, {Gilli},
  {Hasinger}, {Koekemoer}, {Kartaltepe}, {Ilbert}, {Impey}, {Le Floc'h},
  {Lilly}, {Mainieri}, {Martinez-Sansigre}, {McCracken}, {Menci}, {Merloni},
  {Miyaji}, {Sanders}, {Sargent}, {Schinnerer}, {Scoville}, {Silverman},
  {Smolcic}, {Steffen}, {Santini}, {Taniguchi}, {Thompson}, {Trump}, {Vignali},
  {Urry}, \& {Yan}}]{Fiore2009}
{Fiore} F. {et~al.}, 2009, \apj, 693, 447

\bibitem[{{Fischer} {et~al}\mbox{.}(2010){Fischer}, {Sturm},
  {Gonz{\'a}lez-Alfonso}, {Graci{\'a}-Carpio}, {Hailey-Dunsheath}, {Poglitsch},
  {Contursi}, {Lutz}, {Genzel}, {Sternberg}, {Verma}, \&
  {Tacconi}}]{Fischer2010}
{Fischer} J. {et~al.}, 2010, \aap, 518, L41

\bibitem[{{F{\"o}rster Schreiber} {et~al}\mbox{.}(2014){F{\"o}rster Schreiber},
  {Genzel}, {Newman}, {Kurk}, {Lutz}, {Tacconi}, {Wuyts}, {Bandara}, {Burkert},
  {Buschkamp}, {Carollo}, {Cresci}, {Daddi}, {Davies}, {Eisenhauer}, {Hicks},
  {Lang}, {Lilly}, {Mainieri}, {Mancini}, {Naab}, {Peng}, {Renzini}, {Rosario},
  {Shapiro Griffin}, {Shapley}, {Sternberg}, {Tacchella}, {Vergani},
  {Wisnioski}, {Wuyts}, \& {Zamorani}}]{FS2014}
{F{\"o}rster Schreiber} N.~M. {et~al.}, 2014, \apj, 787, 38

\bibitem[{{Freudling} {et~al}\mbox{.}(2013){Freudling}, {Romaniello},
  {Bramich}, {Ballester}, {Forchi}, {Garc{\'{\i}}a-Dabl{\'o}}, {Moehler}, \&
  {Neeser}}]{Freudling13}
{Freudling} W., {Romaniello} M., {Bramich} D.~M., {Ballester} P., {Forchi} V.,
  {Garc{\'{\i}}a-Dabl{\'o}} C.~E., {Moehler} S., {Neeser} M.~J., 2013, \aap,
  559, A96

\bibitem[{{Fu} \& {Stockton}(2009)}]{Fu2009}
{Fu} H., {Stockton} A., 2009, \apj, 690, 953

\bibitem[{{Gebhardt} {et~al}\mbox{.}(2000){Gebhardt}, {Bender}, {Bower},
  {Dressler}, {Faber}, {Filippenko}, {Green}, {Grillmair}, {Ho}, {Kormendy},
  {Lauer}, {Magorrian}, {Pinkney}, {Richstone}, \& {Tremaine}}]{Gebhardt2000}
{Gebhardt} K. {et~al.}, 2000, \apjl, 539, L13

\bibitem[{{Genzel} {et~al}\mbox{.}(2014){Genzel}, {F{\"o}rster Schreiber},
  {Rosario}, {Lang}, {Lutz}, {Wisnioski}, {Wuyts}, {Wuyts}, {Bandara},
  {Bender}, {Berta}, {Kurk}, {Mendel}, {Tacconi}, {Wilman}, {Beifiori},
  {Brammer}, {Burkert}, {Buschkamp}, {Chan}, {Carollo}, {Davies}, {Eisenhauer},
  {Fabricius}, {Fossati}, {Kriek}, {Kulkarni}, {Lilly}, {Mancini}, {Momcheva},
  {Naab}, {Nelson}, {Renzini}, {Saglia}, {Sharples}, {Sternberg}, {Tacchella},
  \& {van Dokkum}}]{Genzel2014}
{Genzel} R. {et~al.}, 2014, ArXiv e-prints

\bibitem[{{Gindilis} \& {Pariiskii}(1961)}]{Gindilis1961}
{Gindilis} L.~M., {Pariiskii} N.~N., 1961, \sovast, 5, 72

\bibitem[{{Glikman} {et~al}\mbox{.}(2012){Glikman}, {Urrutia}, {Lacy},
  {Djorgovski}, {Mahabal}, {Myers}, {Ross}, {Petitjean}, {Ge}, {Schneider}, \&
  {York}}]{Glikman2012}
{Glikman} E. {et~al.}, 2012, \apj, 757, 51

\bibitem[{{Greene} {et~al}\mbox{.}(2011){Greene}, {Zakamska}, {Ho}, \&
  {Barth}}]{Greene2011}
{Greene} J.~E., {Zakamska} N.~L., {Ho} L.~C., {Barth} A.~J., 2011, \apj, 732, 9

\bibitem[{{G{\"u}ltekin} {et~al}\mbox{.}(2009){G{\"u}ltekin}, {Cackett},
  {Miller}, {Di Matteo}, {Markoff}, \& {Richstone}}]{Gultekin2009}
{G{\"u}ltekin} K., {Cackett} E.~M., {Miller} J.~M., {Di Matteo} T., {Markoff}
  S., {Richstone} D.~O., 2009, \apj, 706, 404

\bibitem[{{Hainline} {et~al}\mbox{.}(2013){Hainline}, {Hickox}, {Greene},
  {Myers}, \& {Zakamska}}]{Hainline2013}
{Hainline} K.~N., {Hickox} R., {Greene} J.~E., {Myers} A.~D., {Zakamska} N.~L.,
  2013, \apj, 774, 145

\bibitem[{{Hamann} {et~al}\mbox{.}(2002){Hamann}, {Korista}, {Ferland},
  {Warner}, \& {Baldwin}}]{Hamann2002}
{Hamann} F., {Korista} K.~T., {Ferland} G.~J., {Warner} C., {Baldwin} J., 2002,
  \apj, 564, 592

\bibitem[{{Harrison}(2014)}]{Harrison2014IAU}
{Harrison} C.~M., 2014, in IAU Symposium, Vol. 304, IAU Symposium, pp. 284--290

\bibitem[{{Harrison} {et~al}\mbox{.}(2012{\natexlab{a}}){Harrison},
  {Alexander}, {Mullaney}, {Altieri}, {Coia}, {Charmandaris}, {Daddi},
  {Dannerbauer}, {Dasyra}, {Del Moro}, {Dickinson}, {Hickox}, {Ivison},
  {Kartaltepe}, {Le Floc'h}, {Leiton}, {Magnelli}, {Popesso}, {Rovilos},
  {Rosario}, \& {Swinbank}}]{Harrison2012a}
{Harrison} C.~M. {et~al.}, 2012{\natexlab{a}}, \apjl, 760, L15

\bibitem[{{Harrison} {et~al}\mbox{.}(2014){Harrison}, {Alexander}, {Mullaney},
  \& {Swinbank}}]{Harrison2014}
{Harrison} C.~M., {Alexander} D.~M., {Mullaney} J.~R., {Swinbank} A.~M., 2014,
  \mnras, 441, 3306

\bibitem[{{Harrison} {et~al}\mbox{.}(2012{\natexlab{b}}){Harrison},
  {Alexander}, {Swinbank}, {Smail}, {Alaghband-Zadeh}, {Bauer}, {Chapman}, {Del
  Moro}, {Hickox}, {Ivison}, {Men{\'e}ndez-Delmestre}, {Mullaney}, \&
  {Nesvadba}}]{Harrison2012}
{Harrison} C.~M. {et~al.}, 2012{\natexlab{b}}, \mnras, 426, 1073

\bibitem[{{Hasinger} {et~al}\mbox{.}(2007){Hasinger}, {Cappelluti}, {Brunner},
  {Brusa}, {Comastri}, {Elvis}, {Finoguenov}, {Fiore}, {Franceschini}, {Gilli},
  {Griffiths}, {Lehmann}, {Mainieri}, {Matt}, {Matute}, {Miyaji}, {Molendi},
  {Paltani}, {Sanders}, {Scoville}, {Tresse}, {Urry}, {Vettolani}, \&
  {Zamorani}}]{Hasinger2007}
{Hasinger} G. {et~al.}, 2007, \apjs, 172, 29

\bibitem[{{Heckman} {et~al}\mbox{.}(1990){Heckman}, {Armus}, \&
  {Miley}}]{Heckman1990}
{Heckman} T.~M., {Armus} L., {Miley} G.~K., 1990, \apjs, 74, 833

\bibitem[{{Heckman} {et~al}\mbox{.}(1981){Heckman}, {Miley}, {van Breugel}, \&
  {Butcher}}]{Heckman1981}
{Heckman} T.~M., {Miley} G.~K., {van Breugel} W.~J.~M., {Butcher} H.~R., 1981,
  \apj, 247, 403

\bibitem[{{Hickox} {et~al}\mbox{.}(2014){Hickox}, {Mullaney}, {Alexander},
  {Chen}, {Civano}, {Goulding}, \& {Hainline}}]{Hickox2014}
{Hickox} R.~C., {Mullaney} J.~R., {Alexander} D.~M., {Chen} C.-T.~J., {Civano}
  F.~M., {Goulding} A.~D., {Hainline} K.~N., 2014, \apj, 782, 9

\bibitem[{{Hopkins} {et~al}\mbox{.}(2008){Hopkins}, {Hernquist}, {Cox}, \&
  {Kere{\v s}}}]{Hopkins2008}
{Hopkins} P.~F., {Hernquist} L., {Cox} T.~J., {Kere{\v s}} D., 2008, \apjs,
  175, 356

\bibitem[{{James} \& {Roos}(1975)}]{James1975}
{James} F., {Roos} M., 1975, Computer Physics Communications, 10, 343

\bibitem[{{Karim} {et~al}\mbox{.}(2011){Karim}, {Schinnerer},
  {Mart{\'{\i}}nez-Sansigre}, {Sargent}, {van der Wel}, {Rix}, {Ilbert},
  {Smol{\v c}i{\'c}}, {Carilli}, {Pannella}, {Koekemoer}, {Bell}, \&
  {Salvato}}]{Karim2011}
{Karim} A. {et~al.}, 2011, \apj, 730, 61

\bibitem[{{Kashino} {et~al}\mbox{.}(2013){Kashino}, {Silverman}, {Rodighiero},
  {Renzini}, {Arimoto}, {Daddi}, {Lilly}, {Sanders}, {Kartaltepe}, {Zahid},
  {Nagao}, {Sugiyama}, {Capak}, {Carollo}, {Chu}, {Hasinger}, {Ilbert},
  {Kajisawa}, {Kewley}, {Koekemoer}, {Kova{\v c}}, {Le F{\`e}vre}, {Masters},
  {McCracken}, {Onodera}, {Scoville}, {Strazzullo}, {Symeonidis}, \&
  {Taniguchi}}]{Kashino2013}
{Kashino} D. {et~al.}, 2013, \apjl, 777, L8

\bibitem[{{Kennicutt}(1998)}]{Kennicutt1998}
{Kennicutt}, Jr. R.~C., 1998, \araa, 36, 189

\bibitem[{{King}(2005)}]{King2005}
{King} A., 2005, \apjl, 635, L121

\bibitem[{{King}(2010)}]{King2010}
{King} A.~R., 2010, in Astronomical Society of the Pacific Conference Series,
  Vol. 427, Accretion and Ejection in AGN: a Global View, {Maraschi} L.,
  {Ghisellini} G., {Della Ceca} R., {Tavecchio} F., eds., p. 315

\bibitem[{{Kova{\v c}evi{\'c}} {et~al}\mbox{.}(2010){Kova{\v c}evi{\'c}},
  {Popovi{\'c}}, \& {Dimitrijevi{\'c}}}]{Kovacevic2010}
{Kova{\v c}evi{\'c}} J., {Popovi{\'c}} L.~{\v C}., {Dimitrijevi{\'c}} M.~S.,
  2010, \apjs, 189, 15

\bibitem[{{Lagos} {et~al}\mbox{.}(2013){Lagos}, {Lacey}, \&
  {Baugh}}]{Lagos2013}
{Lagos} C.~d.~P., {Lacey} C.~G., {Baugh} C.~M., 2013, \mnras, 436, 1787

\bibitem[{{LaMassa} {et~al}\mbox{.}(2013){LaMassa}, {Urry}, {Cappelluti},
  {Civano}, {Ranalli}, {Glikman}, {Treister}, {Richards}, {Ballantyne},
  {Stern}, {Comastri}, {Cardamone}, {Schawinski}, {B{\"o}hringer}, {Chon},
  {Murray}, {Green}, \& {Nandra}}]{Lamassa2013}
{LaMassa} S.~M. {et~al.}, 2013, \mnras, 436, 3581

\bibitem[{{Lamastra} {et~al}\mbox{.}(2013){Lamastra}, {Menci}, {Fiore},
  {Santini}, {Bongiorno}, \& {Piconcelli}}]{Lamastra2013}
{Lamastra} A., {Menci} N., {Fiore} F., {Santini} P., {Bongiorno} A.,
  {Piconcelli} E., 2013, \aap, 559, A56

\bibitem[{{L{\'{\i}}pari} \& {Terlevich}(2006)}]{Lipari2006}
{L{\'{\i}}pari} S.~L., {Terlevich} R.~J., 2006, \mnras, 368, 1001

\bibitem[{{Liu} {et~al}\mbox{.}(2014){Liu}, {Zakamska}, \& {Greene}}]{Liu2014}
{Liu} G., {Zakamska} N.~L., {Greene} J.~E., 2014, \mnras, 442, 1303

\bibitem[{{Liu} {et~al}\mbox{.}(2013){Liu}, {Zakamska}, {Greene}, {Nesvadba},
  \& {Liu}}]{Liu2013}
{Liu} G., {Zakamska} N.~L., {Greene} J.~E., {Nesvadba} N.~P.~H., {Liu} X.,
  2013, \mnras, 436, 2576

\bibitem[{{Lusso} {et~al}\mbox{.}(2012){Lusso}, {Comastri}, {Simmons},
  {Mignoli}, {Zamorani}, {Vignali}, {Brusa}, {Shankar}, {Lutz}, {Trump},
  {Maiolino}, {Gilli}, {Bolzonella}, {Puccetti}, {Salvato}, {Impey}, {Civano},
  {Elvis}, {Mainieri}, {Silverman}, {Koekemoer}, {Bongiorno}, {Merloni},
  {Berta}, {Le Floc'h}, {Magnelli}, {Pozzi}, \& {Riguccini}}]{Lusso2012}
{Lusso} E. {et~al.}, 2012, \mnras, 425, 623

\bibitem[{{Lusso} {et~al}\mbox{.}(2013){Lusso}, {Hennawi}, {Comastri},
  {Zamorani}, {Richards}, {Vignali}, {Treister}, {Schawinski}, {Salvato}, \&
  {Gilli}}]{Lusso2013}
{Lusso} E. {et~al.}, 2013, \apj, 777, 86

\bibitem[{{Lutz} {et~al}\mbox{.}(2011){Lutz}, {Poglitsch}, {Altieri},
  {Andreani}, {Aussel}, {Berta}, {Bongiovanni}, {Brisbin}, {Cava}, {Cepa},
  {Cimatti}, {Daddi}, {Dominguez-Sanchez}, {Elbaz}, {F{\"o}rster Schreiber},
  {Genzel}, {Grazian}, {Gruppioni}, {Harwit}, {Le Floc'h}, {Magdis},
  {Magnelli}, {Maiolino}, {Nordon}, {P{\'e}rez Garc{\'{\i}}a}, {Popesso},
  {Pozzi}, {Riguccini}, {Rodighiero}, {Saintonge}, {Sanchez Portal}, {Santini},
  {Shao}, {Sturm}, {Tacconi}, {Valtchanov}, {Wetzstein}, \&
  {Wieprecht}}]{Lutz2011}
{Lutz} D. {et~al.}, 2011, \aap, 532, A90

\bibitem[{{Magorrian} {et~al}\mbox{.}(1998){Magorrian}, {Tremaine},
  {Richstone}, {Bender}, {Bower}, {Dressler}, {Faber}, {Gebhardt}, {Green},
  {Grillmair}, {Kormendy}, \& {Lauer}}]{Magorrian1998}
{Magorrian} J. {et~al.}, 1998, \aj, 115, 2285

\bibitem[{{Mainieri} {et~al}\mbox{.}(2011){Mainieri}, {Bongiorno}, {Merloni},
  {Aller}, {Carollo}, {Iwasawa}, {Koekemoer}, {Mignoli}, {Silverman},
  {Bolzonella}, {Brusa}, {Comastri}, {Gilli}, {Halliday}, {Ilbert}, {Lusso},
  {Salvato}, {Vignali}, {Zamorani}, {Contini}, {Kneib}, {Le F{\`e}vre},
  {Lilly}, {Renzini}, {Scodeggio}, {Balestra}, {Bardelli}, {Caputi}, {Coppa},
  {Cucciati}, {de la Torre}, {de Ravel}, {Franzetti}, {Garilli}, {Iovino},
  {Kampczyk}, {Knobel}, {Kova{\v c}}, {Lamareille}, {Le Borgne}, {Le Brun},
  {Maier}, {Nair}, {Pello}, {Peng}, {Perez Montero}, {Pozzetti},
  {Ricciardelli}, {Tanaka}, {Tasca}, {Tresse}, {Vergani}, {Zucca}, {Aussel},
  {Capak}, {Cappelluti}, {Elvis}, {Fiore}, {Hasinger}, {Impey}, {Le Floc'h},
  {Scoville}, {Taniguchi}, \& {Trump}}]{Mainieri2011}
{Mainieri} V. {et~al.}, 2011, \aap, 535, A80

\bibitem[{{Maiolino} {et~al}\mbox{.}(2012){Maiolino}, {Gallerani}, {Neri},
  {Cicone}, {Ferrara}, {Genzel}, {Lutz}, {Sturm}, {Tacconi}, {Walter},
  {Feruglio}, {Fiore}, \& {Piconcelli}}]{Maiolino2012}
{Maiolino} R. {et~al.}, 2012, \mnras, 425, L66

\bibitem[{{Martin}(1999)}]{Martin1999}
{Martin} C.~L., 1999, \apj, 513, 156

\bibitem[{{Mart{\'{\i}}nez-Sansigre}
  {et~al}\mbox{.}(2006){Mart{\'{\i}}nez-Sansigre}, {Rawlings}, {Lacy}, {Fadda},
  {Jarvis}, {Marleau}, {Simpson}, \& {Willott}}]{Sansigre2006}
{Mart{\'{\i}}nez-Sansigre} A., {Rawlings} S., {Lacy} M., {Fadda} D., {Jarvis}
  M.~J., {Marleau} F.~R., {Simpson} C., {Willott} C.~J., 2006, \mnras, 370,
  1479

\bibitem[{{Matsuoka}(2012)}]{Matsuoka2012}
{Matsuoka} Y., 2012, \apj, 750, 54

\bibitem[{{McElroy} {et~al}\mbox{.}(2014){McElroy}, {Croom}, {Pracy}, {Sharp},
  {Ho}, \& {Medling}}]{McElroy2014}
{McElroy} R., {Croom} S.~M., {Pracy} M., {Sharp} R., {Ho} I., {Medling} A.~M.,
  2014, ArXiv e-prints

\bibitem[{{Menci} {et~al}\mbox{.}(2008){Menci}, {Fiore}, {Puccetti}, \&
  {Cavaliere}}]{Menci2008}
{Menci} N., {Fiore} F., {Puccetti} S., {Cavaliere} A., 2008, \apj, 686, 219

\bibitem[{{Merloni} {et~al}\mbox{.}(2014){Merloni}, {Bongiorno}, {Brusa},
  {Iwasawa}, {Mainieri}, {Magnelli}, {Salvato}, {Berta}, {Cappelluti},
  {Comastri}, {Fiore}, {Gilli}, {Koekemoer}, {Le Floc'h}, {Lusso}, {Lutz},
  {Miyaji}, {Pozzi}, {Riguccini}, {Rosario}, {Silverman}, {Symeonidis},
  {Treister}, {Vignali}, \& {Zamorani}}]{Merloni2014}
{Merloni} A. {et~al.}, 2014, \mnras, 437, 3550

\bibitem[{{Merloni} {et~al}\mbox{.}(2012){Merloni}, {Predehl}, {Becker},
  {B{\"o}hringer}, {Boller}, {Brunner}, {Brusa}, {Dennerl}, {Freyberg},
  {Friedrich}, {Georgakakis}, {Haberl}, {Hasinger}, {Meidinger}, {Mohr},
  {Nandra}, {Rau}, {Reiprich}, {Robrade}, {Salvato}, {Santangelo}, {Sasaki},
  {Schwope}, {Wilms}, \& {German eROSITA Consortium}}]{Merloni2012}
{Merloni} A. {et~al.}, 2012, ArXiv e-prints

\bibitem[{{Mignoli} {et~al}\mbox{.}(2004){Mignoli}, {Pozzetti}, {Comastri},
  {Brusa}, {Ciliegi}, {Cocchia}, {Fiore}, {La Franca}, {Maiolino}, {Matt},
  {Molendi}, {Perola}, {Puccetti}, {Severgnini}, \& {Vignali}}]{Mignoli2004}
{Mignoli} M. {et~al.}, 2004, \aap, 418, 827

\bibitem[{{Moe} {et~al}\mbox{.}(2009){Moe}, {Arav}, {Bautista}, \&
  {Korista}}]{Moe2009}
{Moe} M., {Arav} N., {Bautista} M.~A., {Korista} K.~T., 2009, \apj, 706, 525

\bibitem[{{Mullaney} {et~al}\mbox{.}(2013){Mullaney}, {Alexander}, {Fine},
  {Goulding}, {Harrison}, \& {Hickox}}]{Mullaney2013}
{Mullaney} J.~R., {Alexander} D.~M., {Fine} S., {Goulding} A.~D., {Harrison}
  C.~M., {Hickox} R.~C., 2013, \mnras, 433, 622

\bibitem[{{Mullaney} {et~al}\mbox{.}(2012){Mullaney}, {Pannella}, {Daddi},
  {Alexander}, {Elbaz}, {Hickox}, {Bournaud}, {Altieri}, {Aussel}, {Coia},
  {Dannerbauer}, {Dasyra}, {Dickinson}, {Hwang}, {Kartaltepe}, {Leiton},
  {Magdis}, {Magnelli}, {Popesso}, {Valtchanov}, {Bauer}, {Brandt}, {Del Moro},
  {Hanish}, {Ivison}, {Juneau}, {Luo}, {Lutz}, {Sargent}, {Scott}, \&
  {Xue}}]{Mullaney2012}
{Mullaney} J.~R. {et~al.}, 2012, \mnras, 419, 95

\bibitem[{{Murray} {et~al}\mbox{.}(2005){Murray}, {Quataert}, \&
  {Thompson}}]{Murray2005}
{Murray} N., {Quataert} E., {Thompson} T.~A., 2005, \apj, 618, 569

\bibitem[{{Nesvadba} {et~al}\mbox{.}(2011){Nesvadba}, {De Breuck}, {Lehnert},
  {Best}, {Binette}, \& {Proga}}]{Nesvadba2011}
{Nesvadba} N.~P.~H., {De Breuck} C., {Lehnert} M.~D., {Best} P.~N., {Binette}
  L., {Proga} D., 2011, \aap, 525, A43

\bibitem[{{Nesvadba} {et~al}\mbox{.}(2008){Nesvadba}, {Lehnert}, {De Breuck},
  {Gilbert}, \& {van Breugel}}]{Nesvadba2008}
{Nesvadba} N.~P.~H., {Lehnert} M.~D., {De Breuck} C., {Gilbert} A.~M., {van
  Breugel} W., 2008, \aap, 491, 407

\bibitem[{{Netzer} {et~al}\mbox{.}(2004){Netzer}, {Shemmer}, {Maiolino},
  {Oliva}, {Croom}, {Corbett}, \& {di Fabrizio}}]{Netzer2004}
{Netzer} H., {Shemmer} O., {Maiolino} R., {Oliva} E., {Croom} S., {Corbett} E.,
  {di Fabrizio} L., 2004, \apj, 614, 558

\bibitem[{{Newman} {et~al}\mbox{.}(2012){Newman}, {Genzel},
  {F{\"o}rster-Schreiber}, {Shapiro Griffin}, {Mancini}, {Lilly}, {Renzini},
  {Bouch{\'e}}, {Burkert}, {Buschkamp}, {Carollo}, {Cresci}, {Davies},
  {Eisenhauer}, {Genel}, {Hicks}, {Kurk}, {Lutz}, {Naab}, {Peng}, {Sternberg},
  {Tacconi}, {Vergani}, {Wuyts}, \& {Zamorani}}]{Newman2012}
{Newman} S.~F. {et~al.}, 2012, \apj, 761, 43

\bibitem[{{Osterbrock}(1981)}]{Osterbrock1981}
{Osterbrock} D.~E., 1981, \apj, 249, 462

\bibitem[{{Osterbrock}(1989)}]{Osterbrock1989}
{Osterbrock} D.~E., 1989, {Astrophysics of gaseous nebulae and active galactic
  nuclei}. Research supported by the University of California, John Simon
  Guggenheim Memorial Foundation, University of Minnesota, et al.~Mill Valley,
  CA, University Science Books, 1989, 422 p.

\bibitem[{{Page} {et~al}\mbox{.}(2012){Page}, {Symeonidis}, {Vieira},
  {Altieri}, {Amblard}, {Arumugam}, {Aussel}, {Babbedge}, {Blain}, {Bock},
  {Boselli}, {Buat}, {Castro-Rodr{\'{\i}}guez}, {Cava}, {Chanial}, {Clements},
  {Conley}, {Conversi}, {Cooray}, {Dowell}, {Dubois}, {Dunlop}, {Dwek}, {Dye},
  {Eales}, {Elbaz}, {Farrah}, {Fox}, {Franceschini}, {Gear}, {Glenn},
  {Griffin}, {Halpern}, {Hatziminaoglou}, {Ibar}, {Isaak}, {Ivison}, {Lagache},
  {Levenson}, {Lu}, {Madden}, {Maffei}, {Mainetti}, {Marchetti}, {Nguyen},
  {O'Halloran}, {Oliver}, {Omont}, {Panuzzo}, {Papageorgiou}, {Pearson},
  {P{\'e}rez-Fournon}, {Pohlen}, {Rawlings}, {Rigopoulou}, {Riguccini},
  {Rizzo}, {Rodighiero}, {Roseboom}, {Rowan-Robinson}, {Portal}, {Schulz},
  {Scott}, {Seymour}, {Shupe}, {Smith}, {Stevens}, {Trichas}, {Tugwell},
  {Vaccari}, {Valtchanov}, {Viero}, {Vigroux}, {Wang}, {Ward}, {Wright}, {Xu},
  \& {Zemcov}}]{Page2012}
{Page} M.~J. {et~al.}, 2012, \nat, 485, 213

\bibitem[{{Perna} {et~al}\mbox{.}(2014){Perna}, {Brusa}, {Cresci}, {Comastri},
  {Lanzuisi}, {Lusso}, {Marconi}, {Salvato}, {Zamorani}, {Bongiorno},
  {Mainieri}, {Maiolino}, \& {Mignoli}}]{Perna2014}
{Perna} M. {et~al.}, 2014, ArXiv e-prints

\bibitem[{{Pierre}(2012)}]{Pierre2012}
{Pierre} M., 2012, in Science from the Next Generation Imaging and
  Spectroscopic Surveys

\bibitem[{{Rodr{\'{\i}}guez-Zaur{\'{\i}}n}
  {et~al}\mbox{.}(2013){Rodr{\'{\i}}guez-Zaur{\'{\i}}n}, {Tadhunter}, {Rose},
  \& {Holt}}]{Zaurin2013}
{Rodr{\'{\i}}guez-Zaur{\'{\i}}n} J., {Tadhunter} C.~N., {Rose} M., {Holt} J.,
  2013, \mnras, 432, 138

\bibitem[{{Rosario} {et~al}\mbox{.}(2012){Rosario}, {Santini}, {Lutz}, {Shao},
  {Maiolino}, {Alexander}, {Altieri}, {Andreani}, {Aussel}, {Bauer}, {Berta},
  {Bongiovanni}, {Brandt}, {Brusa}, {Cepa}, {Cimatti}, {Cox}, {Daddi}, {Elbaz},
  {Fontana}, {F{\"o}rster Schreiber}, {Genzel}, {Grazian}, {Le Floch},
  {Magnelli}, {Mainieri}, {Netzer}, {Nordon}, {P{\'e}rez Garcia}, {Poglitsch},
  {Popesso}, {Pozzi}, {Riguccini}, {Rodighiero}, {Salvato}, {Sanchez-Portal},
  {Sturm}, {Tacconi}, {Valtchanov}, \& {Wuyts}}]{Rosario2012}
{Rosario} D.~J. {et~al.}, 2012, \aap, 545, A45

\bibitem[{{Rousselot} {et~al}\mbox{.}(2000){Rousselot}, {Lidman}, {Cuby},
  {Moreels}, \& {Monnet}}]{Rousselot2000}
{Rousselot} P., {Lidman} C., {Cuby} J.-G., {Moreels} G., {Monnet} G., 2000,
  \aap, 354, 1134

\bibitem[{{Rupke} \& {Veilleux}(2011)}]{Rupke2011}
{Rupke} D.~S.~N., {Veilleux} S., 2011, \apjl, 729, L27

\bibitem[{{Rupke} \& {Veilleux}(2013)}]{Rupke2013}
{Rupke} D.~S.~N., {Veilleux} S., 2013, \apjl, 775, L15

\bibitem[{{Salvato} {et~al}\mbox{.}(2011){Salvato}, {Ilbert}, {Hasinger},
  {Rau}, {Civano}, {Zamorani}, {Brusa}, {Elvis}, {Vignali}, {Aussel},
  {Comastri}, {Fiore}, {Le Floc'h}, {Mainieri}, {Bardelli}, {Bolzonella},
  {Bongiorno}, {Capak}, {Caputi}, {Cappelluti}, {Carollo}, {Contini},
  {Garilli}, {Iovino}, {Fotopoulou}, {Fruscione}, {Gilli}, {Halliday}, {Kneib},
  {Kakazu}, {Kartaltepe}, {Koekemoer}, {Kovac}, {Ideue}, {Ikeda}, {Impey}, {Le
  Fevre}, {Lamareille}, {Lanzuisi}, {Le Borgne}, {Le Brun}, {Lilly}, {Maier},
  {Manohar}, {Masters}, {McCracken}, {Messias}, {Mignoli}, {Mobasher}, {Nagao},
  {Pello}, {Puccetti}, {Perez-Montero}, {Renzini}, {Sargent}, {Sanders},
  {Scodeggio}, {Scoville}, {Shopbell}, {Silvermann}, {Taniguchi}, {Tasca},
  {Tresse}, {Trump}, \& {Zucca}}]{Salvato2011}
{Salvato} M. {et~al.}, 2011, \apj, 742, 61

\bibitem[{{Sanders} {et~al}\mbox{.}(1988){Sanders}, {Soifer}, {Elias},
  {Neugebauer}, \& {Matthews}}]{Sanders1988}
{Sanders} D.~B., {Soifer} B.~T., {Elias} J.~H., {Neugebauer} G., {Matthews} K.,
  1988, \apjl, 328, L35

\bibitem[{{Santini} {et~al}\mbox{.}(2009){Santini}, {Fontana}, {Grazian},
  {Salimbeni}, {Fiore}, {Fontanot}, {Boutsia}, {Castellano}, {Cristiani}, {de
  Santis}, {Gallozzi}, {Giallongo}, {Menci}, {Nonino}, {Paris}, {Pentericci},
  \& {Vanzella}}]{Santini2009}
{Santini} P. {et~al.}, 2009, \aap, 504, 751

\bibitem[{{Santini} {et~al}\mbox{.}(2012){Santini}, {Rosario}, {Shao}, {Lutz},
  {Maiolino}, {Alexander}, {Altieri}, {Andreani}, {Aussel}, {Bauer}, {Berta},
  {Bongiovanni}, {Brandt}, {Brusa}, {Cepa}, {Cimatti}, {Daddi}, {Elbaz},
  {Fontana}, {F{\"o}rster Schreiber}, {Genzel}, {Grazian}, {Le Floc'h},
  {Magnelli}, {Mainieri}, {Nordon}, {P{\'e}rez Garcia}, {Poglitsch}, {Popesso},
  {Pozzi}, {Riguccini}, {Rodighiero}, {Salvato}, {Sanchez-Portal}, {Sturm},
  {Tacconi}, {Valtchanov}, \& {Wuyts}}]{Santini2012}
{Santini} P. {et~al.}, 2012, \aap, 540, A109

\bibitem[{{Sarria} {et~al}\mbox{.}(2010){Sarria}, {Maiolino}, {La Franca},
  {Pozzi}, {Fiore}, {Marconi}, {Vignali}, \& {Comastri}}]{Sarria2010}
{Sarria} J.~E., {Maiolino} R., {La Franca} F., {Pozzi} F., {Fiore} F.,
  {Marconi} A., {Vignali} C., {Comastri} A., 2010, \aap, 522, L3+

\bibitem[{{Schinnerer} {et~al}\mbox{.}(2010){Schinnerer}, {Sargent}, {Bondi},
  {Smol{\v c}i{\'c}}, {Datta}, {Carilli}, {Bertoldi}, {Blain}, {Ciliegi},
  {Koekemoer}, \& {Scoville}}]{Schinnerer2010}
{Schinnerer} E. {et~al.}, 2010, \apjs, 188, 384

\bibitem[{{Scoville} {et~al}\mbox{.}(2007){Scoville}, {Aussel}, {Brusa},
  {Capak}, {Carollo}, {Elvis}, {Giavalisco}, {Guzzo}, {Hasinger}, {Impey},
  {Kneib}, {LeFevre}, {Lilly}, {Mobasher}, {Renzini}, {Rich}, {Sanders},
  {Schinnerer}, {Schminovich}, {Shopbell}, {Taniguchi}, \&
  {Tyson}}]{Scoville2007}
{Scoville} N. {et~al.}, 2007, \apjs, 172, 1

\bibitem[{{Soltan}(1982)}]{Soltan1982}
{Soltan} A., 1982, \mnras, 200, 115

\bibitem[{{Soto} \& {Martin}(2012)}]{SotoMartin2012}
{Soto} K.~T., {Martin} C.~L., 2012, \apjs, 203, 3

\bibitem[{{Soto} {et~al}\mbox{.}(2012){Soto}, {Martin}, {Prescott}, \&
  {Armus}}]{Soto2012}
{Soto} K.~T., {Martin} C.~L., {Prescott} M.~K.~M., {Armus} L., 2012, \apj, 757,
  86

\bibitem[{{Spergel} {et~al}\mbox{.}(2003){Spergel}, {Verde}, {Peiris},
  {Komatsu}, {Nolta}, {Bennett}, {Halpern}, {Hinshaw}, {Jarosik}, {Kogut},
  {Limon}, {Meyer}, {Page}, {Tucker}, {Weiland}, {Wollack}, \&
  {Wright}}]{Spergel2003}
{Spergel} D.~N. {et~al.}, 2003, \apjs, 148, 175

\bibitem[{{Stanghellini} \& {Kaler}(1989)}]{Stanghellini1989}
{Stanghellini} L., {Kaler} J.~B., 1989, \apj, 343, 811

\bibitem[{{Tacconi} {et~al}\mbox{.}(2013){Tacconi}, {Neri}, {Genzel}, {Combes},
  {Bolatto}, {Cooper}, {Wuyts}, {Bournaud}, {Burkert}, {Comerford}, {Cox},
  {Davis}, {F{\"o}rster Schreiber}, {Garc{\'{\i}}a-Burillo}, {Gracia-Carpio},
  {Lutz}, {Naab}, {Newman}, {Omont}, {Saintonge}, {Shapiro Griffin}, {Shapley},
  {Sternberg}, \& {Weiner}}]{Tacconi2013}
{Tacconi} L.~J. {et~al.}, 2013, \apj, 768, 74

\bibitem[{{Urrutia} {et~al}\mbox{.}(2009){Urrutia}, {Becker}, {White},
  {Glikman}, {Lacy}, {Hodge}, \& {Gregg}}]{Urrutia2009}
{Urrutia} T., {Becker} R.~H., {White} R.~L., {Glikman} E., {Lacy} M., {Hodge}
  J., {Gregg} M.~D., 2009, \apj, 698, 1095

\bibitem[{{Urrutia} {et~al}\mbox{.}(2008){Urrutia}, {Lacy}, \&
  {Becker}}]{Urrutia2008}
{Urrutia} T., {Lacy} M., {Becker} R.~H., 2008, \apj, 674, 80

\bibitem[{{Urrutia} {et~al}\mbox{.}(2012){Urrutia}, {Lacy}, {Spoon}, {Glikman},
  {Petric}, \& {Schulz}}]{Urrutia2012}
{Urrutia} T., {Lacy} M., {Spoon} H., {Glikman} E., {Petric} A., {Schulz} B.,
  2012, \apj, 757, 125

\bibitem[{{Veilleux} {et~al}\mbox{.}(2005){Veilleux}, {Cecil}, \&
  {Bland-Hawthorn}}]{Veilleux2005}
{Veilleux} S., {Cecil} G., {Bland-Hawthorn} J., 2005, \araa, 43, 769

\bibitem[{{Vernet} {et~al}\mbox{.}(2011){Vernet}, {Dekker}, {D'Odorico},
  {Kaper}, {Kjaergaard}, {Hammer}, {Randich}, {Zerbi}, {Groot}, {Hjorth},
  {Guinouard}, {Navarro}, {Adolfse}, {Albers}, {Amans}, {Andersen}, {Andersen},
  {Binetruy}, {Bristow}, {Castillo}, {Chemla}, {Christensen}, {Conconi},
  {Conzelmann}, {Dam}, {de Caprio}, {de Ugarte Postigo}, {Delabre}, {di
  Marcantonio}, {Downing}, {Elswijk}, {Finger}, {Fischer}, {Flores}, {Fran{\c
  c}ois}, {Goldoni}, {Guglielmi}, {Haigron}, {Hanenburg}, {Hendriks},
  {Horrobin}, {Horville}, {Jessen}, {Kerber}, {Kern}, {Kiekebusch}, {Kleszcz},
  {Klougart}, {Kragt}, {Larsen}, {Lizon}, {Lucuix}, {Mainieri}, {Manuputy},
  {Martayan}, {Mason}, {Mazzoleni}, {Michaelsen}, {Modigliani}, {Moehler},
  {M{\o}ller}, {Norup S{\o}rensen}, {N{\o}rregaard}, {P{\'e}roux}, {Patat},
  {Pena}, {Pragt}, {Reinero}, {Rigal}, {Riva}, {Roelfsema}, {Royer}, {Sacco},
  {Santin}, {Schoenmaker}, {Spano}, {Sweers}, {Ter Horst}, {Tintori}, {Tromp},
  {van Dael}, {van der Vliet}, {Venema}, {Vidali}, {Vinther}, {Vola},
  {Winters}, {Wistisen}, {Wulterkens}, \& {Zacchei}}]{Vernet2011}
{Vernet} J. {et~al.}, 2011, \aap, 536, A105

\bibitem[{{Villar-Mart{\'{\i}}n} {et~al}\mbox{.}(2011){Villar-Mart{\'{\i}}n},
  {Humphrey}, {Delgado}, {Colina}, \& {Arribas}}]{Villar2011a}
{Villar-Mart{\'{\i}}n} M., {Humphrey} A., {Delgado} R.~G., {Colina} L.,
  {Arribas} S., 2011, \mnras, 418, 2032

\bibitem[{{Weedman} {et~al}\mbox{.}(2012){Weedman}, {Sargsyan}, {Lebouteiller},
  {Houck}, \& {Barry}}]{Weedman2012}
{Weedman} D., {Sargsyan} L., {Lebouteiller} V., {Houck} J., {Barry} D., 2012,
  \apj, 761, 184

\bibitem[{{Westmoquette} {et~al}\mbox{.}(2012){Westmoquette}, {Clements},
  {Bendo}, \& {Khan}}]{Westmoquette2012}
{Westmoquette} M.~S., {Clements} D.~L., {Bendo} G.~J., {Khan} S.~A., 2012,
  \mnras, 424, 416

\bibitem[{{Whitaker} {et~al}\mbox{.}(2012){Whitaker}, {van Dokkum}, {Brammer},
  \& {Franx}}]{Whitaker2012}
{Whitaker} K.~E., {van Dokkum} P.~G., {Brammer} G., {Franx} M., 2012, \apjl,
  754, L29

\bibitem[{{Zakamska} \& {Greene}(2014)}]{Zakamska2014}
{Zakamska} N.~L., {Greene} J.~E., 2014, \mnras, 442, 784

\bibitem[{{Zhang} {et~al}\mbox{.}(2011){Zhang}, {Dong}, {Wang}, \&
  {Gaskell}}]{Zhang2011}
{Zhang} K., {Dong} X.-B., {Wang} T.-G., {Gaskell} C.~M., 2011, \apj, 737, 71

\end{thebibliography}
\bibliographystyle{mn2e_mod}

\appendix

\section{Fit to the Urrutia et al. (2012) sample}

In Figure A1, we report the fit to 11 out of 13 objects in the Urrutia et al. (2012) sample. F2M0729+3336 has low S/N, while F2M0825+4726 has double peaked broad lines, that is thought to be due to a disklike BLR seen edge on \citep{Eracleous1994}, but the spectrum doesn't show  very broad H$\beta$ and H$\alpha$ lines, or could also be from two separate AGNs close to merger. However, HST ACS image doesn't show multiple nuclei \citep{Urrutia2008}. These 2 objects are not included in the analysis.

The model is the same as descibed in Section 4.2, but limited to the [OIII] lines. In this case we fixed the redshift at the value given in the \citet{Urrutia2012} paper. Therefore, the rest-frame wavelength reported in Table~\ref{tab_app} with respect to the ``S'' component may be different from the rest frame wavelength of the [OIII]5007 line. Given the better S/N in these spectra, the minimum number of Gaussian components required to fit [OIII]$\lambda\lambda4959,5007$ emission lines ranges from 1 to 4, with four objects (44\%) requiring 3, three objects (33\%) requiring 2, two objects requiring 1 (F2M11135+1244) and 4 (F2M0825+4716) components for an adequate fit. 
In cases where H$\beta$ and [OIII]$\lambda\lambda 4959,5007$ are blended with broad permitted FeII emission lines (indicative of the presence of these, may be the blue [$\lambda\lambda 4400-4700$\AA] and red [$\lambda\lambda 5150-5400$\AA] bumps of the FeII multiplets F, S ang G in the spectra as indicated by \citealt{Kovacevic2010}) the FeII emission was fitted .
Since all FeII lines probably originate in the same region, with same kinematical properties, values of the relative shift and FWHM are the same for the FeII lines; intensities are assumed to be different \citep{Kovacevic2010}.

\begin{table*}
\begin{minipage}{170mm}
\footnotesize
\caption{Fit results of the Urrutia et al. (2012) sample }
\begin{tabular}{lcccccccccccccccc}
\hline
ID & specz & $\lambda$,S &  Flux(S) & FWHM(S) & $\lambda$,B & Flux(B)  & FWHM(B) &  $\Delta$v \\ 
&  &  \AA &   (10$^{-17}$) & km s$^{-1}$ & \AA & (10$^{-17}$) & km s$^{-1}$ & km s$^{-1}$  \\ 
\hline
%
083011+37*& 0.414 &$5005.7\pm0.1$ & $624\pm3$& $285\pm2$&$5003.9\pm0.1$ &$592\pm2$ &$1468\pm11$ & -110 \\
%
083413+35 & 0.470 & $5007.1\pm0.2$&$22\pm1$&$500\pm100$& $5007.9\pm0.1$&$114\pm1$&$1191\pm30$& 50 \\
%
084104+36 & 0.555 & $5012.1\pm1.2$&$2.6\pm0.7$&$594\pm172$&$5001.1\pm1.6$&$11.8\pm0.7$&$1091\pm150$&  -650 \\
%
091501+24*& 0.843& $5010.7\pm0.1$&$178\pm1$&$500\pm100$&$5004.4\pm0.1$ &$1170\pm1$ &$1470\pm3$&   -380 \\ 
111354+12 & 0.681& $5007.6\pm0.1$& $208\pm3$& $640\pm24$& ... & ... & ... & ...\\
%
111811-00*&0.686&$5008.0\pm0.2$&$16\pm1$&$388\pm30$&$5000.9\pm0.4$&$111\pm1$&$1311\pm40$&  -420 \\
%
115152+53*& 0.780 & $5008.0\pm0.1$&$884\pm6$&$500\pm100$&$5006.0\pm0.1$& $3060\pm11$&$1554\pm17$&  -140 \\
%
165647+38& 0.732&$5006.3\pm0.7$&$111\pm27$& $304\pm125$& $5009.1\pm1.2$& $614\pm 33$&$900\pm180$& 170 \\
1012+2825  &  0.937 &  $5010.2\pm0.3$ &   $6.2\pm0.4$ & $520\pm40$  & 
$5003.6\pm0.2$ & $54.4\pm1.5$ &  $1440\pm30$ & -156  \\
1507+3129 &  0.988  & $ 5007.5 \pm0.2$ &  $  17.5\pm  1.4$& $  520 \pm  15 $ &  $   5005.9 \pm0.3$ & $  43 \pm 3 $  & $ 1233 \pm 60$&-96\\
1532+2415 &  0.564 & $5006.4\pm0.1$  & $6.5\pm1.8$  & $510\pm35$   & $5006.7\pm0.2$ & $921\pm 80$ & $7.4\pm1.8$ &18\\ 
\hline

\end{tabular}
\hspace{3.2cm}
Fluxes are in units of 10$^{-17}$ erg cm$^{-2}$ s$^{-1}$. Wavelengths are given in the rest-frame. $\lambda$(S), FWHM(S) and Flux(S) denote the best fit parameters and errors for the ``systemic'' (narrow) component; $\lambda$(B), FWHM(B) and Flux(B) instead refer to the ``broad'' (shifted)  component.
Objects marked with * required additional components for the line fit (see text for details and Figure~\ref{fit_urrutia}). In this case we report the component we most likely associate to a outflow. 
 $\Delta$v is measured from the difference in centroids of the 2 measured components. 

\end{minipage}
\label{tab_app}
\end{table*}

\begin{figure*}
\includegraphics[angle=0,scale=0.42]{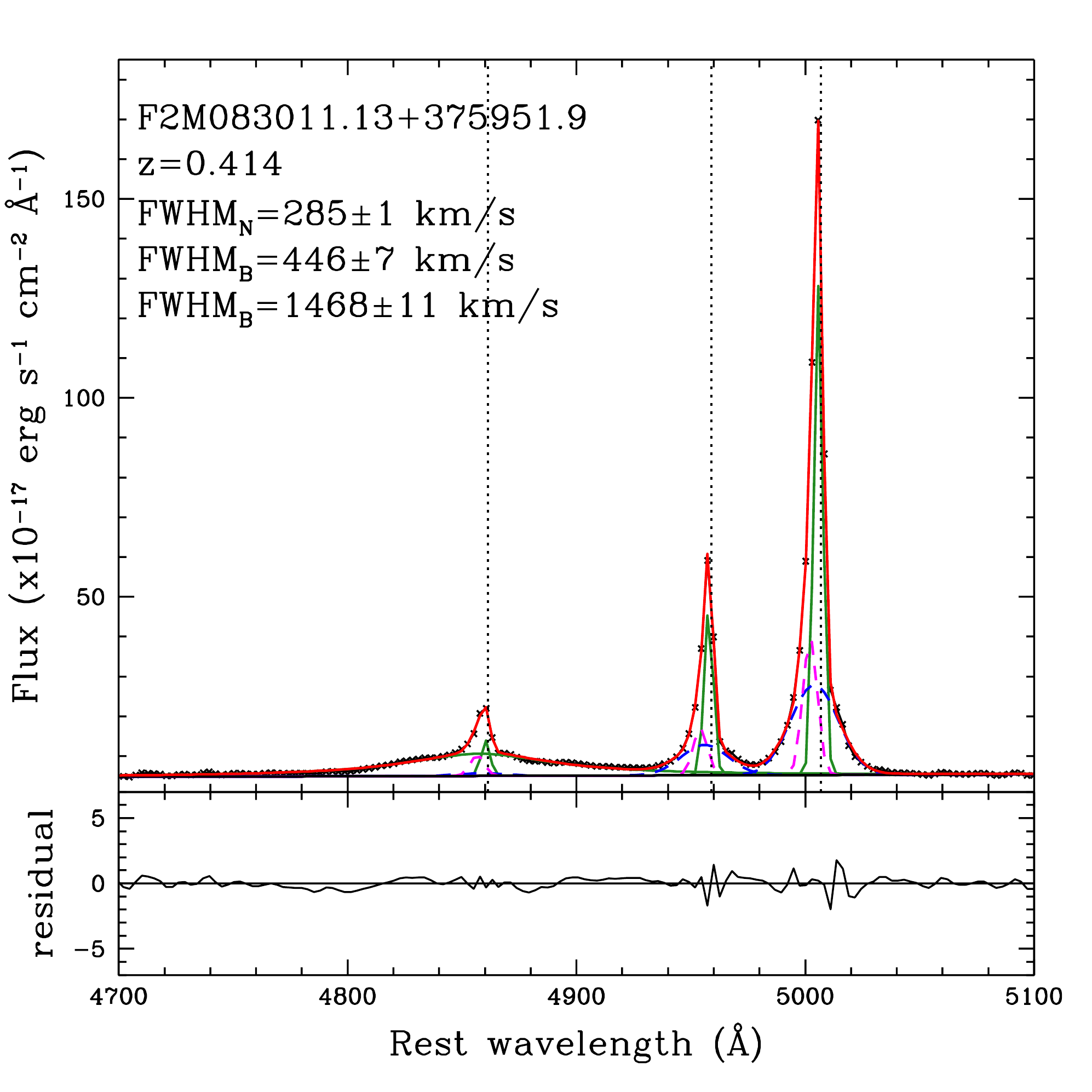}
\includegraphics[angle=0,scale=0.42]{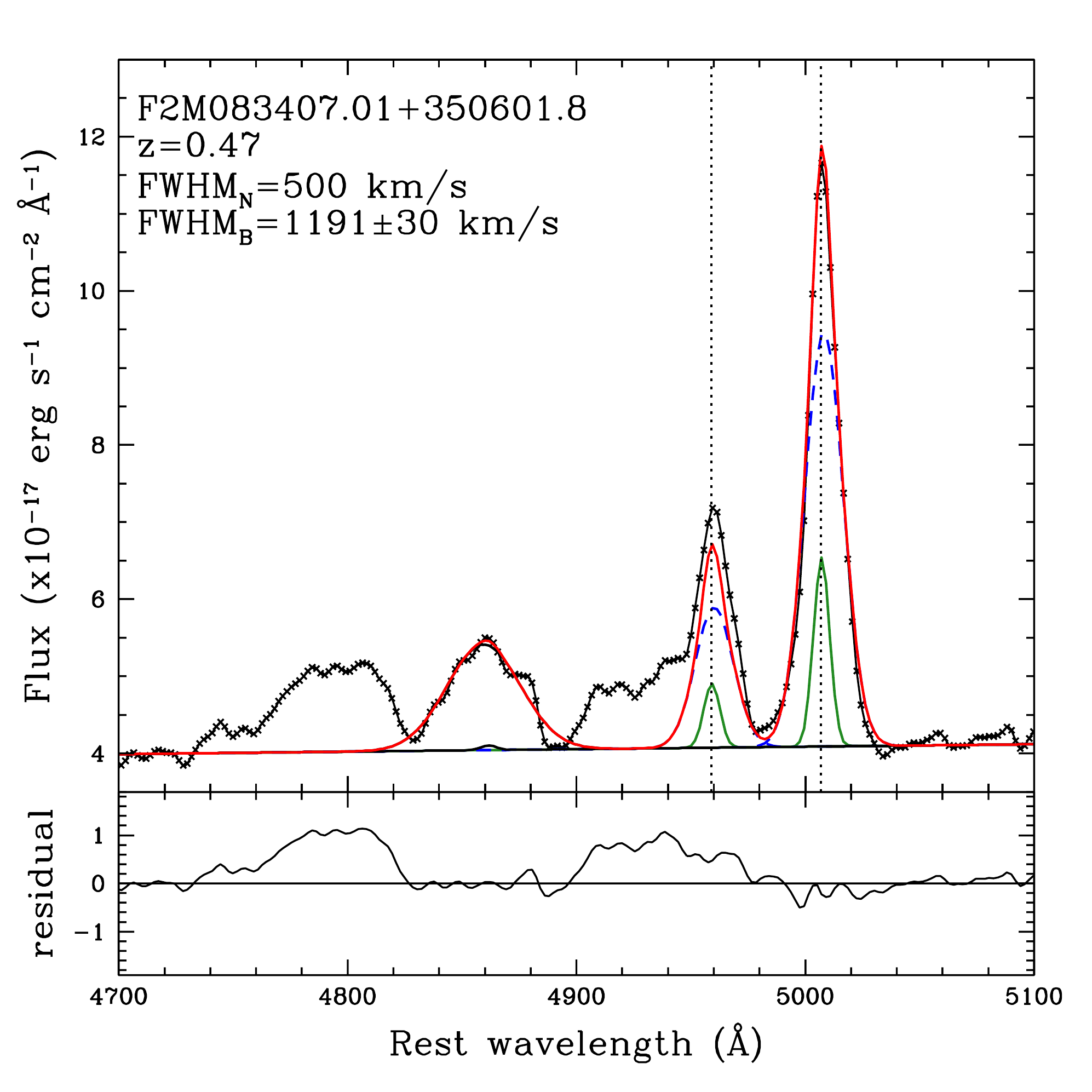}
\includegraphics[angle=0,scale=0.42]{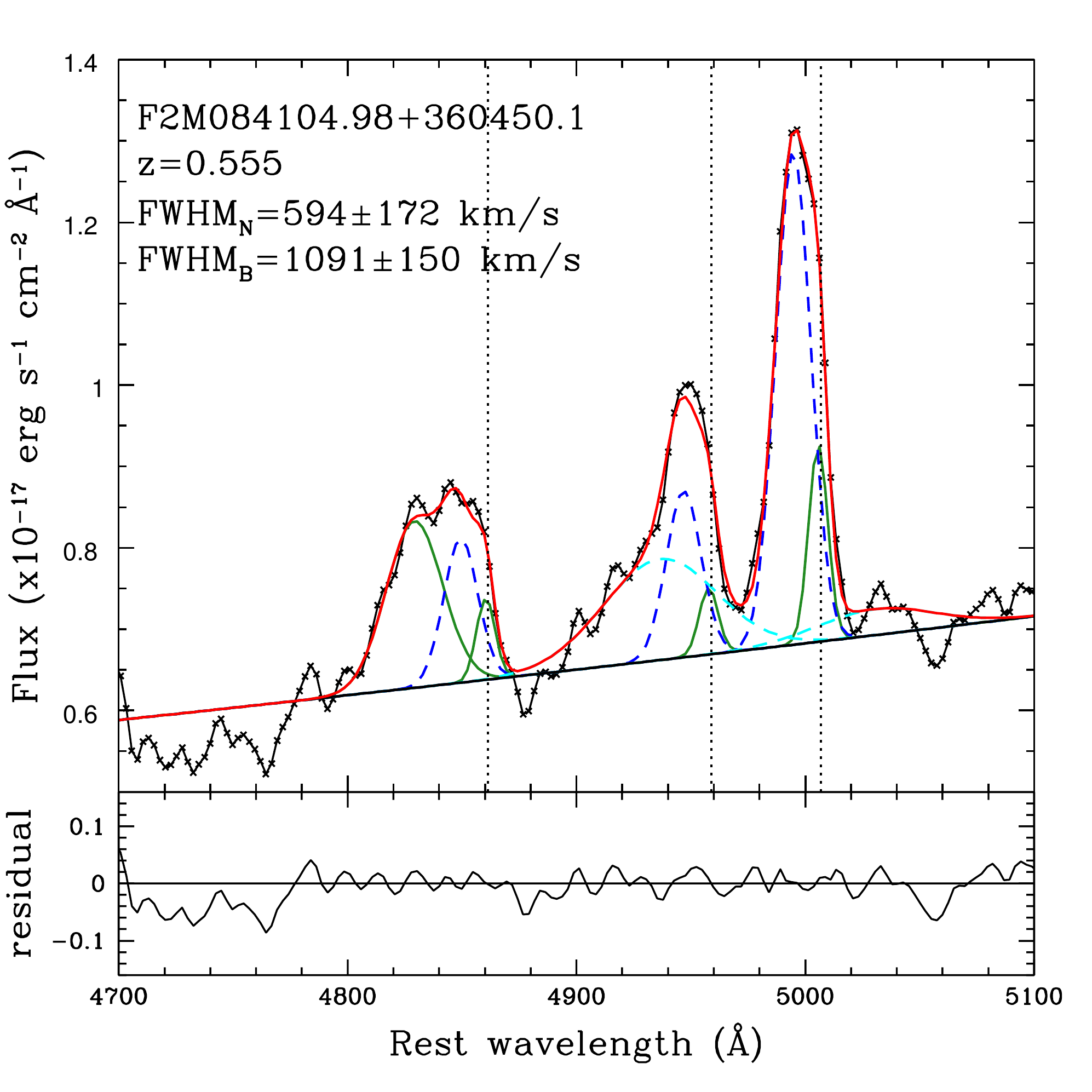}
\includegraphics[angle=0,scale=0.42]{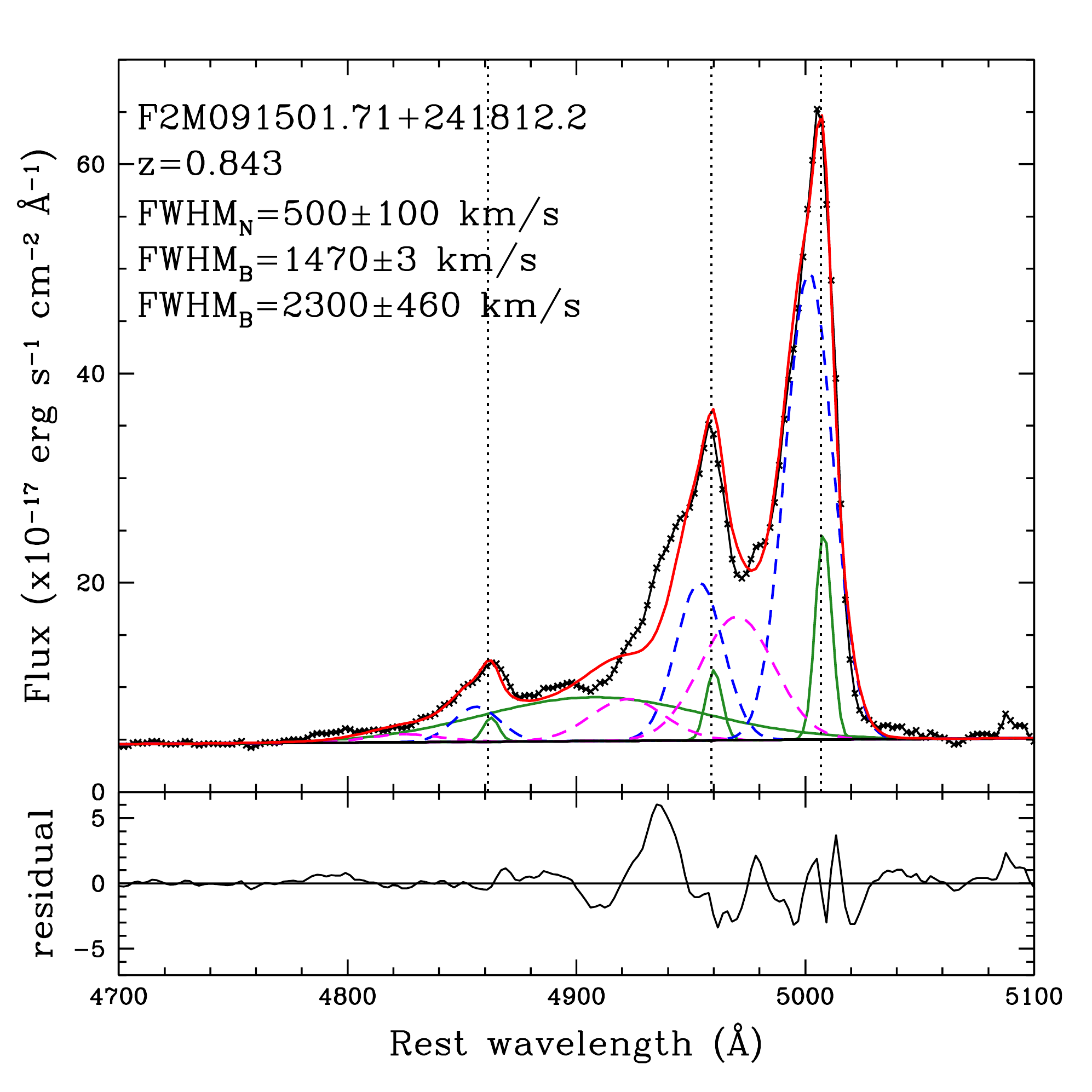}
\caption{ Fits to the $H\beta+[OIII]$ emission lines profile of all the 12 objects in the Urrutia et al. (2012) sample, with the exception of F2M0729+3336 (low S/N).  Solid (green) curves represent the systemic component, dashed blue, magenta, and gold curves the broad, shifted components. The red solid curve shows the sum of all components, and the best fit to the data. In the bottom panel of each fit the residual (differences between data and model) with respect to the best fit are shown. In cases where the lines are blended with prominent broad permitted FeII emission lines (F2M0841+3604,F2M1113+1244,F2M1118-0033), dashed cyan curves displayed represents the fitted FeII components. \label{fit_urrutia}}
\end{figure*}

\begin{figure*}
\includegraphics[angle=0,scale=0.42]{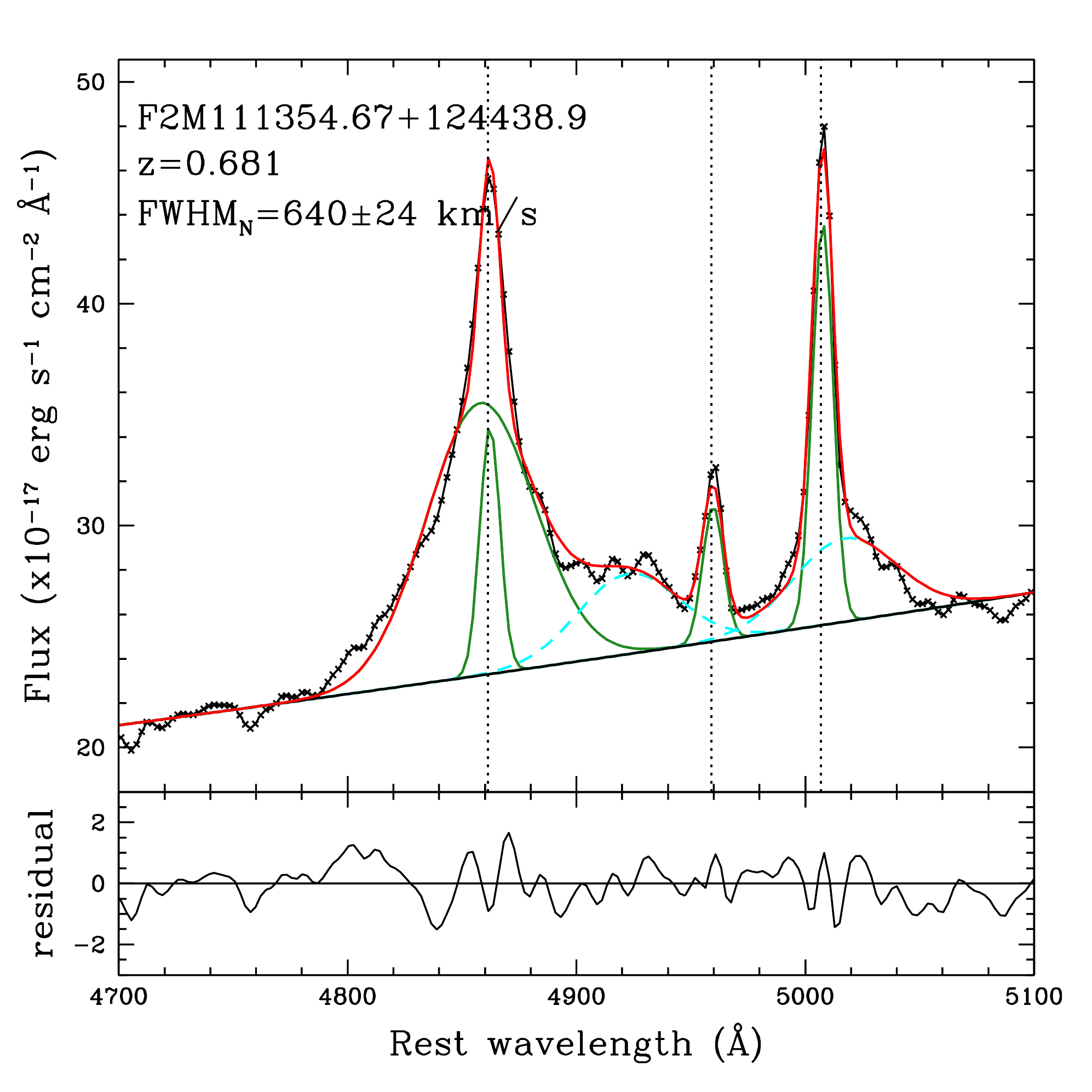}
\includegraphics[angle=0,scale=0.42]{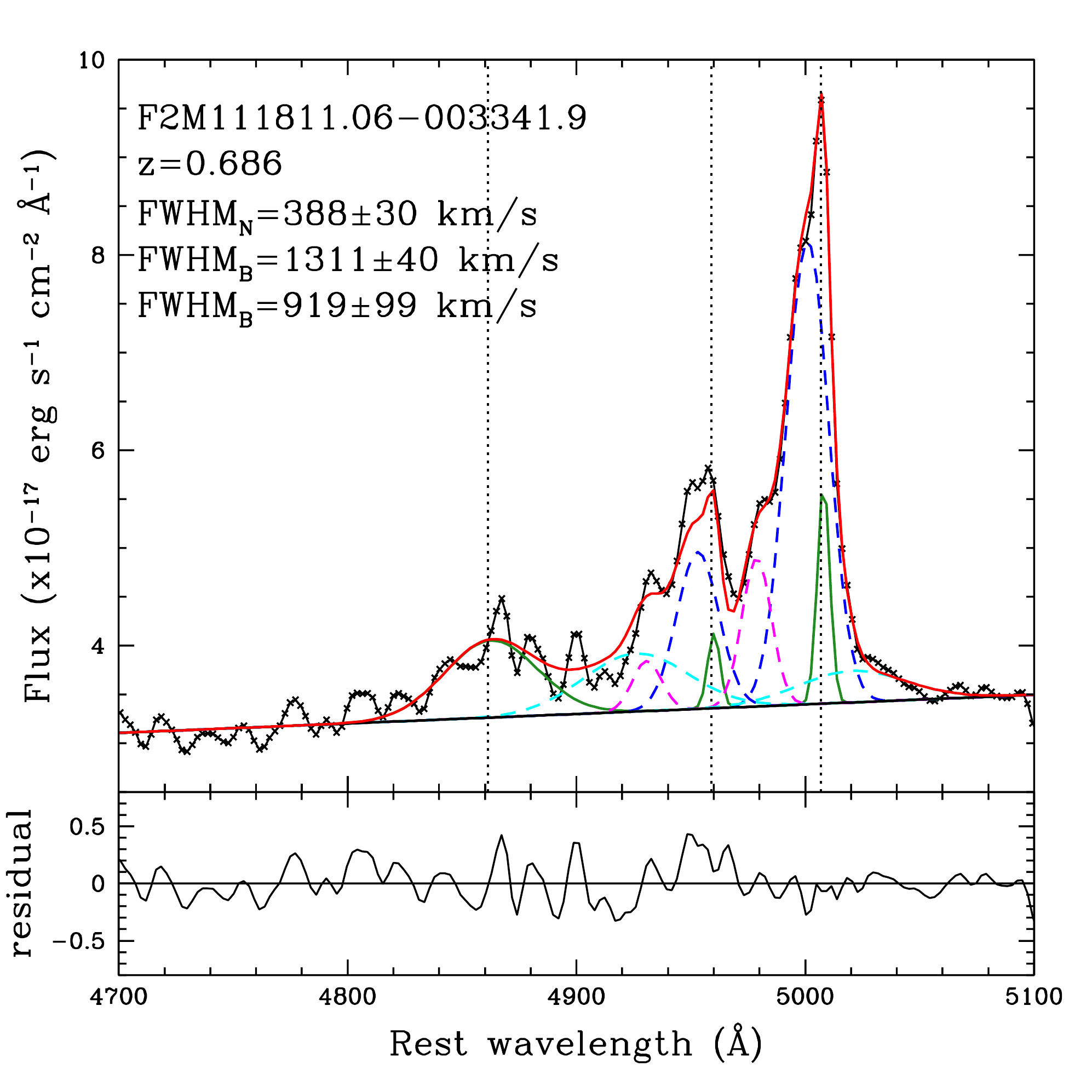}
\includegraphics[angle=0,scale=0.42]{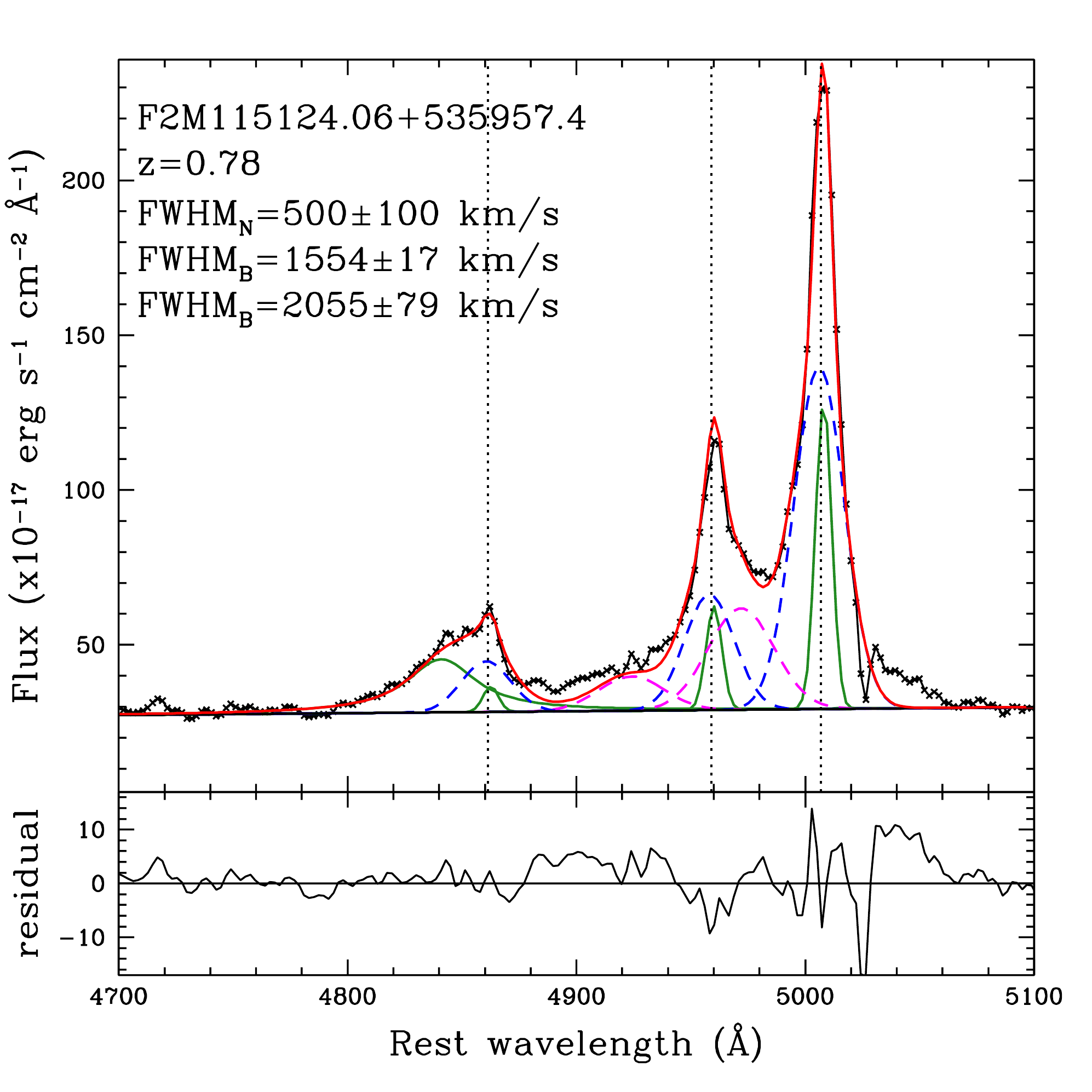}
\includegraphics[angle=0,scale=0.42]{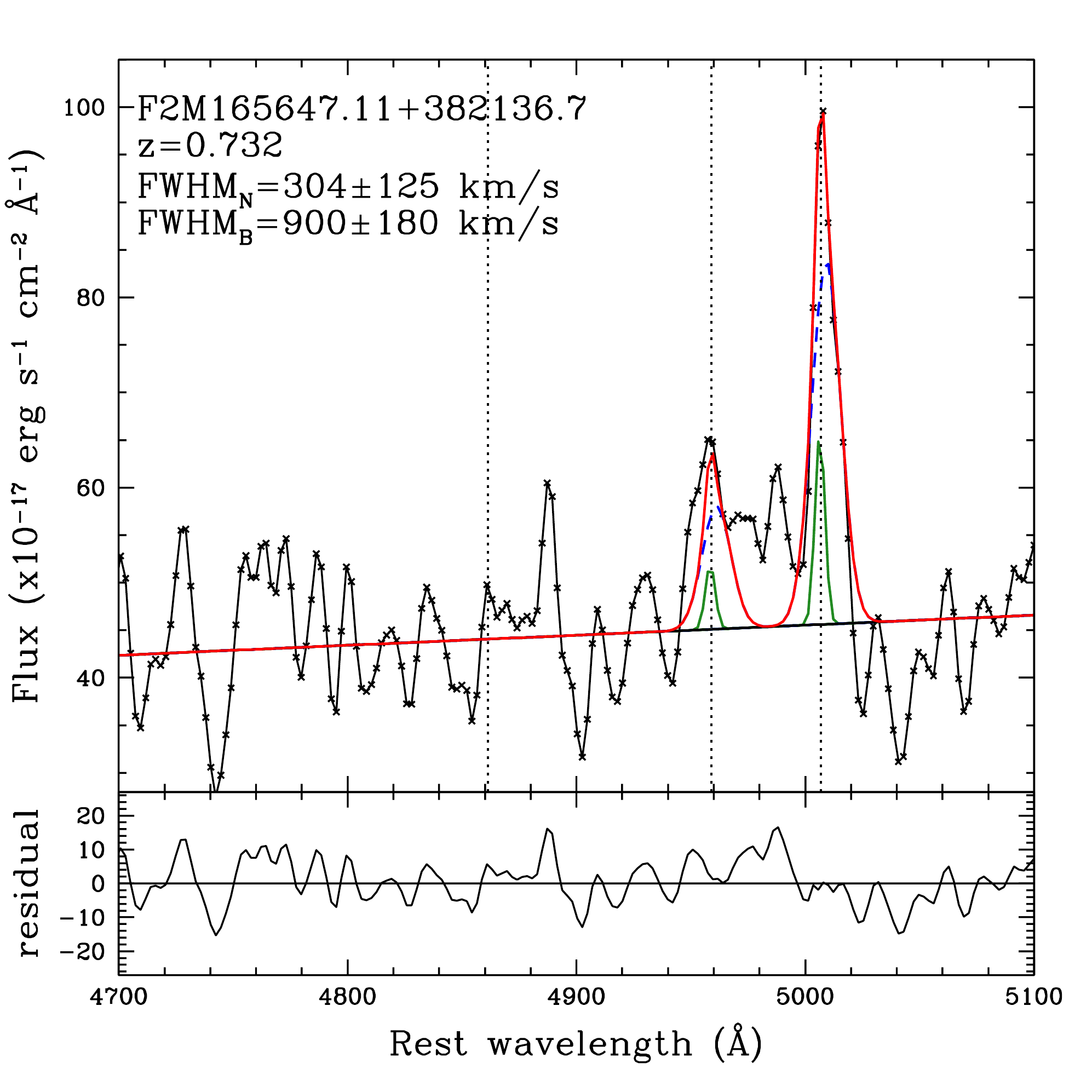}
\contcaption{ }
\end{figure*}

\begin{figure*}
\includegraphics[angle=0,scale=0.42]{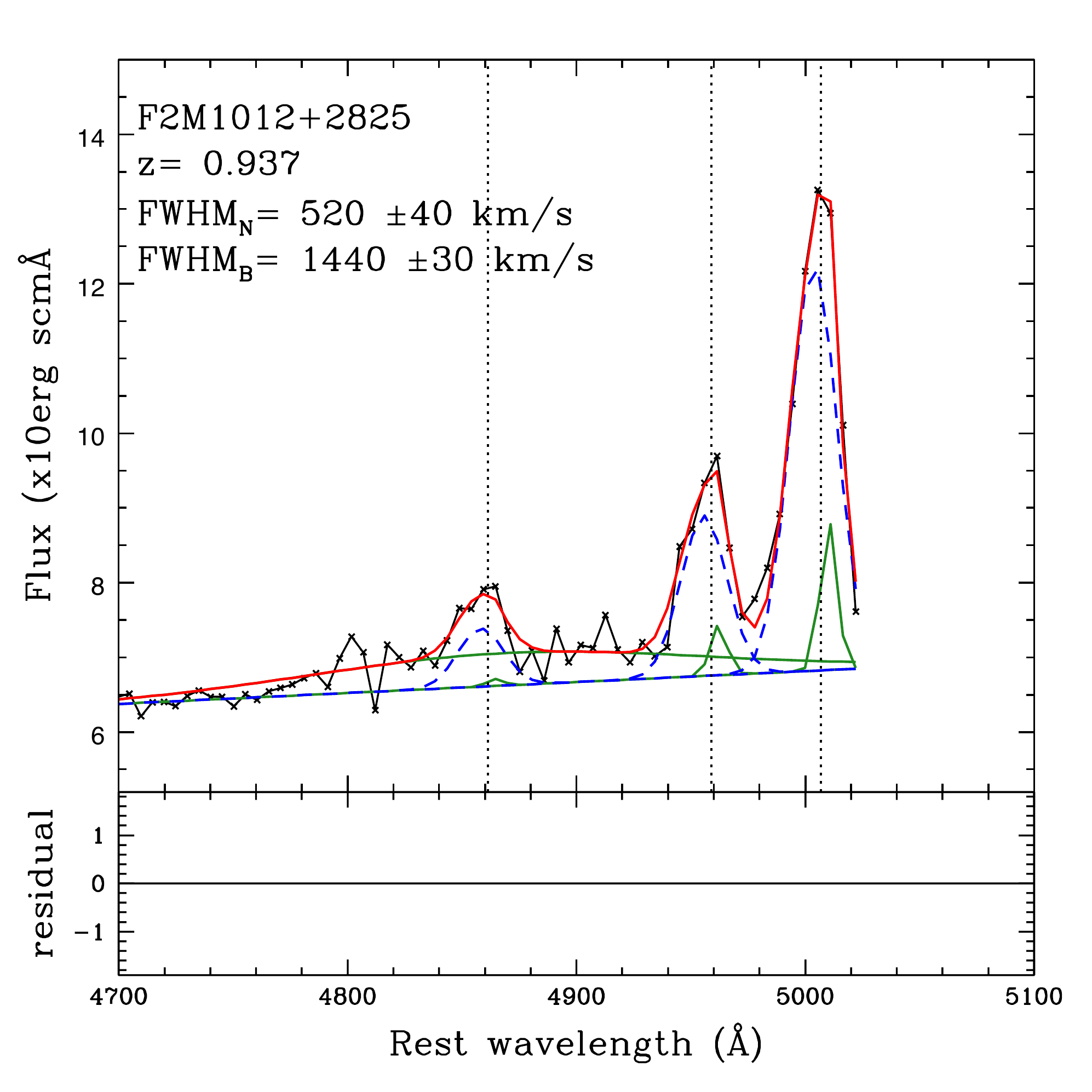}
\includegraphics[angle=0,scale=0.42]{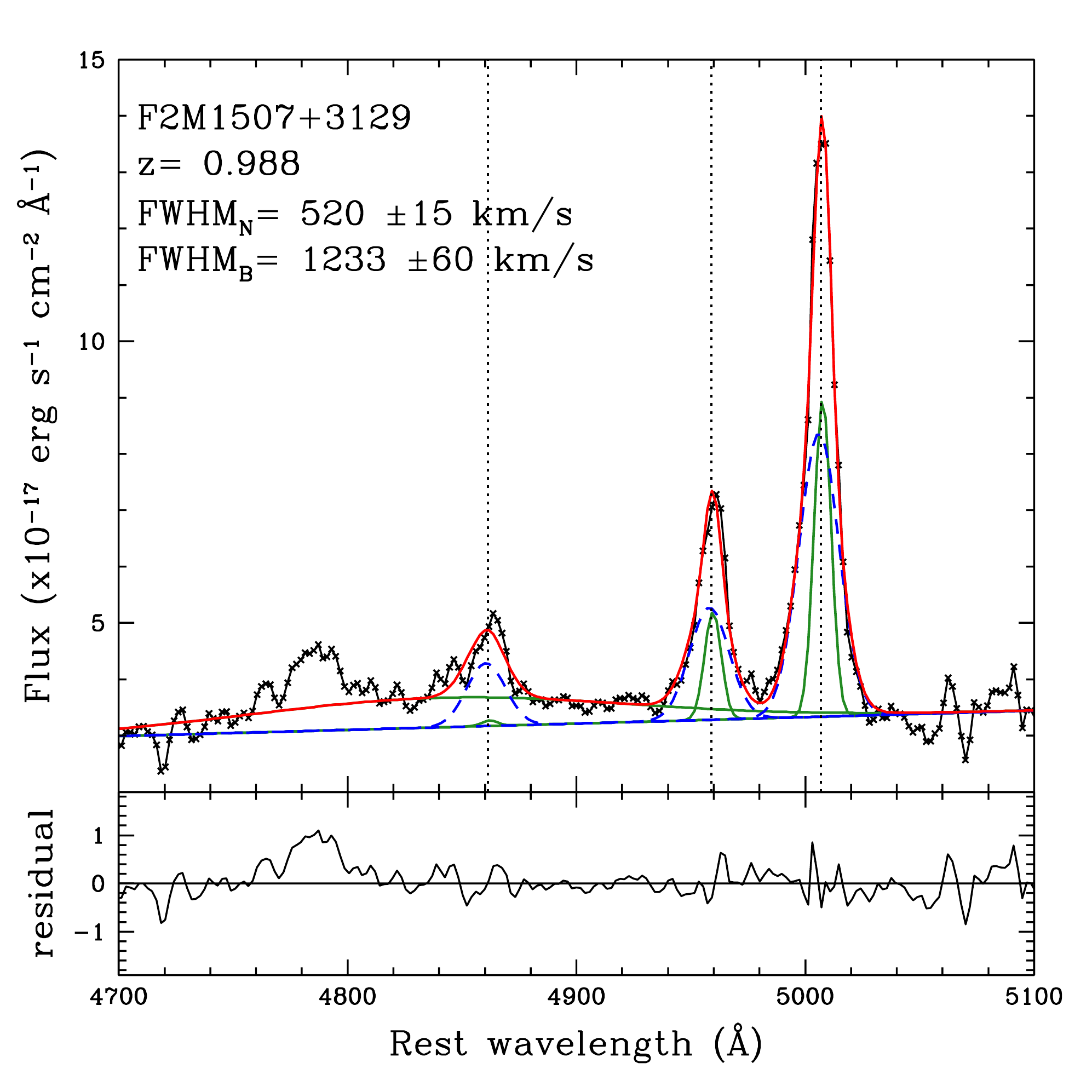}
\includegraphics[angle=0,scale=0.42]{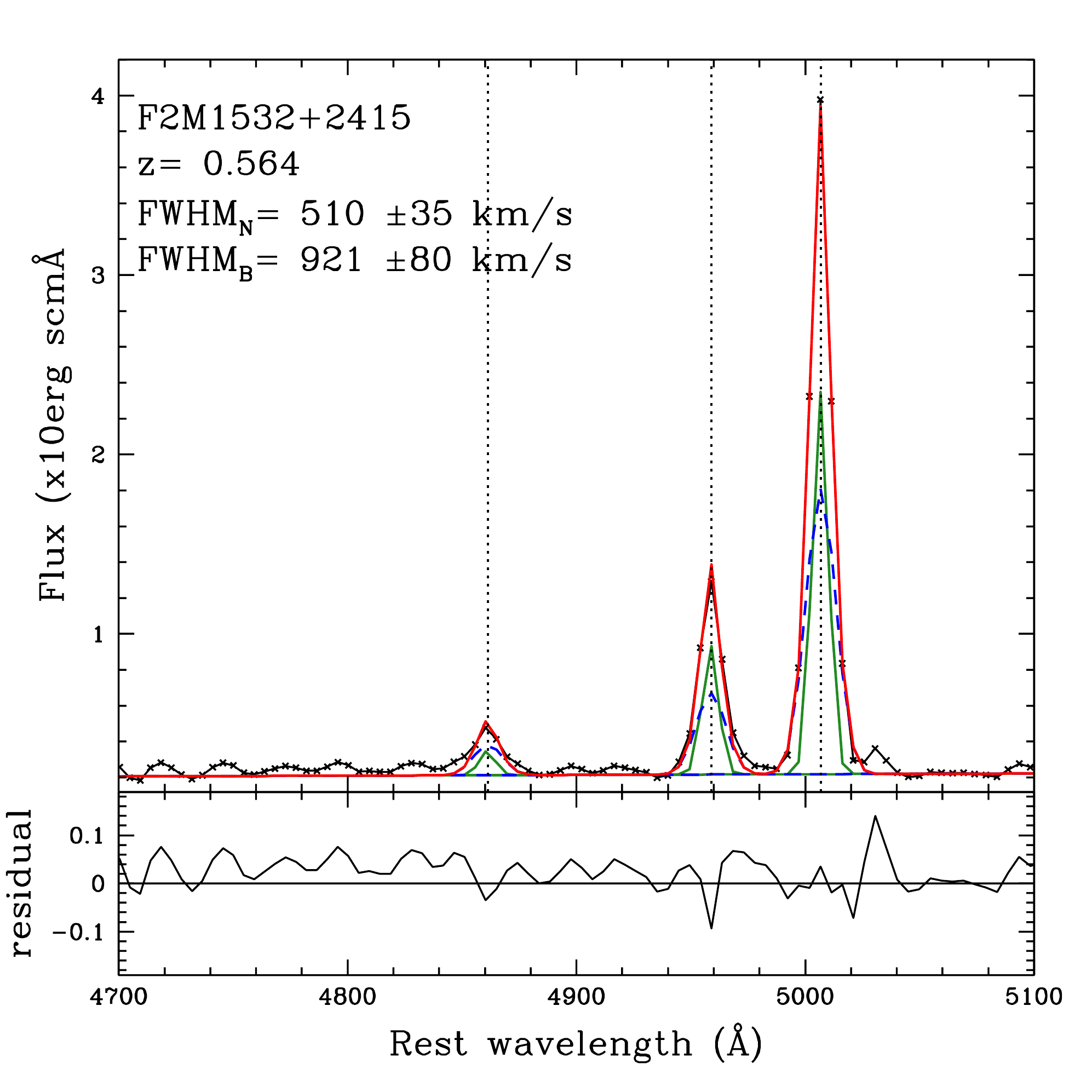}
\contcaption{ }
\end{figure*}

\end{document}